\documentclass[amsmath,amssymb]{revtex4}
\usepackage{graphicx}
\usepackage{amscd,amsmath,amsthm,amsfonts,amssymb}

\def\[{\begin{equation}}
\def\]{\end{equation}}

\begin{document}
\title{General rogue waves in the three-wave resonant interaction systems}
\author{Bo Yang, Jianke Yang}
\affiliation{Department of Mathematics and Statistics, University of Vermont, Burlington, VT 05405, U.S.A}

\begin{abstract}
General rogue waves in (1+1)-dimensional three-wave resonant interaction systems are derived by the bilinear method. These solutions are divided into three families, which correspond to a simple root, two simple roots and a double root of a certain quartic equation arising from the dimension reduction respectively. It is shown that while the first family of solutions associated with a simple root exist for all signs of the nonlinear coefficients in the three-wave interaction equations, the other two families of solutions associated with two simple roots and a double root can only exist in the so-called soliton-exchange case, where the nonlinear coefficients have certain signs. Many of these rogue wave solutions, such as those associated with two simple roots, the ones generated by a $2\times 2$ block determinant in the double-root case, and higher-order solutions associated with a simple root, are new solutions which have not been reported before. Technically, our bilinear derivation of rogue waves for the double-root case is achieved by a generalization to the previous dimension reduction procedure in the bilinear method, and this generalized procedure allows us to treat roots of arbitrary multiplicities. Dynamics of the derived rogue waves is also examined, and new rogue-wave patterns are presented. Connection between these bilinear rogue waves and those derived earlier by Darboux transformation is also explained.
\end{abstract}

\maketitle

\section{Introduction}
Three-wave interaction is a common phenomenon in water waves, nonlinear optics, plasma physics and other nonlinear physical systems \cite{Bloembergen1965,Benney_Newell,Kaupreview1979,Ablowitz_book,3wave_wateropt1,3wave_wateropt2,3wave_wateropt3,3wave_wateropt4,3wave_wateropt5}. When the wavenumbers and frequencies of the three waves form a resonant triad (i.e., exact phase matching), this interaction is the strongest. In this case, the governing equations for this interaction are integrable \cite{Zakharov1973,Ablowitz_Haberman1975,Zakharov1975,Zakharov1976,Kaup1976,Kaup1980,Kaup1981,Kaup1981b}. As a consequence, multi-solitons in one spatial dimension and multi-lumps in two spatial dimensions of this system have been derived \cite{Zakharov1975,Zakharov1976,Kaup1976,Craik1978,Ablowitz_book,Zakharov_book,Kaup1981b,Gilson1998,Yang_Stud,Yang_JMP}.

In the past decade, rogue waves attracted a lot of attention in the physical and mathematical communities
\cite{Ocean_rogue_review,Pelinovsky_book,Solli_Nature,Wabnitz_book}. These waves are large and spontaneous local excitations that ``come from nowhere and disappear with no trace" \cite{Akhmediev_2009}. In oceanography, rogue waves are a threat to ships and even ocean liners. In optics, rogue waves can induce pulse compression. Thus, understanding of rogue waves is clearly desirable. If a nonlinear wave system is integrable, its rogue waves would admit explicit analytical expressions. Because of this, rogue waves have been derived in a large number of integrable equations, such as the nonlinear Schr\"odinger (NLS) equation \cite{Peregrine,AAS2009,DGKM2010,ACA2010,KAAN2011,GLML2012,OhtaJY2012}, the derivative NLS equations \cite{KN_rogue_2011,KN_rogue_2013,CCL_rogue_Chow_Grimshaw2014,CCL_rogue_2017}, the Manakov equations \cite{BDCW2012,ManakovDark,LingGuoZhaoCNLS2014,Chen_Shihua2015,ZhaoGuoLingCNLS2016}, the Davey-Stewartson equations \cite{OhtaJKY2012,OhtaJKY2013}, and many others \cite{AANJM2010,OhtaJKY2014,ASAN2010,Chow,MuQin2016,ClarksonDowie2017,YangYang2019Nonloc,JCChen2018LS,XiaoeYong2018}.
These explicit solutions of rogue waves significantly enhance our understanding of rogue-wave phenomena in the physical systems governed by the underlying integrable equations. Indeed, rogue-wave predictions based on these analytical solutions have been confirmed in both water-wave and optics experiments \cite{Tank1,Tank2,Fiber1,Fiber2,Fiber3}.

Rogue waves in the three-wave resonant interaction systems have also received a fair amount of investigation, all by Darboux transformation \cite{BaroDegas2013,DegasLomba2013,ChenSCrespo2015,WangXChenY2015,ZhangYanWen2018}. In \cite{BaroDegas2013,DegasLomba2013,ZhangYanWen2018}, fundamental rogue waves for double and triple eigenvalues of the scattering matrix were explicitly calculated. In \cite{ChenSCrespo2015}, second-order rogue waves for triple eigenvalues of the scattering matrix were presented. In \cite{WangXChenY2015}, higher-order rogue waves for triple eigenvalues of the scattering matrix were derived. However, many other rogue solutions in the three-wave systems have been missed, such as the ones arising from two double eigenvalues of the scattering matrix, and the ones generated by a $2\times 2$ block determinant for a triple eigenvalue of the scattering matrix. Thus, a full picture of rogue wave solutions in the three-wave resonant interaction systems is still lacking.

From the point of view of mathematical methodology, earlier studies of rogue waves on these three-wave systems all used Darboux transformation. It is known that the bilinear method can produce rogue-wave expressions that are more explicit and compact. But this bilinear rogue derivation has not been done on the three-wave systems yet. In particular, what are the counterparts of double- and triple-eigenvalue rogue waves of Darboux transformation in the bilinear framework and how to derive them bilinearly has remained an intriguing question.

In this article, we derive general rogue waves in three-wave resonant interaction systems by the bilinear method, and our solutions are presented  as determinants with Schur-polynomial matrix elements. These rogue waves are divided into three families, which correspond to a simple root, two simple roots and a double root of a certain quartic equation arising from the dimension reduction respectively. We show that these three families of bilinear rogue waves are the counterparts of rogue waves for a double eigenvalue, two double eigenvalues and a triple eigenvalue of the scattering matrix in Darboux transformation respectively. Among these rogue waves, the ones associated with two simple roots, the ones generated by a $2\times 2$ block determinant for a double root, and the higher-order solutions associated with a simple root are new solutions which have not been reported before. We also show that while the first family of solutions for a simple root exist for all signs of the nonlinear coefficients in the three-wave interaction equations, the other two families of solutions for two simple roots and a double root can only exist in the so-called soliton-exchange case, where the nonlinear coefficients have certain signs. Technically, we find that the bilinear derivation of rogue waves for a double root requires a nontrivial generalization of the previous bilinear method, and our generalization makes it clear how to treat roots of arbitrary multiplicities should they arise during the dimension reduction in other integrable systems. Dynamics of the derived rogue waves is also examined, and new rogue-wave patterns are reported.

The outline of the paper is as follows. In Sec. 2, we introduce three-wave interaction systems and boundary conditions of their rogue wave solutions. In Sec. 3, we present our bilinear rogue wave solutions to these three-wave systems, expressed as determinants with Schur-polynomial elements, and show how our solutions relate to and extend previous rogue solutions derived by Darboux transformation. In Sec. 4, we graphically illustrate these bilinear rogue solutions, and present new rogue patterns. In Sec. 5, we derive the rogue solutions of Sec. 3 by the bilinear Kadomtsev-Petviashvili-hierarchy reduction method. Sec. 6 concludes the paper and highlights the generality of our dimension-reduction procedure for rogue waves in general integrable systems.

\section{Preliminaries} \label{sec:pre}

The general (1+1)-dimensional three-wave resonant interaction system is given by \cite{Kaupreview1979}
\begin{eqnarray}
&& \left(\partial_t+c_1 \partial_x\right)u_1= \epsilon_{1} u_{2}^* u_{3}^*, \nonumber \\
&& \left(\partial_t+c_2 \partial_x\right)u_2= \epsilon_{2} u_{1}^* u_{3}^*,  \label{3WRIModel}\\
&& \left(\partial_t+c_3 \partial_x\right)u_3= \epsilon_{3} u_{1}^* u_{2}^*, \nonumber
\end{eqnarray}
where $(c_1, c_2, c_3)$ are group velocities of the three waves, $(\epsilon_1, \epsilon_2, \epsilon_3)$ are real-valued nonlinear coefficients, and the asterisk `*' represents complex conjugation. To remove ambiguity, we order the three group velocities as $c_{1} > c_{2} > c_{3}$, and make $c_{3}=0$ by choosing a coordinate system that moves with velocity $c_3$. The nonlinear coefficients $\epsilon_n$ can be normalized to $\pm 1$ by variable scalings. In addition, we can fix $\epsilon_{1}=1$ without loss of generality.

This interaction system (with $c_3=0$) is invariant under the gauge transformation
\begin{eqnarray}
&& u_{1}(x,t) \to  u_1(x,t) \hspace{0.07cm} e^{{\rm{i}} (kx -kc_1 t)},  \nonumber \\
&& u_{2}(x,t) \to u_2(x,t)  \hspace{0.07cm} e^{{\rm{i}} [-(kc_1/c_2)x +kc_1t]},   \label{Gauge}\\
&& u_{3}(x,t) \to u_3(x,t)  \hspace{0.07cm} e^{-{\rm{i}} (k-kc_1/c_2)x},   \nonumber
\end{eqnarray}
where $k$ is an arbitrary real constant. In addition, it is invariant under the phase transformation
\begin{eqnarray}
&& u_{k}(x,t) \to  u_k(x,t) \hspace{0.07cm} e^{{\rm{i}} \theta_k},  \quad k=1, 2, 3, \label{Phase}
\end{eqnarray}
where $\theta_3=-(\theta_1+\theta_2)$, and $\theta_1$, $\theta_2$ are arbitrary real constants. These two invariances can help us reduce free parameters in the system, as we will see below.

There are three types of three-wave interaction models, which are termed the soliton exchange case, the explosive case, and the stimulated backscatter case in Ref. \cite{Kaupreview1979}. These three cases correspond to the following signs of the nonlinear coefficients,
\begin{eqnarray}
&& (\epsilon_1, \epsilon_2, \epsilon_3)=(1, -1, 1), \hspace{0.34cm} (\mbox{soliton-exchange case}) \label{solexchange} \\
&& (\epsilon_1, \epsilon_2, \epsilon_3)=(1, 1, 1),  \hspace{0.55cm} (\mbox{explosive case}) \label{explosive} \\
&& (\epsilon_1, \epsilon_2, \epsilon_3)=(1, -1, -1),  \hspace{0.15cm} (\mbox{stimulated backscatter case}) \label{scatter1} \\
&& (\epsilon_1, \epsilon_2, \epsilon_3)=(1, 1, -1). \hspace{0.42cm} (\mbox{stimulated backscatter case}) \label{scatter2}
\end{eqnarray}
Note that the $(1, -1, -1)$ case can be converted to the $(1, 1, -1)$ case by flipping the sign of $x$, reordering the $(u_1, u_2, u_3)$ equations in decreasing order of their group velocities, and renormalizing the nonlinear coefficients; thus these two cases belong to the same stimulated backscatter case. In this article, we will treat all these cases by allowing $(\epsilon_1, \epsilon_2, \epsilon_3)$ to be arbitrary real parameters.

The above three-wave interaction system (\ref{3WRIModel}) admits plane wave solutions
\begin{eqnarray}
&& u_{1,0}(x,t)= \rho_{1}  e^{{\rm{i}} (k_{1}x + \omega_{1} t)},   \nonumber \\
&& u_{2,0}(x,t)= \rho_{2}  e^{{\rm{i}} (k_{2}x + \omega_{2} t)},   \label{PlanewaveSolu}\\
&& u_{3,0}(x,t)= {\rm{i}}\rho_{3}  e^{-{\rm{i}} [ (k_1+k_2)x +(\omega_1+\omega_2) t]},  \nonumber
\end{eqnarray}
where $(k_1, k_2)$ and $(\omega_1, \omega_2)$ are the wavenumbers and frequencies of the first two waves, and $(\rho_1, \rho_2, \rho_3)$ are the complex amplitudes of the three waves. Parameters of these plane waves satisfy the following relations,
\begin{eqnarray}
&& \rho_{1} \left( \omega_{1} + c_{1}k_{1} \right) = -\epsilon_{1} \rho_{2}^* \rho_{3}^*,  \nonumber \\
&& \rho_{2} \left( \omega_{2} + c_{2}k_{2} \right) = -\epsilon_{2} \rho_{1}^* \rho_{3}^*,  \label{Pararlation}\\
&& \rho_{3} \left( \omega_{1} +\omega_{2} \right) = \epsilon_{3} \rho_{1}^* \rho_{2}^*.  \nonumber
\end{eqnarray}
In this article, we assume $\rho_1$, $\rho_2$ and $\rho_3$ are all non-zero. In view of the phase invariance (\ref{Phase}), we can normalize $\rho_1$ and $\rho_2$ to be real. Then the above relations show that $\rho_3$ is real as well. In addition, the gauge invariance (\ref{Gauge}) allows us to impose a restriction on the four parameters $(k_1, k_2, \omega_1, \omega_2)$, such as fixing one of them as zero, or equating $k_1=k_2$, or equating $\omega_1=\omega_2$, without any loss of generality. Under such a restriction, wavenumber and frequency parameters $(k_1, k_2, \omega_1, \omega_2)$ would be fully determined from the three real background-amplitude parameters $(\rho_1, \rho_2, \rho_3)$ through equations (\ref{Pararlation}).

Rogue waves in the three-wave interaction system (\ref{3WRIModel}) are rational solutions which approach plane-wave solutions (\ref{PlanewaveSolu}) as $x, t\to \pm \infty$. From the above discussions on plane-wave solutions, we can set the boundary conditions for these rogue waves as
\begin{eqnarray}
&& u_{1}(x,t)\rightarrow  \rho_{1}  e^{{\rm{i}} (k_{1}x + \omega_{1} t)},  \hspace{1.8cm}  x, t\to \pm \infty, \nonumber \\
&& u_{2}(x,t)\rightarrow  \rho_{2}  e^{{\rm{i}} (k_{2}x + \omega_{2} t)},  \hspace{1.8cm}  x, t\to \pm \infty, \label{BoundaryCond}\\
&& u_{3}(x,t)\rightarrow  \textrm{i}\rho_{3}  e^{-{\rm{i}} [ (k_1+k_2)x +(\omega_1+\omega_2) t]}, \hspace{0.4cm} x, t\to \pm \infty,   \nonumber
\end{eqnarray}
where $(\rho_1, \rho_2, \rho_3)$ are free real amplitudes, and the other parameters $(k_1, k_{2}, \omega_{1}, \omega_{2})$ are determined by these real amplitudes through equations (\ref{Pararlation}) and an extra restriction on them from the gauge invariance (\ref{Gauge}).

In this article, we will present rogue waves of the three-wave resonant interaction system (\ref{3WRIModel}) through elementary Schur polynomials. These Schur polynomials $S_j(\mbox{\boldmath $x$})$ are defined by
\begin{equation} \label{defSchur}
\sum_{j=0}^{\infty}S_j(\mbox{\boldmath $x$})\lambda^j
=\exp\left(\sum_{i=1}^{\infty}x_i\lambda^i\right),
\end{equation}
or more explicitly,
\begin{equation}
S_0(\mbox{\boldmath $x$})=1, \quad S_1(\mbox{\boldmath $x$})=x_1,
\quad S_2(\mbox{\boldmath $x$})=\frac{1}{2}x_1^2+x_2, \quad \cdots, \quad
S_{j}(\mbox{\boldmath $x$}) =\sum_{l_{1}+2l_{2}+\cdots+ml_{m}=j} \left( \ \prod _{i=1}^{m} \frac{x_{i}^{l_{i}}}{l_{i}!}\right),
\end{equation}
where $\mbox{\boldmath $x$}=(x_1,x_2,\cdots)$.

\section{General rogue wave solutions}

\subsection{Root structure of an algebraic equation} \label{subsection_root}
In our bilinear framework, rogue-wave expressions will depend on the root structure of the following algebraic equation
\[ \label{QurticeqQ1dp}
\mathcal{Q}'_{1}(p)= 0,
\]
where
\[\label{Q1polynomial}
\mathcal{Q}_{1}(p)= \left( \frac{  \gamma_{1} c_{2}}{\gamma_3(c_{2}-c_{1})} \right) \frac{1 }{p} - \left( \frac{\gamma_{2} c_{1}}{\gamma_3(c_{2}-c_{1})}  \right) \frac{1 }{p-{\rm i}} - p,
\]
\[ \label{gamma123}
\gamma_{1} \equiv \epsilon_1 \frac{ \rho_{2} \rho_{3} }{ \rho_{1}} ,\  \gamma_{2} \equiv \epsilon_2 \frac{\rho_{1} \rho_{3} }{ \rho_{2}}, \
\gamma_3\equiv \epsilon_3 \frac{ \rho_{1} \rho_{2} }{ \rho_{3}},
\]
and the prime in $\mathcal{Q}'_{1}(p)$ represents the derivative. This $\mathcal{Q}_{1}(p)$ function and the associated algebraic equation (\ref{QurticeqQ1dp}) will appear in the dimension reduction of our bilinear derivation of rogue waves, which will be explained in more detail in Sec. \ref{sec:dimred}.

The algebraic equation (\ref{QurticeqQ1dp}) can be rewritten as
\[\label{Q1polynomial2}
\gamma_3(c_{1}-c_{2})p^2 (p-{\rm i})^2- \gamma_{1} c_{2} (p-{\rm i})^2 + \gamma_{2} c_{1} p^2=0,
\]
which is a quartic equation for $p$. Thus, it has four roots (counting multiplicity). These roots are dependent on the parameters in the three-wave interaction system (\ref{3WRIModel}) and in the boundary conditions (\ref{BoundaryCond}). Notice that if $p$ is a root, so is $-p^*$. Thus, non-imaginary roots appear as pairs of $(p, -p^*)$. Writing $p={\rm{i}}\tilde{p}$, Eq. (\ref{Q1polynomial2}) becomes a quartic equation for $\tilde{p}$ with real coefficients, whose root structure depends only on the sign of its discriminant
\[ \label{eDelta}
\Delta =-16 c_{1} c_{2} \left(c_{1}-c_{2}\right) \gamma_{1} \gamma_{2} \gamma_{3} \left\{ \left[\gamma_{1}c_{2}+\gamma_{3}(c_{1}-c_{2})-\gamma_{2}c_{1} \right]^3 + 27 c_{1} c_{2} \left(c_{1}-c_{2}\right) \gamma_{1} \gamma_{2} \gamma_{3} \right\}.
\]
Below, we delineate this root structure for the four cases of $(\epsilon_1, \epsilon_2, \epsilon_3)$ values in Eqs. (\ref{solexchange})-(\ref{scatter2}).

(1) In the soliton-exchange case (\ref{solexchange}), $(\epsilon_1, \epsilon_2, \epsilon_3)=(1, -1, 1)$. In this case, it is easy to see that $\gamma_1$ and $\gamma_3$ have the same sign, and $\gamma_2$ has the opposite sign of $(\gamma_1, \gamma_3)$. Then, in view of our velocity arrangement of $c_1>c_2>0$ and the inequality of $(a+b+c)^3\ge 27abc$ for any non-negative real values of $a, b$ and $c$, with the equal sign realized if and only if $a=b=c$, we see that $\Delta \ge 0$, and $\Delta=0$ if and only if
\[ \label{CubicRestrict}
\rho_{2}=\pm \sqrt{\frac{c_{1}}{c_{2}}}\rho_{1},\ \ \ \rho_{3}= \pm \sqrt{\frac{c_{1}-c_{2}}{c_{2}}}\rho_{1}.
\]

When $\Delta=0$, i.e., under the above parameter conditions (\ref{CubicRestrict}), Eq. (\ref{Q1polynomial2}) simplifies to
\[
\frac{1}{p^2}+\frac{1}{(p-{\rm i})^2}-1=0,
\]
whose roots are
\[ \label{roots2}
(\hat{p}_{0}, \hat{p}_{0}, -\hat{p}_{0}^*, -\hat{p}_{0}^*),
\]
where
\[ \label{p0}
\hat{p}_{0}=(\sqrt{3}+\rm{i})/2.
\]
Thus, there is a pair of double roots here.

When $\Delta> 0$, i.e., the parameter conditions (\ref{CubicRestrict}) are not met, there cannot be any repeated root. In addition, Eq. (\ref{Q1polynomial2}) cannot admit any purely-imaginary root, because such a root would make all terms on the left side of Eq. (\ref{Q1polynomial2}) to have the same sign, whose sum cannot be zero. Thus, the root structure in this case is
\[ \label{roots1}
(p_{0,1}, p_{0,2}, -p_{0,1}^*, -p_{0,2}^*),
\]
where $p_{0,1}\ne p_{0,2}$, i.e., there are two pairs of non-imaginary simple roots here.

(2) In the explosive and stimulated backscatter cases with $(\epsilon_1, \epsilon_2, \epsilon_3)$ values given in Eqs. (\ref{explosive})-(\ref{scatter2}), Eq. (\ref{Q1polynomial2}) always admits at least two simple imaginary roots. The reason can be seen by dividing that equation with $p^2 (p-{\rm i})^2$ and setting $p={\rm{i}}\tilde{p}$, which results in a real equation for $\tilde{p}$ with two rational terms and one constant term. By examining the signs of these terms at $\tilde{p}=\pm \infty$ and near the singularities at $\tilde{p}=0$ and $1$, and utilizing the intermediate value theorem, we can readily see that this real $\tilde{p}$ equation has at least two simple real roots, and thus the $p$ equation (\ref{Q1polynomial2}) admits at least two simple imaginary roots.
The nature of the other two roots of $p$ can be obtained by putting $p={\rm{i}}\tilde{p}$ into Eq. (\ref{Q1polynomial2}), which results in a real quartic equation for $\tilde{p}$. Combining the classical results on the root structure of a real quartic equation with the current information of $\tilde{p}$ admitting at least two simple real roots, we see that the nature of the other two roots of $\tilde{p}$ (and hence $p$) depends only on the sign of the discriminant $\Delta$ in Eq. (\ref{eDelta}). Putting these results together, root structures of the $p$ equation (\ref{Q1polynomial2}) in the explosive and stimulated backscatter cases are summarized as follows.
\begin{eqnarray}
&&\Delta>0: \hspace{0.3cm} \mbox{four imaginary simple roots;}   \\
&&\Delta<0: \hspace{0.3cm} \mbox{a pair of non-imaginary simple roots $(p_0, -p_{0}^*)$ and two imaginary simple roots;}  \label{doublecase2} \\
&&\Delta=0: \hspace{0.3cm} \mbox{one imaginary double root and two imaginary simple roots.}
\end{eqnarray}

\subsection{Rogue wave solutions}
Now, we present our general rogue-wave solutions in the three-wave interaction system (\ref{3WRIModel}) according to the root structure of the algebraic equation (\ref{Q1polynomial2}).

\begin{quote}
\textbf{Theorem 1} \hspace{0.05cm} \emph{If the algebraic equation (\ref{Q1polynomial2}) admits a non-imaginary simple root $p_0$, then the three-wave interaction system (\ref{3WRIModel}) under boundary conditions (\ref{BoundaryCond}) admits bounded $N$-th order rogue-wave solutions}
\begin{eqnarray}
  && u_{1,N}(x,t)= \rho_{1}\frac{g_{1,N}}{f_{N}} e^{{\rm{i}} (k_1x+\omega_{1} t)}, \label{Schpolysolu1} \\
  && u_{2,N}(x,t)= \rho_{2}\frac{g_{2,N}}{f_{N}} e^{{\rm{i}} (k_{2}x + \omega_{2} t)}, \label{Schpolysolu2} \\
  && u_{3,N}(x,t)= {\rm{i}}\rho_{3}\frac{g_{3,N}}{f_{N}} e^{-{\rm{i}} [(k_1+k_{2})x + (\omega_{1}+\omega_{2}) t]}, \label{Schpolysolu3}
\end{eqnarray}
\emph{where $N$ is an arbitrary positive integer,}
\[ \label{SchpolysolufN}
f_{N}=\sigma_{0,0}, \quad g_{1,N}=\sigma_{1,0}, \quad  g_{2,N}=\sigma_{0,-1}, \quad g_{3,N}=\sigma_{-1,1},
\]
\[ \label{sigmank1}
\sigma_{n,k}=
\det_{
\begin{subarray}{l}
1\leq i, j \leq N
\end{subarray}
}
\left(
\begin{array}{c}
m_{2i-1,2j-1}^{(n,k)}
\end{array}
\right),
\]
\emph{the matrix elements in $\sigma_{n,k}$ are defined by}
\[ \label{Schmatrimnij}
m_{i,j}^{(n,k)}=\sum_{\nu=0}^{\min(i,j)} \left[ \frac{|p_{1}|^2 }{(p_{0}+p_{0}^*)^2}  \right]^{\nu} \hspace{0.06cm} S_{i-\nu}(\textbf{\emph{x}}^{+}(n,k) +\nu \textbf{\emph{s}})  \hspace{0.06cm} S_{j-\nu}(\textbf{\emph{x}}^{-}(n,k) + \nu \textbf{\emph{s}}^*),
\]
\emph{vectors} $\textbf{\emph{x}}^{\pm}(n,k)=\left( x_{1}^{\pm}, x_{2}^{\pm},\cdots \right)$ \emph{are defined by}
\begin{eqnarray}
&&x_{r}^{+}(n,k)= \left( \alpha_{r} - \beta_{r} \right) x +\left( c_{1}\beta_{r}-c_{2}\alpha_{r} \right)t + n \theta_{r} + k \lambda_{r} + a_{r},  \label{defxrp} \\
&&x_{r}^{-}(n,k)=   \left( \alpha^*_{r} - \beta^*_{r} \right) x +\left( c_{1}\beta^*_{r}-c_{2}\alpha^*_{r} \right)t - n \theta^*_{r} - k \lambda^*_{r} +a^*_{r},  \label{defxrm}
\end{eqnarray}
\emph{$\alpha_{r}$, $\beta_{r}$, $\theta_{r}$ and $\lambda_{r}$ are coefficients from the expansions}
\begin{eqnarray}
&& \frac{ \gamma_{1}}{c_{1}-c_{2}} \left(\frac{1 }{p \left( \kappa \right)} -\frac{1}{p_{0}} \right)= \sum_{r=1}^{\infty} \alpha_{r}\kappa^{r}, \label{schucoefalpha} \\
&& \frac{ \gamma_{2}}{c_{2}-c_{1}} \left(\frac{1}{p \left( \kappa \right)-\rm{i}} -\frac{1}{p_{0}-\rm{i}} \right)= \sum_{r=1}^{\infty} \beta_{r}\kappa^{r},\\
&& \ln \frac{ p \left( \kappa \right)}{p_{0}}  =\sum_{r=1}^{\infty} \lambda_{r}\kappa^{r},  \quad \ln \frac{ p \left( \kappa \right)-\rm{i}}{p_{0}-\rm{i}}  =\sum_{r=1}^{\infty} \theta_{r}\kappa^{r},   \label{schucoeflambda}
\end{eqnarray}
\emph{the vector} $\textbf{\emph{s}}=(s_1, s_2, \cdots)$ \emph{is defined by the expansion}
\begin{equation}
\ln \left[\frac{1}{\kappa} \left(\frac{p_{0}+p_{0}^*}{p_{1}} \right) \left( \frac{ p \left( \kappa \right)-p_{0}}{p \left( \kappa \right)+p_{0}^*} \right)  \right] = \sum_{r=1}^{\infty}s_{r} \kappa^r,  \label{schurcoeffsr}
\end{equation}
\emph{the function $p \left(\kappa\right)$ is defined by the equation
\[\label{defpk}
\mathcal{Q}_{1}\left[p \left( \kappa \right)\right] = \mathcal{Q}_{1}(p_{0}) \cosh(\kappa),
\]
with $\mathcal{Q}_{1}(p)$ given in Eq. (\ref{Q1polynomial}), $p_1\equiv (dp/d\kappa)|_{\kappa=0}$, and $a_{r} \hspace{0.05cm} (r=1, 2, \dots)$ are free complex constants. }
\end{quote}

\begin{quote}
\textbf{Theorem 2} \hspace{0.05cm} \emph{If the algebraic equation (\ref{Q1polynomial2}) admits two non-imaginary simple roots $p_{0,1}$ and $p_{0,2}$ with $p_{0,2}\ne -p_{0,1}^*$, which is only possible in the soliton-exchange case (\ref{solexchange}) with background amplitudes not satisfying conditions (\ref{CubicRestrict}), then the three-wave interaction system (\ref{3WRIModel}) under boundary conditions (\ref{BoundaryCond}) admits bounded $(N_1,N_2)$-th order rogue-wave solutions}
\begin{eqnarray}
  && u_{1, N_1,N_2}(x,t)= \rho_{1}\frac{g_{1, N_1,N_2}}{f_{N_1,N_2}} e^{{\rm{i}} (k_1x+\omega_{1} t)}, \label{Schpolysolu1b} \\
  && u_{2, N_1,N_2}(x,t)= \rho_{2}\frac{g_{2, N_1,N_2}}{f_{N_1,N_2}} e^{{\rm{i}} (k_{2}x + \omega_{2} t)}, \label{Schpolysolu2b} \\
  && u_{3, N_1,N_2}(x,t)= {\rm{i}}\rho_{3}\frac{g_{3, N_1,N_2}}{f_{N_1,N_2}} e^{-{\rm{i}} [(k_1+k_{2})x + (\omega_{1}+\omega_{2}) t]},  \label{Schpolysolu3b}
\end{eqnarray}
\emph{where $N_1, N_2$ are arbitrary positive integers,}
\[ \label{SchpolysolufN2}
f_{N_1,N_2}=\sigma_{0,0}, \quad g_{1, N_1,N_2}=\sigma_{1,0}, \quad  g_{2, N_1,N_2}=\sigma_{0,-1}, \quad g_{3, N_1,N_2}=\sigma_{-1,1},
\]
\emph{$\sigma_{n,k}$ is a $2 \times 2 $ block determinant}
\[ \label{sigmaTheorem2}
\sigma_{n,k} = \det \left( \begin{array}{cc}
                           \sigma_{n,k}^{\left[1,1\right]} & \sigma_{n,k}^{\left[1,2\right]} \\
                           \sigma_{n,k}^{\left[2,1\right]} & \sigma_{n,k}^{\left[2,2\right]}
                         \end{array}
\right),
\]
\[ \label{sigmank2}
\sigma_{n,k}^{\left[I,J\right]} = \left( m_{2i-1,2j-1}^{(n,k,I,J)}   \right)_{1\leq i \le N_{I}, 1\leq j\leq N_{J}},
\]
\emph{the matrix elements in $\sigma_{n,k}^{\left[I,J\right]}$ are defined by}
\[ \label{Schmatrimnij2}
m_{i,j}^{(n,k,I, J)}=\sum_{\nu=0}^{\min(i,j)} \left( \frac{1}{p_{0,I}+p^*_{0,J}} \right) \left[ \frac{p_{1,I} p^*_{1,J} }{(p_{0,I}+p^*_{0,J})^2}  \right]^{\nu} \hspace{0.06cm} S_{i-\nu}\left(\textbf{\emph{x}}^{+}_{I,J}(n,k) +\nu \textbf{\emph{s}}_{I,J}\right)  \hspace{0.06cm} S_{j-\nu}\left(\textbf{\emph{x}}^{-}_{I,J}(n,k) + \nu \textbf{\emph{s}}_{J,I}^*\right),
\]
\emph{vectors} $\textbf{\emph{x}}^{\pm}_{I,J}(n,k)=\left( x_{1,I,J}^{\pm}, x_{2,I,J}^{\pm},\cdots \right)$ \emph{and}  $\textbf{\emph{s}}_{I,J}=\left( s_{1,I,J}, s_{2,I,J},\cdots \right)$ \emph{are defined by}
\begin{eqnarray}
&&x_{r,I,J}^{+}(n,k)= \left( \alpha_{r,I} - \beta_{r,I} \right) x +\left( c_{1}\beta_{r,I}-c_{2}\alpha_{r,I} \right)t + n \theta_{r,I}+ k \lambda_{r,I}-b_{r,I,J} + a_{r,I},\\
&&x_{r,I,J}^{-}(n,k)= \left( \alpha^*_{r,J} - \beta^*_{r,J} \right) x +\left( c_{1}\beta^*_{r,J}-c_{2}\alpha^*_{r,J} \right)t -n \theta^*_{r,J}- k \lambda^*_{r,J}-b^*_{r,J,I}+a^*_{r,J},
\end{eqnarray}
\emph{$\alpha_{r,I}$, $\beta_{r,I}$, $\theta_{r,I}$, $\lambda_{r,I}$ and $s_{r,I,J}$ are coefficients from the expansions (\ref{schucoefalpha})-(\ref{schurcoeffsr}) with $p_0$ replaced by $p_{0,I}$, $p_1$ replaced by $p_{1,I}$, $p_0^*$ replaced by $p^*_{0,J}$, $p(\kappa)$ replaced by $p_I(\kappa)$ which is defined by Eq. (\ref{defpk}) with $p_0$ replaced by $p_{0,I}$, $p_{1,I}\equiv (dp_I/d\kappa)|_{\kappa=0}$, $b_{r,I,J}$ is the coefficient from the expansion}
\[
\ln \left[ \frac{ p_I \left( \kappa \right) + p^*_{0,J}}{p_{0,I}+p^*_{0,J}}  \right] = \sum_{r=1}^{\infty} b_{r,I,J} \kappa^r, \label{schurcoeffcc}
\]
\emph{and $a_{r,1}, a_{r,2}$ $(r=1, 2, \dots)$ are free complex constants.}
\end{quote}

\begin{quote}
\textbf{Theorem 3} \hspace{0.05cm} \emph{If the algebraic equation (\ref{Q1polynomial2}) admits a non-imaginary double root $p_{0}$, which is only possible in the soliton-exchange case (\ref{solexchange}) with background amplitudes satisfying conditions (\ref{CubicRestrict}), and $p_{0}=(\sqrt{3}+\rm{i})/2$ or $(-\sqrt{3}+\rm{i})/2$, then the three-wave interaction system (\ref{3WRIModel}) under boundary conditions (\ref{BoundaryCond}) admits bounded $(N_1, N_2)$-th order rogue-wave solutions $u_{i, N_1, N_2}(x,t)$ $(1\le i\le 3)$, where $N_1$ and $N_2$ are arbitrary non-negative integers, and $u_{i, N_1, N_2}(x,t)$ are of the same forms as (\ref{Schpolysolu1b})-(\ref{SchpolysolufN2}), except that their $\sigma_{n,k}$ is given by the following $2\times 2$ block determinant}
\[ \label{cubicrwstype3}
\sigma_{n,k}=
\det \left(
\begin{array}{cc}
  \sigma^{[1,1]}_{n,k} & \sigma^{[1,2]}_{n,k} \\
  \sigma^{[2,1]}_{n,k} & \sigma^{[2,2]}_{n,k}
\end{array}
\right),
\]
\emph{where}
\[\label{Blockmatrix}
\sigma^{[I, J]}_{n,k}=
\left(
m_{3i-I, \, 3j-J}^{(n,k, \hspace{0.04cm} I, J)}
\right)_{1\leq i \leq N_{I}, \, 1\leq j \leq N_{J}},
\]
\emph{the matrix elements in $\sigma^{[I, J]}_{n,k}$ are defined by}
\[ \label{Schmatrimnij9a}
m_{i,j}^{(n,k, I, J)}=\sum_{\nu=0}^{\min(i,j)} \left[ \frac{|p_{1}|^2 }{(p_{0}+p_{0}^*)^2}  \right]^{\nu} \hspace{0.06cm} S_{i-\nu}(\textbf{\emph{x}}_I^{+}(n,k) +\nu \textbf{\emph{s}})  \hspace{0.06cm} S_{j-\nu}(\textbf{\emph{x}}_J^{-}(n,k) + \nu \textbf{\emph{s}}^*),
\]
\emph{vectors} $\textbf{\emph{x}}^{\pm}_{I}(n,k)=\left( x_{1,I}^{\pm}, x_{2,I}^{\pm},\cdots \right)$ $(I=1, 2)$ \emph{are given by}
\begin{eqnarray}
&&x_{r,I}^{+}(n,k)= \left( \alpha_{r} - \beta_{r} \right) x +\left( c_{1}\beta_{r}-c_{2}\alpha_{r} \right)t + n \theta_{r} + k \lambda_{r} + a_{r,I},  \label{defxrp2} \\
&&x_{r,I}^{-}(n,k)=   \left( \alpha^*_{r} - \beta^*_{r} \right) x +\left( c_{1}\beta^*_{r}-c_{2}\alpha^*_{r} \right)t - n \theta^*_{r} - k \lambda^*_{r} +a^*_{r,I},  \label{defxrm2}
\end{eqnarray}
\emph{$\alpha_{r}$, $\beta_{r}$, $\theta_{r}$ and $\lambda_{r}$ are defined in Eqs. (\ref{schucoefalpha})-(\ref{schucoeflambda}),}
$\textbf{\emph{s}}=(s_1, s_2, \cdots)$ \emph{is defined in Eq. (\ref{schurcoeffsr})}, \emph{the function $p\left( \kappa \right)$ which appears in Eqs. (\ref{schucoefalpha})-(\ref{schurcoeffsr}) is defined by the equation}
\[  \label{Q1ptriple}
\mathcal{Q}_{1}\left[p \left( \kappa \right)\right]= \frac{\mathcal{Q}_{1}(p_{0})}{3} \left[ e^{\kappa} +2 e^{-\kappa/2}
\cos\left(\frac{\sqrt{3}}{2} \kappa \right) \right],
\]
\emph{$\mathcal{Q}_{1}(p)$ is given by Eq. (\ref{Q1polynomial}), or equivalently}
\[ \label{Q1ptriple9}
\mathcal{Q}_{1}(p)=-\left(\frac{1}{p}+\frac{1}{p-\rm{i}}+p \right)
\]
\emph{in view of the parameter restrictions (\ref{CubicRestrict}), $p_1\equiv (dp/d\kappa)|_{\kappa=0}$, and $a_{r,1}, a_{r,2}$ $(r=1, 2, \dots)$ are free complex constants.}
\end{quote}

These theorems will be proved in Sec. \ref{sec:derivation}.

\textbf{Remark 1} \hspace{0.03cm} In Theorem 1, the algebraic equation (\ref{Q1polynomial2}) admits a non-imaginary simple root $p_0$ in two situations. One is the soliton-exchange case (\ref{solexchange}) when the background-amplitude conditions (\ref{CubicRestrict}) are not met [see Eq. (\ref{roots1})]. The other is the explosive and stimulated backscatter cases (\ref{explosive})-(\ref{scatter2}) when the discriminant $\Delta$ in Eq. (\ref{eDelta}) is negative [see Eq. (\ref{doublecase2})].

\textbf{Remark 2} \hspace{0.03cm} In Theorems 1 and 3, out of a non-imaginary root pair $(\hat{p}_0, -\hat{p}_0^*)$, we can pick $p_0$ to be either one of them, and keep complex parameters $a_r$ and $a_{r,I}$ free, without any loss of generality. The reason is that the function $\mathcal{Q}_{1}(p)$ in these theorems satisfies the symmetry $\mathcal{Q}_{1}(-p^*)=-\mathcal{Q}_{1}^*(p)$. Thus, both equations (\ref{defpk}) and (\ref{Q1ptriple}) show that when $p_0 \to -p_0^*$, $p(\kappa) \to -p^*(\kappa)$. As a result, Eqs. (\ref{Schmatrimnij})-(\ref{schurcoeffsr}) show that in Theorem 1, when $p_0 \to -p_0^*$,
\begin{equation*}
p_1\to -p_1^*, \quad \alpha_r \to -\alpha_r^*, \quad \beta_r\to -\beta_r^*, \quad \theta_r \to \theta_r^*, \quad
\lambda_r\to \lambda_r^*, \quad s_r \to s_r^*.
\end{equation*}
Together with the parameter change of $a_r \to a_r^*$, then
\begin{equation*}
\textbf{\emph{x}}^{\pm}(n,k; x, t) \to [\textbf{\emph{x}}^{\pm}]^*(n,k; -x, -t), \quad m_{i,j}^{(n,k)}(x, t)\to \left[m_{i,j}^{(n,k)}\right]^*(-x, -t).
\end{equation*}
Hence,
\[  \label{utransform}
u_{1, N}(x, t) \to u_{1, N}^*(-x, -t), \quad u_{2, N}(x, t) \to u_{2, N}^*(-x, -t), \quad u_{3, N}(x, t) \to -u_{3, N}^*(-x, -t)
\]
for solutions in Theorem 1. Similar relations also hold for the solutions in Theorem 3. But the three-wave interaction system (\ref{3WRIModel}) is invariant under the variable transformation (\ref{utransform}). Thus, different choices of $p_0$ from the root pair $(\hat{p}_0, -\hat{p}_0^*)$ in Theorems 1 and 3 yield equivalent rogue wave solutions under appropriate parameter connections.
Regarding rogue waves in Theorem 2, if one chooses $\left(p_{0,1}, p_{0,2}\right)$ as $\left(\hat{p}_{0,1}, \hat{p}_{0,2} \right)$ or $\left(-\hat{p}_{0,1}^*, -\hat{p}_{0,2}^* \right)$, then the two resulting solutions are also related by Eq.~(\ref{utransform}) under parameter changes of $a_{r,1} \to a_{r,1}^*$ and $a_{r,2} \to a_{r,2}^*$. However, if one chooses $\left(p_{0,1}, p_{0,2}\right)$ as $\left(\hat{p}_{0,1}, \hat{p}_{0,2}\right)$, or $\left(\hat{p}_{0,1}, -\hat{p}_{0,2}^* \right)$, or $\left(-\hat{p}_{0,1}^*, \hat{p}_{0,2}\right)$, relations between the three resulting solutions would be more difficult to establish in general. In the fundamental case, with $N_1=N_2=1$ in Theorem 2, we have verified that these three solutions are still equivalent under simple linear transformations between their parameters $(a_{1,1}, a_{1,2})$. This suggests that these three solutions may still be equivalent for higher-order rogue waves in Theorem 2.

\textbf{Remark 3} \hspace{0.03cm} In all these theorems, there are multiple $p(\kappa)$ functions which satisfy Eq. (\ref{defpk}) or (\ref{Q1ptriple}), and those multiple $p(\kappa)$ functions are related to each other by simple symmetries. We can choose any one of those multiple functions, and keep complex parameters $a_r$ and $a_{r,I}$ free, without any loss of generality. The reason is as follows. In Theorem 1, there are two functions of $p(\kappa)$ which satisfy Eq. (\ref{defpk}), because in the $\kappa\to 0$ limit, $p=p_0$ is a double root of Eq. (\ref{defpk}) in view that $\mathcal{Q}'_{1}(p_0)=0$ [see Eq. (\ref{QurticeqQ1dp})]. It is easy to see that if $p(\kappa)$ satisfies Eq. (\ref{defpk}), so does $p(-\kappa)$. Thus, these two functions are related as $p(\pm \kappa)$. Using this connection, we can relate the expansion coefficients ($\alpha_{r}$, $\beta_{r}$, $\theta_{r}$, $\lambda_{r}$, $s_{r}$), and hence $x_{r}^{\pm}(n,k)$, for these $p(\pm \kappa)$ functions. Then, using Lemma 2 of Ref. \cite{YangYang2019Nonloc}, we can show that the solutions $u_{i,N}(x,t)$ in Theorem 1 for the function $p(\kappa)$ and free complex parameters $a_r$, and such solutions for the function $p(-\kappa)$ and complex parameters $(-1)^ra_r$, are equal to each other. This means that we can choose either of the two functions $p(\pm \kappa)$ from Eq. (\ref{defpk}), and keep $a_r$ parameters free, without loss of generality. Similarly, in Theorem 2, we can choose either of the two functions $p_I(\pm \kappa)$ and keep $a_{r,I}$ parameters free without loss of generality. In Theorem 3, there are three functions of $p(\kappa)$ which satisfy Eq. (\ref{Q1ptriple}), because in the $\kappa\to 0$ limit, $p=p_0$ is a triple root of Eq. (\ref{Q1ptriple}) in view that $p_0$ is a double root of equation $\mathcal{Q}'_{1}(p)=0$. Since the right side of Eq. (\ref{Q1ptriple}) can be rewritten as $\mathcal{Q}_{1}(p_{0})[\exp(\kappa)+\exp(\kappa e^{\rm{i}2\pi/3})+\exp(\kappa e^{\rm{i}4\pi/3})]/3$, which is invariant when $\kappa$ changes to $\kappa e^{\rm{i}2\pi/3}$, we see that if $p(\kappa)$ is a solution to this equation, so are $p(\kappa e^{\rm{i}2\pi/3})$ and $p(\kappa e^{\rm{i}4\pi/3})$. Thus, these three $p(\kappa)$ functions are related as $p(\kappa e^{{\rm{i}}2j\pi/3})$, where $j=0, 1, 2$. Using this symmetry and similar arguments, we can show that the $u_i(x,t)$ solutions with the functional branch $p(\kappa)$ and complex parameters ($a_{r,1}$, $a_{r,2}$), and such solutions with the functional branches $p(\kappa e^{{\rm{i}}2j\pi/3})$ $(j=1,2)$ and complex parameters ($e^{{\rm{i}}2rj\pi/3}a_{r,1}$, $e^{{\rm{i}}2rj\pi/3}a_{r,2}$), are equal to each other. Thus, we can pick any of these three $p(\kappa e^{{\rm{i}}2j\pi/3})$ functions, and keep ($a_{r,1}$, $a_{r,2}$) parameters free, without loss of generality.

\textbf{Remark 4} \hspace{0.03cm} The series expansions of these $p(\kappa)$ and $p_I(\kappa)$ functions can be obtained by performing Taylor expansions to both sides of Eq. (\ref{defpk}) or (\ref{Q1ptriple}) and then solving the resulting algebraic equations at each order of the Taylor series. These $p(\kappa)$ and $p_I(\kappa)$ expansions can then be used to determine the coefficients in the expansions of Eqs. (\ref{schucoefalpha})-(\ref{schurcoeffsr}) and (\ref{schurcoeffcc}). For Eq. (\ref{Q1ptriple}) in Theorem 3, the series expansion for $p(\kappa)$ can be found as
\begin{equation*}
p(\kappa)=p_0+p_1\kappa+p_2\kappa^2+p_3\kappa^3+\cdots,
\end{equation*}
where $p_{0}=(\pm \sqrt{3}+\rm{i})/2$, $p_1$ is any one of the three cubic roots of $(\pm 3\sqrt{3}+\rm{i})/12$, $p_2=(9\pm \rm{i}\sqrt{3})/(36p_1)$, and so on. For Eq. (\ref{defpk}), the $p(\kappa)$ expansion will depend on the velocity and background parameters $(c_1, c_2, \rho_1, \rho_2, \rho_3)$.

\textbf{Remark 5} \hspace{0.03cm} Here, we discuss the degrees of polynomials for rogue solutions in the above three theorems. For the $N$-th order rogue waves in Theorem 1, by rewriting its $\sigma_{n,k}$ into a larger $3N \times 3N$ determinant as was done in Ref.~\cite{OhtaJY2012}, we can show that the polynomial degree of its $\sigma_{n,k}$ is $N(N+1)$ in both $x$ and $t$ variables. Using similar techniques, we can show that for the $(N_1, N_2)$-th order rogue wave in Theorem 3, the polynomial degree of its $\sigma_{n,k}$ is $2[N_{1}^2+N_{2}^2 - N_{1}(N_{2}-1)]$ in both $x$ and $t$. For the $(N_1,N_2)$-th order rogue wave in Theorem 2, the polynomial degree of its $\sigma_{n,k}$ turns out to be $N_{1}(N_{1}+1)+N_{2}(N_{2}+1)$ in both $x$ and $t$. The proof for it is a bit longer and is given in Appendix A. We note that this polynomial degree for the $2\times 2$ block determinant $\sigma_{n,k}$ in Theorem 2 is the same as that for the product between its two diagonal block determinants $\det(\sigma_{n,k}^{\left[1,1\right]})$ and $\det(\sigma_{n,k}^{\left[2,2\right]})$, whose polynomial degrees can be obtained from those of $\sigma_{n,k}$ determinants in Theorem 1 as $N_{1}(N_{1}+1)$ and $N_{2}(N_{2}+1)$ individually.

\textbf{Remark 6} \hspace{0.03cm} Now, we discuss the number of irreducible free parameters in rogue wave solutions of these theorems. In Theorem 1, the rogue waves of order $N$ contain $2N-1$ free complex parameters $a_1, a_2, \dots, a_{2N-1}$. However, applying the method of Ref.~\cite{YangYangDNLS}, we can show that all even-indexed parameters $a_{even}$ are dummy parameters which cancel out automatically from the solution. Thus, we will set $a_{2}=a_{4}=\cdots=a_{even}=0$ throughout this article. Of the remaining parameters, we can normalize $a_{1}=0$ through a shift of $x$ and $t$. Then, the $N$-th order rogue waves in Theorem 1 contain $N-1$ free irreducible complex parameters, $a_3, a_5, \dots, a_{2N-1}$. Rogue wave solutions in Theorem 2 contain $2(N_1+N_2-1)$ free complex parameters $(a_{1, 1}, a_{2,1}, \dots, a_{2N_1-1, 1})$ and $(a_{1, 2}, a_{2,2}, \dots, a_{2N_2-1, 2})$. We can also show that all the even-indexed parameters $a_{even, 1}$ and $a_{even, 2}$ can be set as zero. In addition, we can set $a_{1, 1}$ to zero through a shift of $x$ and $t$. Then, rogue solutions of order $(N_1, N_2)$ in Theorem 2 contain $N_1+N_2-1$ free irreducible complex parameters. Rogue solutions of order $(N_1, N_2)$ in Theorem 3 contain $3(N_1+N_2-1)$ free complex parameters $(a_{1, 1}, a_{2,1}, \dots, a_{3N_1-1,\hspace{0.05cm} 1})$ and $(a_{1, 2}, a_{2,2}, \dots, a_{3N_2-2, \hspace{0.05cm} 2})$. Using a method modified from Ref.~\cite{YangYangDNLS}, we can show that the parameters $(a_{3k,1}, a_{3k,2})\ (k=1,2,3,\cdots)$ cancel out automatically from the solutions, and thus we will set them as zero. In addition, we can normalize $a_{1,1}$ to be zero through a shift of $x$ and $t$. Then, in the special cases of $N_1=0$ or $N_2=0$ where the $2\times 2$ block determinant (\ref{cubicrwstype3}) degenerates to a single block determinant, the number of irreducible free complex parameters would be $2N_2-2$ when $N_1=0$ and $2N_1-1$ when $N_2=0$. If both $N_1$ and $N_2$ are positive so that (\ref{cubicrwstype3}) is a true $2\times 2$ block determinant, the same considerations above would readily reduce the the number of free parameters from the original $3(N_1+N_2-1)$ to $2(N_1+N_2-1)$. However, this number may be further reduced. For example, when $(N_1, N_2)=(1,1)$, we can reduce rogue waves of Theorem 3 to one with $a_{1,1}=a_{1,2}=0$ through determinant manipulations and $(x,t)$ shifts, leaving it with a single irreducible complex parameter $a_{2,1}$. When $(N_1, N_2)=(1,2)$, we can reduce rogue waves of Theorem 3 to one with $a_{1,1}=a_{1,2}=0$ and $a_{2,1}=a_{2,2}$ through determinant manipulations and $(x,t)$ shifts, leaving it with two irreducible complex parameters ($a_{2,2}$, $a_{4,2}$). The true number of irreducible free parameters in $2\times 2$ block rogue waves of Theorem 3 merits further investigation.

\textbf{Remark 7} \hspace{0.03cm} In the case of a non-imaginary double root (as in Theorem 3), two other types of the $2\times 2$ block determinant for $\sigma_{n,k}$ also yield valid rogue wave solutions to the three-wave system. These two types of block determinants are also in the form of Eq.~(\ref{cubicrwstype3}), but the matrix elements in $\sigma^{[I,J]}_{n,k}$ are now
\[\label{Blockmatrix01}
\sigma^{[I,J]}_{n,k}=
\left(
m_{3i-1, \, 3j-1}^{(n,k, \hspace{0.04cm} I, J)}
\right)_{1\leq i \leq N_{I}, \, 1\leq j \leq N_{J}}
\]
and
\[\label{Blockmatrix02}
\sigma^{[I,J]}_{n,k}=
\left(
m_{3i-2, \, 3j-2}^{(n,k, \hspace{0.04cm} I, J)}
\right)_{1\leq i \leq N_{I}, \, 1\leq j \leq N_{J}}
\]
respectively, where $m_{i,j}^{(n,k, \hspace{0.04cm} I, J)}$ is as given in Eq.~(\ref{Schmatrimnij9a}). However, we can show that rogue waves from these additional block determinants can be reduced to those given in Theorem 3 through determinant manipulations and parameter redefinitions.

\subsection{Connection with rogue waves from Darboux transformation}
In this subsection, we relate our bilinear rogue waves in Theorems 1-3 to those derived earlier by Darboux transformation in \cite{BaroDegas2013,DegasLomba2013,ChenSCrespo2015,WangXChenY2015}.

In the Darboux transformation framework \cite{DegasLomba2013}, derivation of rogue waves needs the underlying $3\times 3$ scattering matrix to admit a double or triple eigenvalue. Since the eigenvalues satisfy a cubic equation, for double or triple eigenvalues to appear, the discriminant of this cubic equation must be zero. This zero-discriminant condition, which turns out to be a quartic equation for the spectral parameter in the scattering matrix, selects the appropriate spectral-parameter values and scattering-matrix eigenvalues in the Darboux transformation.

To relate those eigenvalue conditions of Darboux transformation to our root conditions of Eq. (\ref{QurticeqQ1dp}) in Sec. \ref{subsection_root}, we consider the equation
\[ \label{Q1pp0}
\mathcal{Q}_{1}(p)=\mathcal{Q}_{1}(p_0),
\]
where $\mathcal{Q}_{1}(p)$ is defined in Eq. (\ref{Q1polynomial}), and $p_0$ is a root of Eq. (\ref{QurticeqQ1dp}). This equation can be rewritten as a cubic equation for $p$. Notice that if $p_0$ is a simple root of Eq. (\ref{QurticeqQ1dp}), then it will be a double root of Eq. (\ref{Q1pp0}); and if $p_0$ is a double root of Eq. (\ref{QurticeqQ1dp}), then it will be a triple root of Eq. (\ref{Q1pp0}).

The connection between eigenvalue conditions in Darboux transformation and root conditions in our bilinear method is that, our equation (\ref{Q1pp0}) is the counterpart of the cubic eigenvalue equation of Darboux transformation, and our equation (\ref{QurticeqQ1dp}) [i.e., (\ref{Q1polynomial2})] is the counterpart of the quartic zero-discriminant equation of Darboux transformation. In addition, our requirement of a non-imaginary root $p_0$ for rogue waves corresponds to the requirement of a non-real spectral parameter in Darboux transformation. Notice that our parameter conditions (\ref{CubicRestrict}) for a triple root in Eq. (\ref{Q1pp0}) are exactly the same as the triple-eigenvalue condition of Darboux transformation in \cite{BaroDegas2013,ChenSCrespo2015}.

In view of the above connections between the Darboux and bilinear methods for rogue waves, we see that our rogue waves in Theorem 1, which correspond to a single simple root $p_0$ in Eq. (\ref{QurticeqQ1dp}), are rogue waves corresponding to a single double eigenvalue of the scattering matrix in Darboux transformation; and our rogue waves in Theorem 3, which correspond to a double root $p_0$ in Eq. (\ref{QurticeqQ1dp}), are rogue waves corresponding to a triple eigenvalue of the scattering matrix in Darboux transformation. Thus, fundamental rogue waves for double and triple eigenvalues of the scattering matrix derived by Darboux transformation in \cite{BaroDegas2013,DegasLomba2013,ZhangYanWen2018} are special cases of our Theorems 1 and 3; and higher-order rogue waves for triple eigenvalues of the scattering matrix derived by Darboux transformation in \cite{ChenSCrespo2015,WangXChenY2015} correspond to degenerate single-block cases of our Theorem 3 (where $N_1=0$ or $N_2=0$). However, our three theorems contain many new rogue solutions to the three-wave system. The first new rogue solutions are the $2\times 2$ block determinant solutions in Theorem 3, in the case of a triple eigenvalue of the scattering matrix in Darboux transformation. The second new solutions are higher-order rogue waves in our Theorem 1, in the case of a single double eigenvalue of the scattering matrix in Darboux transformation. The third new solutions are rogue waves in our Theorem 2, in the case of two double eigenvalues of the scattering matrix in Darboux transformation.

\section{Dynamics of rogue wave solutions}
In this section, we examine the dynamics of rogue waves presented in Theorems 1-3. For this purpose, it is helpful to recall
from the previous section that rogue waves of Theorem 1, corresponding to a non-imaginary simple root in Eq. (\ref{Q1polynomial2}), could exist for all signs of the nonlinear coefficients $(\epsilon_1, \epsilon_2, \epsilon_3)$; but rogue waves of Theorems 2 and 3, corresponding to two non-imaginary simple roots and a non-imaginary double root in Eq. (\ref{Q1polynomial2}), could only exist in the soliton-exchange case (\ref{solexchange}) where $\epsilon_1=-\epsilon_2=\epsilon_3=1$.

\subsection{Rogue waves for a non-imaginary simple root}
We first consider rogue waves in Theorem 1, which are associated with a non-imaginary simple root in Eq. (\ref{Q1polynomial2}).
To get the fundamental rogue wave in this solution family, we take $N=1$ in Theorem 1. In addition, we normalize $a_1=0$. Then, we readily find that
\[ \label{Fundamplitu1}
|u_{i,1}(x,t)|=\left| \rho_{i} \frac{g_{i,1}}{f_{1}} \right|,\ \ \ i=1,2,3,
\]
where
\begin{eqnarray}
&& f_{1} =m_{1,1}^{(0,0)} = \left| \left(\alpha_{1}-\beta_{1}\right) x+(c_{1}\beta_{1}-c_{2}\alpha_{1})t \right|^2+\zeta_{0},  \label{f10}
\\
&& g_{1,1}=m_{1,1}^{(1,0)} = \left[\left(\alpha_{1}-\beta_{1}\right) x+(c_{1}\beta_{1}-c_{2}\alpha_{1})t + \theta_{1}\right] \left[\left(\alpha_{1}^*-\beta_{1}^*\right) x+(c_{1}\beta_{1}^*-c_{2}\alpha_{1}^*)t - \theta_{1}^* \right]+\zeta_{0},\\
&& g_{2,1}=m_{1,1}^{(0,-1)} =  \left[\left(\alpha_{1}-\beta_{1}\right) x+(c_{1}\beta_{1}-c_{2}\alpha_{1})t - \lambda_{1}\right] \left[\left(\alpha_{1}^*-\beta_{1}^*\right) x+(c_{1}\beta_{1}^*-c_{2}\alpha_{1}^*)t + \lambda_{1}^* \right]+\zeta_{0}, \\
&& g_{3,1}=m_{1,1}^{(-1,1)} =  \left[\left(\alpha_{1}-\beta_{1}\right) x+(c_{1}\beta_{1}-c_{2}\alpha_{1})t - \theta_{1}+\lambda_{1}\right] \left[\left(\alpha_{1}^*-\beta_{1}^*\right) x+(c_{1}\beta_{1}^*-c_{2}\alpha_{1}^*)t + \theta_{1}^*-\lambda_{1}^* \right]+\zeta_{0},  \label{g310}
\end{eqnarray}
and
\begin{eqnarray} \label{SpecialNum1}
\alpha_1=-\frac{p_1 \epsilon_1 \rho_{2} \rho_{3}}{p_{0}^2  (c_{1}-c_{2})\rho_{1}},\ \beta_{1}=-\frac{p_1 \epsilon_2 \rho_{1} \rho_{3}}{(p_{0}-\rm{i})^2  (c_{1}-c_{2})\rho_{2}}, \ \theta_{1}=\frac{p_{1}}{p_{0}-\rm{i}}, \ \lambda_{1}=\frac{p_{1}}{p_{0}}, \ \zeta_{0} =\frac{|p_{1}|^2 }{(p_{0}+p_{0}^*)^2}.
\end{eqnarray}
Notice that $p_1$ cancels out in these $u_{i,1}$ solutions, and thus its formula is not needed here. In these fundamental rogue waves, $f_1$ and $g_{i,1}$ are all quadratic functions of $x$ and $t$, and there are no free parameters.

To get second-order rogue waves, we take $N=2$ in Theorem 1. Normalizing $a_1=0$, then these second-order rogue waves have a single free complex parameter $a_3$. In these solutions, $f_2$ and $g_{i,2}$ are degree-6 polynomials in both $x$ and $t$, and their expressions are displayed in Appendix B.

To illustrate the dynamics of these rogue waves, we first consider the soliton-exchange case (\ref{solexchange}), i.e., $\epsilon_1=-\epsilon_2=\epsilon_3=1$. For the background and velocity values of
\[ \label{DoubleRootpara1}
c_{1}=1,\ c_{2}=0.5,\ \rho_{1}=1,\ \rho_{2}=2,\ \rho_{3}=1,
\]
the roots of Eq. (\ref{Q1polynomial2}) are $(p_{0,1}, p_{0,2}, -p_{0,1}^*, -p_{0,2}^*)$, where $p_{0,1} \approx 0.521005 +0.853553 \rm{i}$, and $p_{0,2} \approx 0.989219+0.146447 \rm{i}$. Choosing $p_0=p_{0,1}$, the fundamental rogue wave is displayed in Fig. 1 (top row). We see that the intensity variation of each component in this rogue wave is along a slanted angle in the $(x,t)$ plane. In addition,  while the first and third components peak at the origin $x=t=0$, the second component bottoms there. Because of this, we can say the first and third components of this rogue wave are bright, but the second component is dark. If we choose $p_0=p_{0,2}$, the intensity pattern of the resulting rogue wave would also be slanted, but extremely slender, like a needle, in all three components. In addition, the first and third components are now dark, while the second component bright, in this latter case.

The second-order rogue waves involve $p_1$ and the free parameter $a_3$. For the chosen $p_0$ value, we find that $p_1\approx \pm (0.550798-0.289323 \rm{i})$, and choose the plus sign. Then, at two $a_3$ values of $10+10 \rm{i}$ and $0$, the corresponding rogue waves are displayed in Fig. 1 (middle and bottom rows respectively). The rogue wave at $a_3=10+10 \rm{i}$ comprises three separate fundamental rogue waves --- a phenomenon common in other integrable systems, such as the NLS equation \cite{AAS2009,DGKM2010,GLML2012,OhtaJY2012}. The rogue wave at $a_3=0$ cannot be decomposed into separate fundamental rogue waves. It exhibits new patterns and higher peak amplitudes, and is the counterpart of the so-called super rogue waves in other integrable systems \cite{AAS2009,ACA2010,GLML2012,OhtaJY2012,PRX2012}. But the present super rogue wave has a distinctive structure that is very different from those reported before for other integrable equations.

\begin{figure}[htb]
\begin{center}
\vspace{-1.5cm}
\includegraphics[scale=0.326, bb=0 0 385 567]{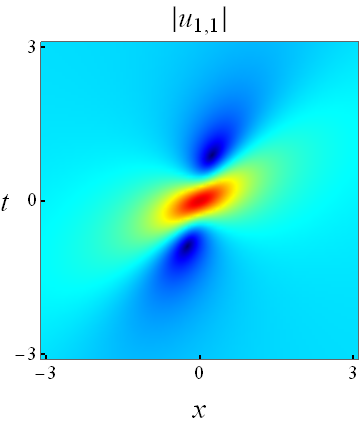}
      \includegraphics[scale=0.320, bb=0 0 385 567]{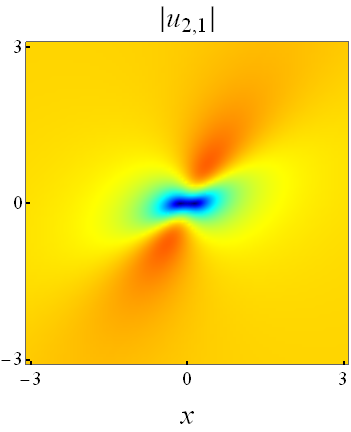}
         \includegraphics[scale=0.320, bb=0 0 385 567]{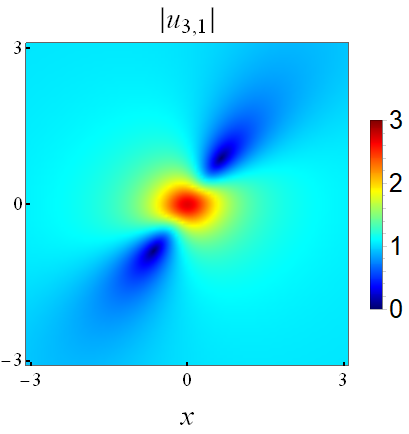}

         \vspace{-1.5cm}
         \includegraphics[scale=0.326, bb=0 0 385 567]{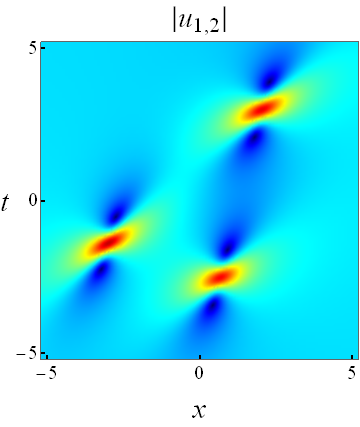}
      \includegraphics[scale=0.320, bb=0 0 385 567]{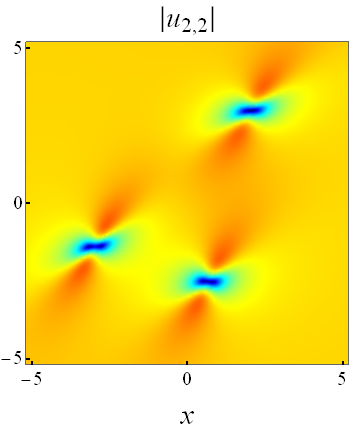}
         \includegraphics[scale=0.320, bb=0 0 385 567]{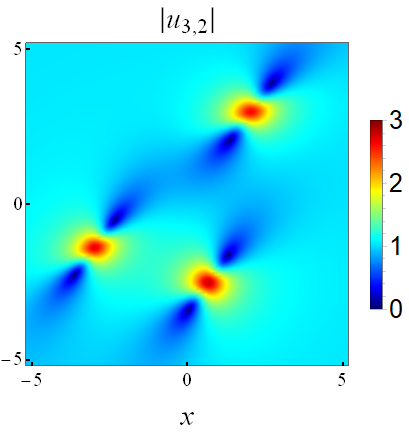}

         \vspace{-1.5cm}
    \includegraphics[scale=0.326, bb=0 0 385 567]{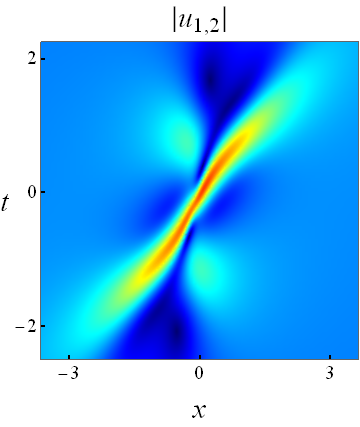}
      \includegraphics[scale=0.320, bb=0 0 385 567]{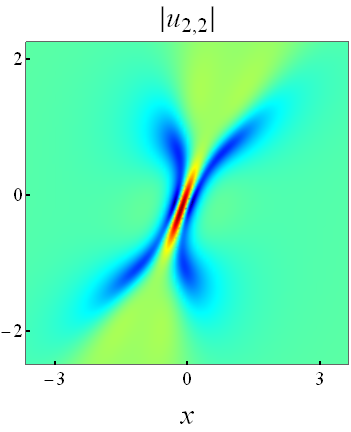}
         \includegraphics[scale=0.320, bb=0 0 385 567]{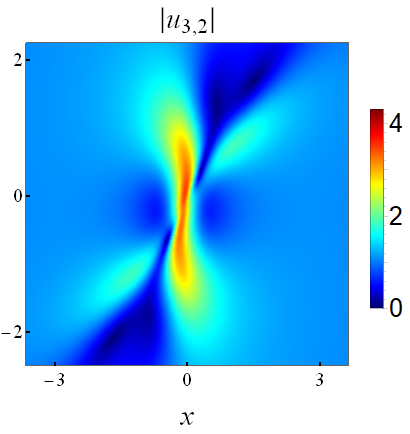}

\caption{Rogue waves of Theorem 1 which correspond to a non-imaginary simple root of Eq. (\ref{Q1polynomial2}), in the soliton exchange case (\ref{solexchange}) with background and velocity values (\ref{DoubleRootpara1}). Top row: the fundamental rogue wave; middle row: a second-order rogue wave with $a_3=10+10 \rm{i}$; bottom row: the second-order super rogue wave with $a_3=0$.   }
\end{center}
\end{figure}

It is important to recognize that rogue wave patterns in the three-wave interaction system are far more diverse than those in most other integrable systems due to its many free physical parameters such as wave velocities and background amplitudes. To appreciate this diversity, we still consider the soliton-exchange case (\ref{solexchange}), but choose a different set of background and velocity values as
\[ \label{DoubleRootpara2}
c_{1}=6,\ c_{2}=5,\ \rho_{1}=\rho_{2}=3,\ \rho_{3}=2.
\]
In this case, Eq. (\ref{Q1polynomial2}) admits four non-imaginary roots, one of them being $p_{0,1} \approx 0.557458\, +0.441122 \rm{i}$. Choosing $p_0=p_{0,1}$, the fundamental rogue wave is displayed in Fig. 2 (upper row). We can see that this fundamental rogue wave looks very different from that in Fig. 1 (top row). In particular, this rogue wave does not have dark components. Instead, centers of intensity fields for the first and second wave components here are saddle-like --- along the bright direction, the center is a local intensity minimum, but along the dark direction, the center is a local intensity maximum.

Under this latter set of background and velocity values (\ref{DoubleRootpara2}), the second-order rogue wave at $a_3=10+10 \rm{i}$ consists of three separate fundamental rogue waves --- a phenomenon similar to the former case. At $a_3=0$, however, we get a super rogue wave which is shown in Fig. 2 (lower row). This super rogue wave has a more delicate structure and looks entirely different from that in Fig. 1 (bottom row) under the former set of parameters (\ref{DoubleRootpara1}).

\begin{figure}[htb]
\begin{center}
\vspace{-1.5cm}
       \includegraphics[scale=0.326, bb=0 0 385 567]{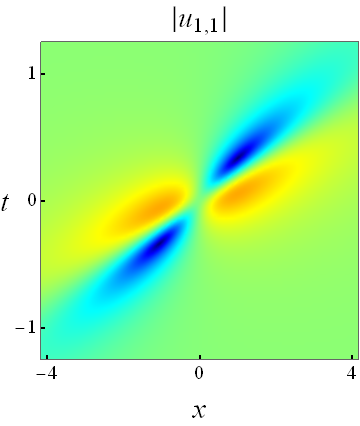}
      \includegraphics[scale=0.320, bb=0 0 385 567]{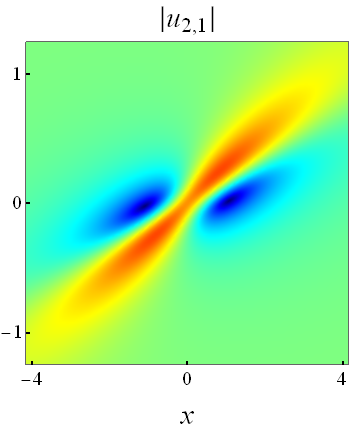}
         \includegraphics[scale=0.320, bb=0 0 385 567]{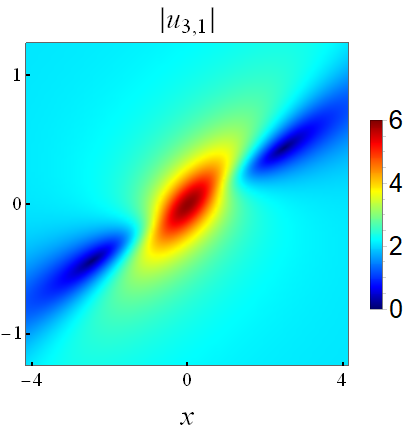}

         \vspace{-1.5cm}
         \includegraphics[scale=0.326, bb=0 0 385 567]{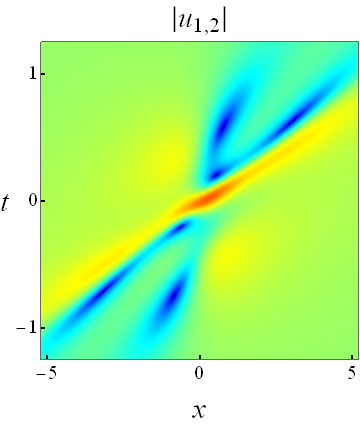}
      \includegraphics[scale=0.320, bb=0 0 385 567]{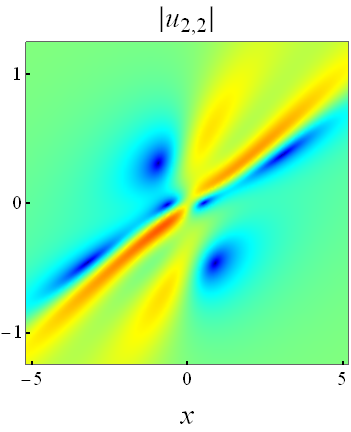}
         \includegraphics[scale=0.320, bb=0 0 385 567]{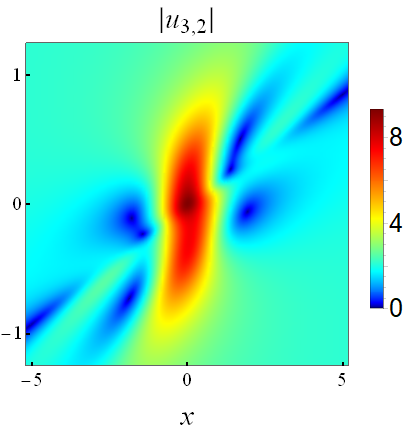}
\caption{Rogue waves of Theorem 1 which correspond to a non-imaginary simple root of Eq. (\ref{Q1polynomial2}), in the soliton exchange case (\ref{solexchange}) with background and velocity values (\ref{DoubleRootpara2}). Upper row: the fundamental rogue wave; lower row: the second-order super rogue wave with $a_3=0$.}
\end{center}
\end{figure}

In the above two sets of parameters (\ref{DoubleRootpara1})-(\ref{DoubleRootpara2}), two of $\rho_{1}$, $\rho_{2}$ and $\rho_{3}$ have been chosen to be equal. If they are all distinct or all equal, we have found that the fundamental rogue waves would remain qualitatively similar to those shown in the top rows of Figs. 1 and 2, except that the bright, dark and saddle components can switch among the three waves, and the slanting slopes of their intensity variations can be positive or negative. Higher-order rogue waves, especially super rogue waves, for general choices of $(\rho_1, \rho_2, \rho_3)$ values, can display additional intricate patterns, as bottom rows of Figs. 1 and 2 have already implied.

Next, we illustrate dynamics of rogue waves in Theorem 1 for the non-soliton-exchange cases. For brevity, we only consider the stimulated backscatter cases, where the $(\epsilon_1, \epsilon_2, \epsilon_3)$ values are given in Eqs. (\ref{scatter1})-(\ref{scatter2}). Since these two sets of $(\epsilon_1, \epsilon_2, \epsilon_3)$ values are equivalent [see the discussion below Eq. (\ref{scatter2})], we choose the first set, i.e., $\epsilon_1=-\epsilon_2=-\epsilon_3=1$. For the background and velocity values of
\[\label{scatterpara}
c_{1}=5,\ c_{2}=2,\ \rho_{1}=\rho_{2}=\rho_{3}=2,
\]
Eq. (\ref{Q1polynomial2}) admits a non-imaginary simple root $p_0\approx  0.391016 + 0.338012 \textrm{i}$. The corresponding fundamental rogue wave (\ref{Fundamplitu1}) is plotted in Fig.~3 (upper row). In this rogue wave, the first component is dark, the second a saddle, and the third bright. In addition, slanting slopes of bright-intensity variations are negative in the second and third components. In second-order rogue waves, if we choose $a_3=5+5\textrm{i}$, the resulting solution comprises three separate fundamental rogue waves. If we choose $a_3=0$, we get a second-order super rogue wave, which is displayed in Fig.~3 (lower row). This super rogue wave develops strong dips in its first and second components and a strong peak in its third component at the wave center, and its pattern is rich and different from those in Figs. 1-2.

\begin{figure}[htb]
\begin{center}
\vspace{-1.5cm}
    \includegraphics[scale=0.326, bb=0 0 385 567]{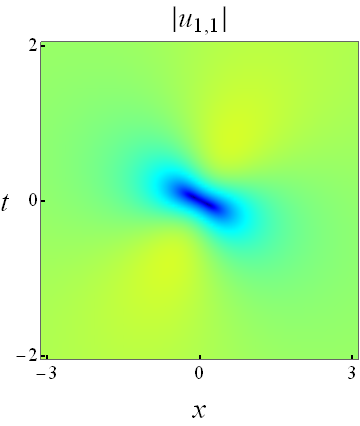}
      \includegraphics[scale=0.320, bb=0 0 385 567]{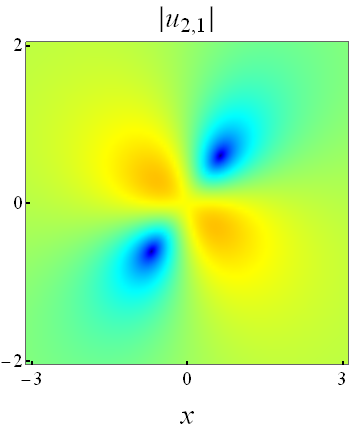}
         \includegraphics[scale=0.320, bb=0 0 385 567]{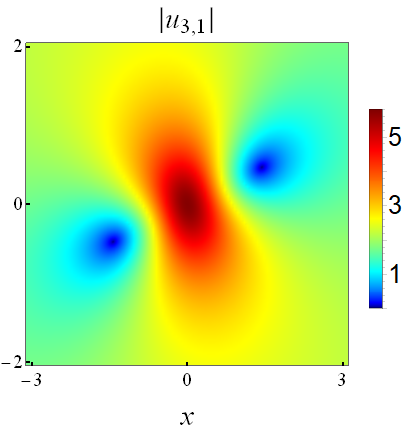}

         \vspace{-1.5cm}
         \includegraphics[scale=0.326, bb=0 0 385 567]{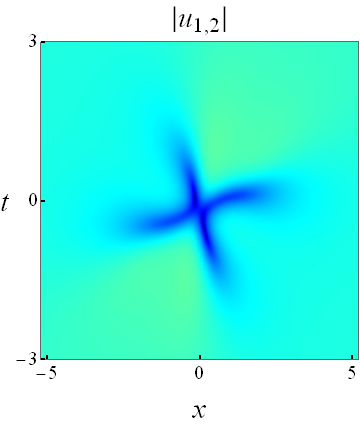}
      \includegraphics[scale=0.320, bb=0 0 385 567]{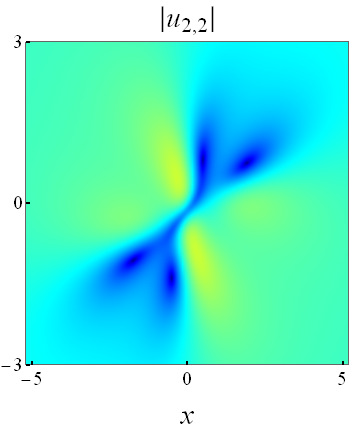}
         \includegraphics[scale=0.320, bb=0 0 385 567]{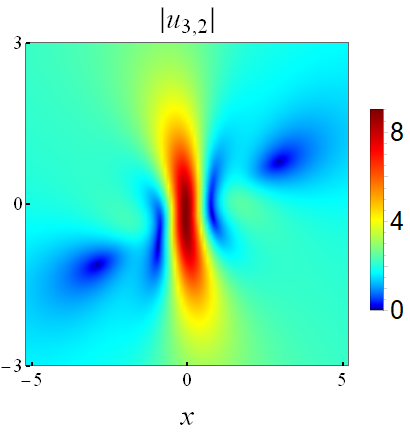}

\caption{Rogue waves of Theorem 1 which correspond to a non-imaginary simple root of Eq. (\ref{Q1polynomial2}), in the stimulated backscatter case (\ref{scatter1}) with background and velocity values (\ref{scatterpara}). Upper row: the fundamental rogue wave; lower row: the second-order super rogue wave with $a_{3}=0$. }
\end{center}
\end{figure}

\subsection{Rogue waves for two non-imaginary simple roots}
Rogue waves in Theorem 2 are associated with two non-imaginary simple roots $p_{0,1}$ and $p_{0,2}$ in Eq. (\ref{Q1polynomial2}), with $p_{0,2}\ne -p_{0,1}^*$. These solutions only appear in the soliton-exchange case of $\epsilon_1=-\epsilon_2=\epsilon_3=1$ when the background amplitudes do not satisfy conditions (\ref{CubicRestrict}). The fundamental rogue waves in this family correspond to $N_{1}=N_{2}=1$, and their expressions are
\[ \label{Fundamplitu11}
|u_{i,1,1}(x,t)|=\left| \rho_{i} \frac{g_{i,1,1}}{f_{1,1}} \right|,\ \ \ i=1,2,3,
\]
where
\begin{eqnarray*}
&& f_{1,1} =m_{1,1}^{(0,0,1,1)}m_{1,1}^{(0,0,2,2)}-m_{1,1}^{(0,0,1,2)}m_{1,1}^{(0,0,2,1)},  \\
&& g_{1,1,1}=m_{1,1}^{(1,0,1,1)}m_{1,1}^{(1,0,2,2)}-m_{1,1}^{(1,0,1,2)}m_{1,1}^{(1,0,2,1)}, \\
&& g_{2,1,1}=m_{1,1}^{(0,-1,1,1)}m_{1,1}^{(0,-1,2,2)}-m_{1,1}^{(0,-1,1,2)}m_{1,1}^{(0,-1,2,1)},\\
&& g_{3,1,1}=m_{1,1}^{(-1,1,1,1)}m_{1,1}^{(-1,1,2,2)}-m_{1,1}^{(-1,1,1,2)}m_{1,1}^{(-1,1,2,1)},
\end{eqnarray*}
\begin{equation*}
m_{1,1}^{(n,k,I,J)}=\frac{1}{p_{0,I}+p^*_{0,J}} \left[x_{1,I,J}^+(n,k) x_{1,I,J}^-(n,k) +\frac{p_{1,I} p^*_{1,J} }{(p_{0,I}+p^*_{0,J})^2}\right],
\end{equation*}
\begin{equation*}
x_{1,I,J}^{+}(n,k)= \left( \alpha_{1,I} - \beta_{1,I} \right) x +\left( c_{1}\beta_{1,I}-c_{2}\alpha_{1,I} \right)t + n \theta_{1,I}+ k \lambda_{1,I}-b_{1,I,J} + p_{1,I} \tilde{a}_{1,I},
\end{equation*}
\begin{equation*}
x_{1,I,J}^{-}(n,k)= \left( \alpha^*_{1,J} - \beta^*_{1,J} \right) x +\left( c_{1}\beta^*_{1,J}-c_{2}\alpha^*_{1,J} \right)t -n \theta^*_{1,J}- k \lambda^*_{1,J}-b^*_{1,J,I}+p_{1,J}^* \tilde{a}^*_{1,J},
\end{equation*}
$\alpha_{1,I}$, $\beta_{1,I}$, $\theta_{1,I}$ and $\lambda_{1,I}$ are given by Eq. (\ref{SpecialNum1}) with $(p_0, p_1)$ replaced by $(p_{0,I}, p_{1,I})$, $b_{1,I,J}$ is given by
\begin{equation*}
b_{1,I,J}=\frac{p_{1,I}}{p_{0,I}+p_{0,J}^*},
\end{equation*}
and $(\tilde{a}_{1,1},\tilde{a}_{1,2})$ are free complex constants. These $\tilde{a}_{1,I}$ constants are related to $a_{1,I}$ in Theorem 2 as $a_{1,I}=p_{1,I} \tilde{a}_{1,I}$. These scaled $\tilde{a}_{1,I}$ constants are chosen because in this case, parameters $p_{1,I} \ (I=1,2)$ would cancel out in these $u_{i,1,1}$ solutions. These $f_{1,1}$ and $g_{i,1,1}$ functions are degree-4 polynomials in both $x$ and $t$.

To illustrate these fundamental rogue waves in this family, we choose background and velocity values of
\[ \label{DoubleRootpara3}
c_{1}=1,\ c_{2}=0.5,\ \rho_{1}=\rho_{2}=\rho_{3}=\sqrt{2}.
\]
The roots of Eq. (\ref{Q1polynomial2}) for this set of values are $(p_{0,1}, p_{0,2}, -p_{0,1}^*, -p_{0,2}^*)$, where
\[ \label{TwoRoots2b2Block}
p_{0,1} \approx 0.529086\, +0.257066 \rm{i},\ \ \  p_{0,2} \approx 1.52909\, +0.742934 \rm{i}.
\]
Regarding free complex parameters $\tilde{a}_{1,1}$ and $\tilde{a}_{1,2}$, one of them can be normalized to zero by a shift of $x$ and $t$, and the other is irreducible. We will normalize $\tilde{a}_{1,1}=0$. Then, at two $\tilde{a}_{1,2}$ values of $2-\textrm{i}$ and 0, the resulting rogue waves are displayed in Fig.~4. The rogue wave at $\tilde{a}_{1,2}=2-\textrm{i}$ (upper row) comprises two separate simpler rogue waves, which turn out to be fundamental rogues of Theorem 1 for the two individual $p_0$ values in Eq. (\ref{TwoRoots2b2Block}). Thus, rogue waves in Theorem 2 can be viewed as a nonlinear superposition of rogue waves of Theorem 1 with two different $p_0$ values. The rogue wave at $\tilde{a}_{1,2}=0$ (lower row) is a super rogue wave formed by merging the two simpler rogue waves in the upper row. It has a new composite structure and higher peak amplitude.

\begin{figure}[htb]
\begin{center}
\vspace{-1.25cm}
     \includegraphics[scale=0.326, bb=0 0 385 567]{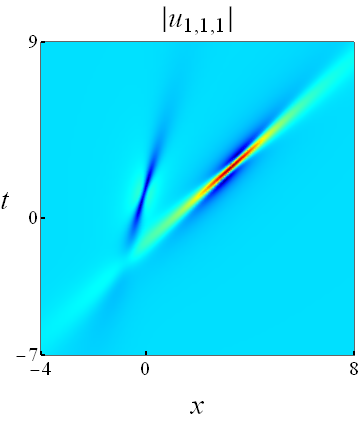}
      \includegraphics[scale=0.320, bb=0 0 385 567]{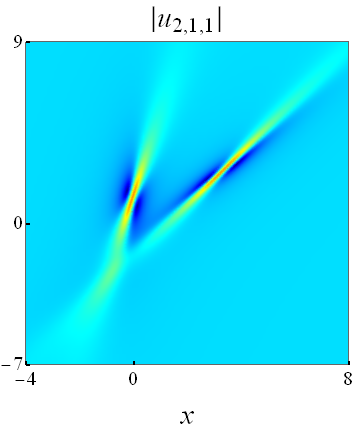}
         \includegraphics[scale=0.320, bb=0 0 385 567]{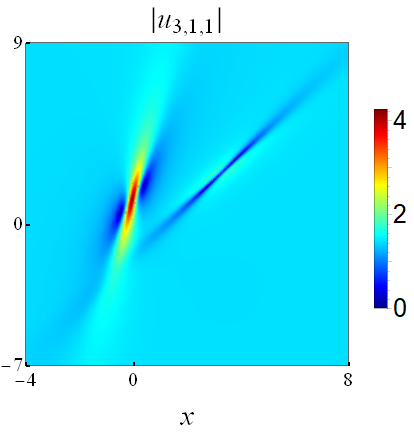}

\vspace{-1.5cm}
 \includegraphics[scale=0.326, bb=0 0 385 567]{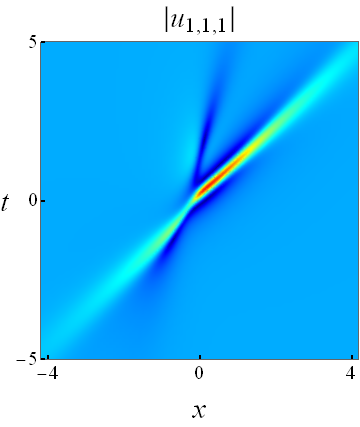}
      \includegraphics[scale=0.320, bb=0 0 385 567]{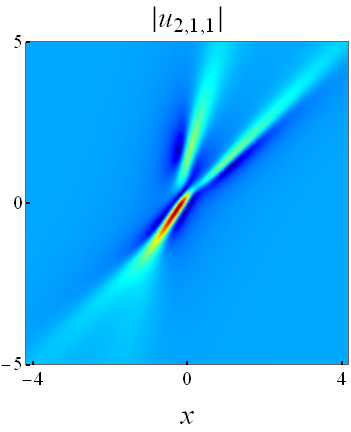}
         \includegraphics[scale=0.320, bb=0 0 385 567]{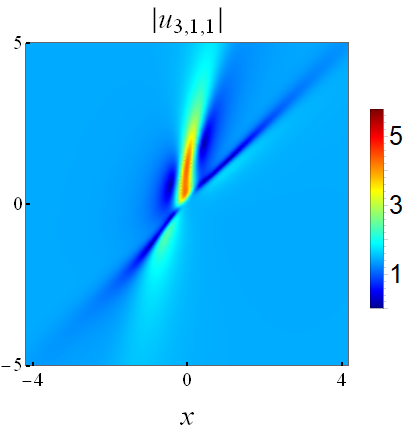}
\caption{Fundamental rogue waves (\ref{Fundamplitu11}) of Theorem 2, which correspond to two non-imaginary simple roots of Eq. (\ref{Q1polynomial2}) in the soliton exchange case (\ref{solexchange}), with background and velocity values (\ref{DoubleRootpara3}). Upper row: $\tilde{a}_{1,2}=2-\textrm{i}$; lower row: $\tilde{a}_{1,2}=0$.}
\end{center}
\end{figure}

\subsection{\textbf{Rogue waves for a non-imaginary double root}}
Rogue waves in Theorem 3 only arise in the soliton-exchange case of $\epsilon_1=-\epsilon_2=\epsilon_3=1$ when the background amplitudes satisfy conditions (\ref{CubicRestrict}), i.e.,
\[ \label{CubicRestrict2}
\rho_{2}=\pm \sqrt{\frac{c_{1}}{c_{2}}}\rho_{1},\ \ \ \rho_{3}= \pm \sqrt{\frac{c_{1}-c_{2}}{c_{2}}}\rho_{1}.
\]
In this case, Eq. (\ref{Q1polynomial2}) admits a pair of non-imaginary double roots $p_{0}=(\pm \sqrt{3}+\rm{i})/2$, see Eq. (\ref{roots2}). We will choose $p_{0}=(\sqrt{3}+\rm{i})/2$. Regarding $p_1$, which is any one of the three cubic roots of $(3\sqrt{3}+\rm{i})/12$ (see Remarks 3 and 4), we pick the one in the first quadrant, which is $p_1\approx 0.759614 + 0.0482053 \rm{i}$. We also normalize $a_{1,1}=0$ through a shift in $(x, t)$. In our illustrations, we choose the background and velocity values as
\[ \label{para_triple}
c_{1}=1, \quad c_{2}=0.5, \quad \rho_{1}=1.
\]

Rogue waves in Theorem 3 are given through a $2\times 2$ block determinant. Unlike the $2\times 2$ block determinant in Theorem~2, the current $2\times 2$ block determinant is allowed to degenerate into a single-block determinant if we choose $N_1$ or $N_2$ to be zero. We will consider these degenerate single-block solutions and non-degenerate $2\times 2$ block solutions separately below.

\subsubsection{Degenerate single-block rogue waves with $N_1=0$}
If $N_1=0$, rogue waves $u_{i, 0, N_2}(x,t)$ in Theorem 3 are given by Eqs. (\ref{Schpolysolu1b})-(\ref{SchpolysolufN2}), where $\sigma_{n,k}$ in Eq.~(\ref{cubicrwstype3}) degenerates to
\[
\sigma_{n,k}=\sigma^{[2,2]}_{n,k}=\left(
m_{3i-2, \, 3j-2}^{(n,k, \hspace{0.04cm}2, 2)}
\right)_{1\leq i, j \leq N_{2}},
\]
and $m_{i,j}^{(n,k, \hspace{0.04cm}2, 2)}$ is given in Eq.~(\ref{Schmatrimnij9a}). These $u_{i, 0, N_2}(x,t)$ rogue waves contain $2N_2-2$ irreducible complex parameters, $a_{2,2}, a_{4,2}, a_{5,2}, a_{7,2}, \dots, a_{3N_2-2, \hspace{0.05cm} 2}$. Fundamental rogue waves of this type, with $N_2=1$, are
\[ \label{Fundamplitu2}
|u_{i,0,1}(x,t)|=\left| \rho_{i} \frac{g_{i,1}}{f_{1}} \right|,\ \ \ 1\le i \le 3,
\]
where $f_1(x,t)$ and $g_{i,1}(x,t)$ are given in Eqs. (\ref{f10})-(\ref{g310}), with parameter values of $\alpha_1, \beta_1, \theta_1, \lambda_1$ and $\zeta_0$ provided by Eq. (\ref{SpecialNum1}) under the parameter constraint (\ref{CubicRestrict2}). For the background and velocity choices (\ref{para_triple}), this fundamental rogue wave is plotted in Fig. 5 (top row). This is a rogue wave with all three components bright at the wave center $x=t=0$. Second-order rogue waves of this type, with $N_2=2$, contain two free complex parameters, $a_{2,2}$ and $a_{4,2}$. Two such solutions, with $(a_{2,2}, a_{4,2})=(0, 50\textrm{i})$ and $(0,0)$, are displayed in the middle and bottom rows of Fig. 5, respectively. It is seen that at $(a_{2,2}, a_{4,2})=(0, 50\textrm{i})$, this second-order rogue wave splits into four fundamental ones, unlike Fig. 1 where the second-order rogue wave in the middle row splits into three fundamental ones. The reason for the current four-splitting is that the polynomial degree of the present second-order rogue waves is eight (see Remark 5), which is four times that of the fundamental rogue waves given in Eq. (\ref{Fundamplitu2}). When $(a_{2,2}, a_{4,2})=(0, 0)$, we get a second-order super rogue wave, which can be viewed as coalescing of those four constituent fundamental rogue waves. This super rogue wave has higher amplitudes, and a superposition of its three components forms a three-needle structure, which was called ``watch-hand-like" in Ref.~\cite{ChenSCrespo2015}.

\begin{figure}[htb]
\begin{center}
\vspace{-1.5cm}
 \includegraphics[scale=0.326, bb=0 0 385 567]{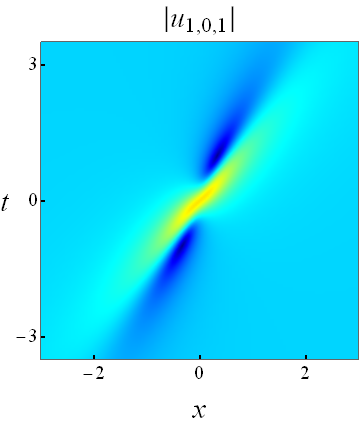}
      \includegraphics[scale=0.320, bb=0 0 385 567]{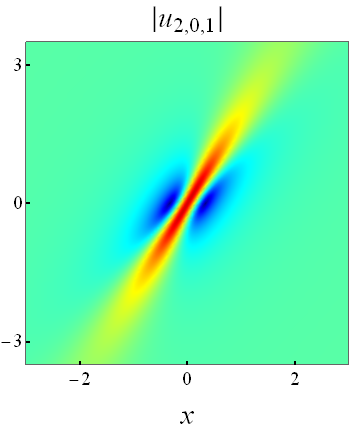}
         \includegraphics[scale=0.320, bb=0 0 385 567]{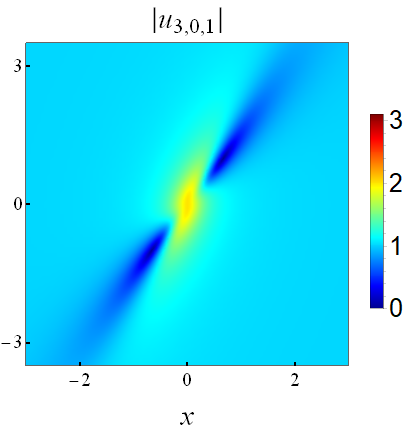}

          \vspace{-1.55cm}
         \includegraphics[scale=0.326, bb=0 0 385 567]{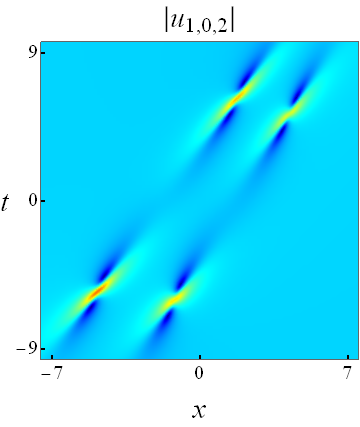}
      \includegraphics[scale=0.320, bb=0 0 385 567]{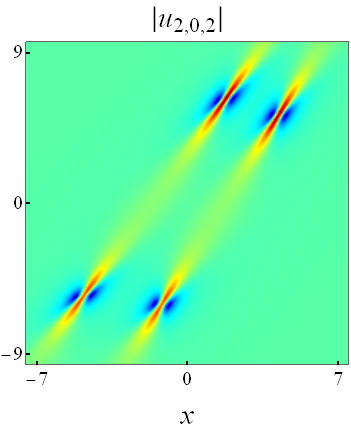}
         \includegraphics[scale=0.320, bb=0 0 385 567]{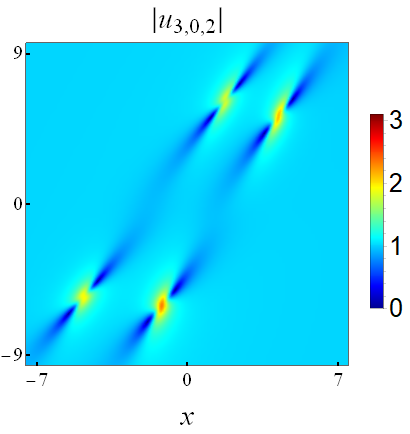}

         \vspace{-1.5cm}
         \includegraphics[scale=0.326, bb=0 0 385 567]{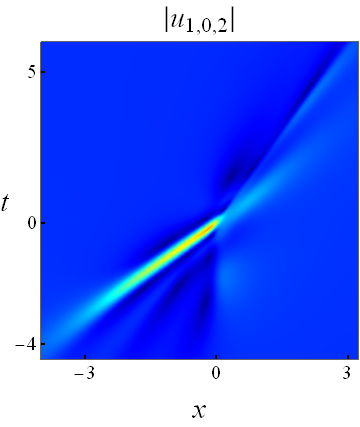}
      \includegraphics[scale=0.320, bb=0 0 385 567]{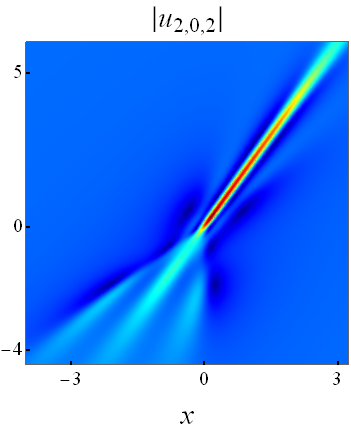}
         \includegraphics[scale=0.320, bb=0 0 385 567]{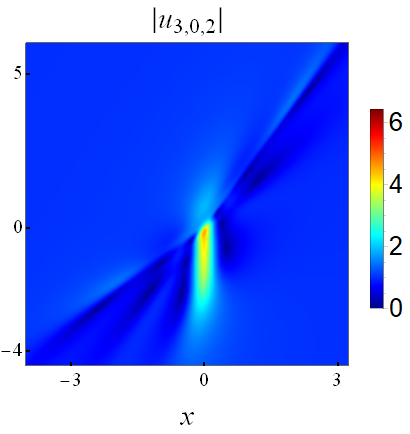}
\caption{Degenerate rogue waves $|u_{i,N_1,N_2}|$ in Theorem 3 with $N_1=0$, for a non-imaginary double root of Eq. (\ref{Q1polynomial2}) in the soliton exchange case (\ref{solexchange}), with background and velocity values (\ref{para_triple}) under relations (\ref{CubicRestrict2}). Top row: the fundamental rogue wave ($N_2=1$); middle row: a second-order rogue wave ($N_2=2$) with $a_{2,2}=0$ and $a_{4,2}=50\textrm{i}$; bottom row: the second-order super rogue wave with $a_{2,2}=a_{4,2}=0$. }
\end{center}
\end{figure}

\subsubsection{Degenerate single-block rogue waves with $N_2=0$}
If $N_2=0$, rogue waves $u_{i, N_1, 0}(x,t)$ in Theorem 3 are given by Eqs. (\ref{Schpolysolu1b})-(\ref{SchpolysolufN2}), where $\sigma_{n,k}$ in Eq.~(\ref{cubicrwstype3}) degenerates to
\[
\sigma_{n,k}=\sigma^{[1,1]}_{n,k}=\left(
m_{3i-1, \, 3j-1}^{(n,k, \hspace{0.04cm}1, 1)}
\right)_{1\leq i, j \leq N_{1}},
\]
and $m_{i,j}^{(n,k, \hspace{0.04cm}1, 1)}$ is given in Eq.~(\ref{Schmatrimnij9a}). These $u_{i, N_1, 0}(x,t)$ rogue waves contain $2N_1-1$ irreducible complex parameters, $a_{2,1}, a_{4,1}, a_{5,1}, a_{7,1}, \dots, a_{3N_1-1, \hspace{0.05cm} 1}$. Fundamental rogue waves of this type, with $N_1=1$, are
\[ \label{Fundamplitu3}
|u_{i,1,0}(x,t)|=\left| \rho_{i} \frac{g_{i,1}}{f_{1}} \right|,\ \ \ 1\le i \le 3,
\]
where
\begin{equation*}
f_{1} =m_{2,2}^{(0,0,1,1)},\quad g_{1,1} =m_{2,2}^{(1,0,1,1)}, \quad g_{2,1} =m_{2,2}^{(0,-1,1,1)}, \quad  g_{3,1}=m_{2,2}^{(-1,1,1,1)},
\end{equation*}
and $m_{2,2}^{(n,k,1,1)}$ is given in Eq.~(\ref{Schmatrimnij9a}). The degrees of polynomials $f_1$ and $g_{i,1}$ are four in both $x$ and $t$, and these functions contain a single free complex parameter $a_{2,1}$. When $a_{2,1}=10+10\textrm{i}$, this rogue wave is plotted in Fig. 6 (upper row). It is seen that this $u_{i, 1, 0}(x,t)$ wave splits into two fundamental rogue waves $u_{i, 0, 1}(x,t)$ of Eq. (\ref{Fundamplitu2}) [see Fig. 5 (top row)]. When $a_{2,1}=0$, we get a super rogue wave where those two constituent $u_{i, 0, 1}(x,t)$ waves merge together. Second-order rogue waves of the present type, $u_{i, 2, 0}(x,t)$, contain three free irreducible complex parameters, $a_{2,1}$, $a_{4,1}$ and $a_{5,1}$. This solution, with $a_{2,1}=10+10\textrm{i}$, $a_{4,1}=0$ and $a_{5,1}=20+20\textrm{i}$, is displayed in the lower row of Fig. 6. This solution splits into six $u_{i, 0, 1}(x,t)$ waves of Eq. (\ref{Fundamplitu2}), because the polynomial degree of the $u_{i, 2, 0}(x,t)$ solution is twelve (see Remark 5), which is six times that of $u_{i, 0, 1}(x,t)$. When $a_{2,1}=a_{4,1}=a_{5,1}=0$, those six $u_{i, 0, 1}(x,t)$ rogue waves merge to form a super rogue wave, which also has a ``watch-hand-like" structure.

\begin{figure}[htb]
\begin{center}
\vspace{-1.5cm}
         \includegraphics[scale=0.326, bb=0 0 385 567]{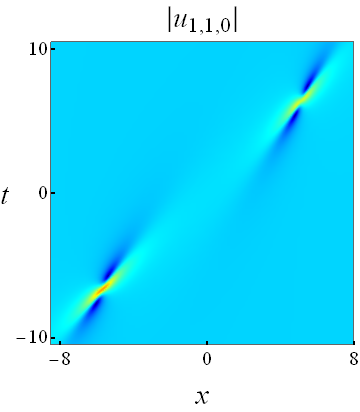}
      \includegraphics[scale=0.320, bb=0 0 385 567]{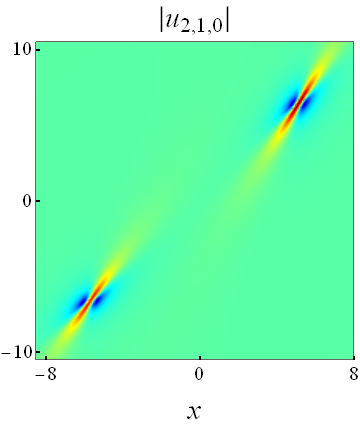}
         \includegraphics[scale=0.320, bb=0 0 385 567]{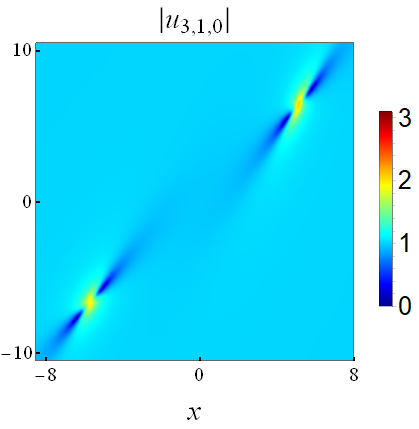}

          \vspace{-1.5cm}
         \includegraphics[scale=0.326, bb=0 0 385 567]{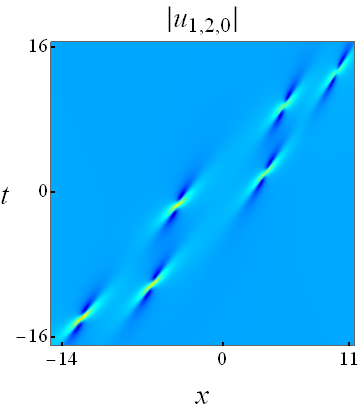}
      \includegraphics[scale=0.320, bb=0 0 385 567]{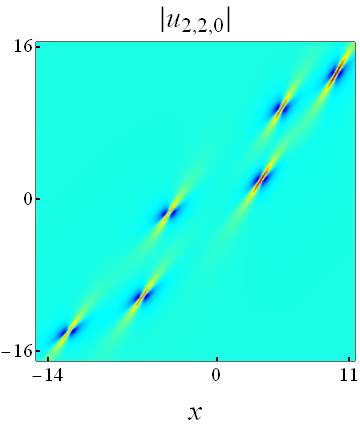}
         \includegraphics[scale=0.320, bb=0 0 385 567]{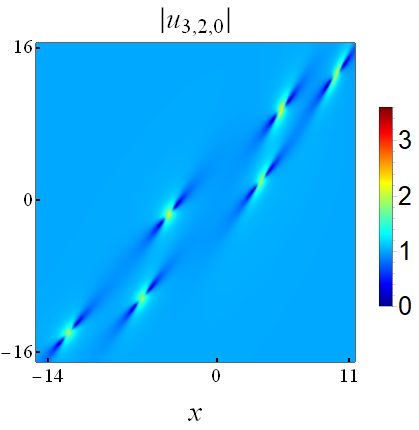}
\caption{Degenerate rogue waves $|u_{i,N_1,N_2}|$ in Theorem 3 with $N_2=0$, for a non-imaginary double root of Eq. (\ref{Q1polynomial2}) in the soliton exchange case (\ref{solexchange}), with background and velocity values (\ref{para_triple}) under relations (\ref{CubicRestrict2}). Upper row: a fundamental rogue wave ($N_1=1$) with $a_{2,1}=10+10\textrm{i}$; lower row: a second-order rogue wave ($N_1=2$) with $a_{2,1}=10+10\textrm{i}$, $a_{4,1}=0$ and $a_{5,1}=20+20\textrm{i}$. }
\end{center}
\end{figure}

\subsubsection{Non-degenerate $2\times 2$ block rogue waves}
If both $N_1>0$ and $N_2>0$, rogue waves $u_{i, N_1, N_2}(x,t)$ given by the $2\times 2$ block determinant (\ref{cubicrwstype3}) in Theorem 3 are new types of rogue solutions to the three wave system (\ref{3WRIModel}). To illustrate these new solutions, we choose $N_{1}=2$ with $N_{2}=1$. This $u_{i, 2, 1}(x,t)$ solution contains free parameters $a_{1,1}, a_{2,1}, a_{4,1}, a_{5,1}$ and $a_{1,2}$. When we choose $a_{1,1}=a_{2,1}=a_{4,1}=a_{1,2}=0$ and $a_{5,1}=30$, the corresponding solution graphs are displayed in Fig. 7 (upper row). It is seen that this rogue wave splits into five $u_{i, 0, 1}(x,t)$ waves of Eq. (\ref{Fundamplitu2}), because the polynomial degree of this $u_{i, 2, 1}(x,t)$ solution is ten (see Remark 5), which is five times that of the $u_{i, 0, 1}(x,t)$ wave. If we choose all parameters to be zero, i.e., $a_{1,1}=a_{2,1}=a_{4,1}=a_{5,1}=a_{1,2}=0$, then we get a super rogue wave which is plotted in the lower row of Fig. 7. It is seen that this super rogue wave does not exhibit a ``watch-hand-like" structure. Instead, a superposition of its three components forms a six-needle, star-like structure.

\begin{figure}[htb]
\begin{center}
\vspace{-1.5cm}
\includegraphics[scale=0.332, bb=0 0 385 567]{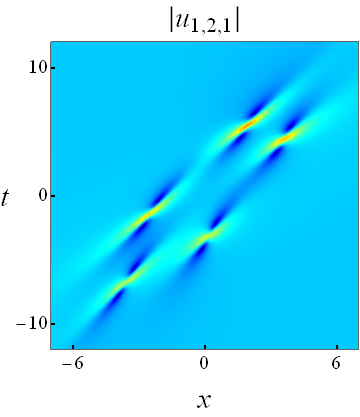}
\includegraphics[scale=0.320, bb=0 0 385 567]{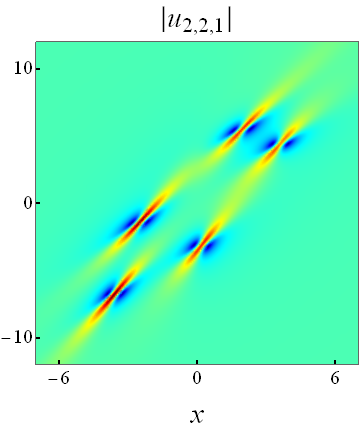}
\includegraphics[scale=0.320, bb=0 0 385 567]{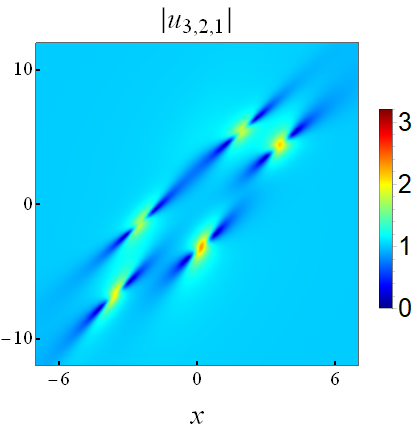}\\
\vspace{-1.5cm}
\includegraphics[scale=0.327, bb=0 0 385 567]{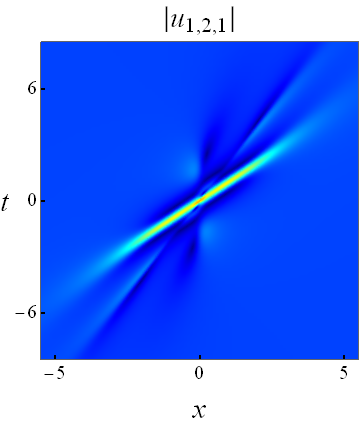}
\includegraphics[scale=0.320, bb=0 0 385 567]{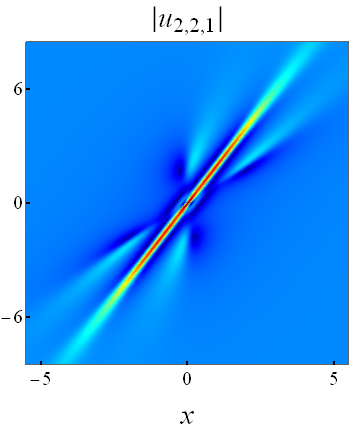}
\includegraphics[scale=0.320, bb=0 0 385 567]{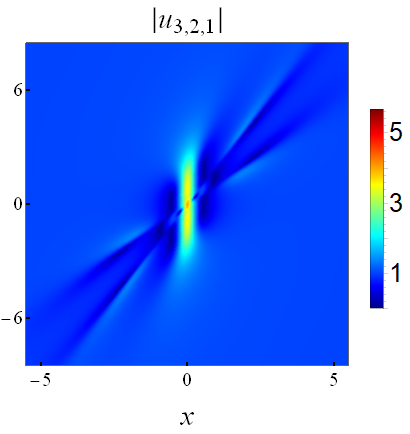}
\caption{Non-degenerate rogue waves $|u_{i,N_1,N_2}|$ with $N_1=2$ and $N_2=1$ in Theorem 3, for a non-imaginary double root of Eq. (\ref{Q1polynomial2}) in the soliton exchange case (\ref{solexchange}), with background and velocity values (\ref{para_triple}) under relations (\ref{CubicRestrict2}). Upper row: the solution for parameters $a_{1,1}=a_{2,1}=a_{4,1}=a_{1,2}=0$ and $a_{5,1}=30$. Lower row: the solution for parameters $a_{1,1}=a_{2,1}=a_{4,1}=a_{5,1}=a_{1,2}=0$. }
\end{center}
\end{figure}

\section{Derivation of rogue-wave solutions}\label{sec:derivation}
In this section, we derive the general rogue-wave solutions given in Theorems 1-3. This derivation uses the bilinear method in the soliton theory \cite{Hirota_book,Jimbo_Miwa}. The bilinear method has been used to derive rogue waves in some other integrable equations before \cite{OhtaJY2012,OhtaJKY2012,OhtaJKY2013,OhtaJKY2014,JCChen2018LS,XiaoeYong2018,YangYang2019Nonloc,YangYangBoussi}. However, bilinear rogue waves in all previous (1+1)-dimensional wave equations only correspond to a simple root of a certain algebraic equation $\mathcal{Q}'_{1}(p)= 0$, where the function $\mathcal{Q}_{1}(p)$ arises in the dimension reduction step of the derivation. The reason was that in all previous cases, the algebraic equation $\mathcal{Q}'_{1}(p)= 0$ only admitted simple roots. For instance, in the NLS equation, $\mathcal{Q}_{1}(p)=p+p^{-1}$; and in the Boussinesq equation, $\mathcal{Q}_{1}(p)=p^3-3p$ \cite{OhtaJY2012,YangYangBoussi}. In both cases, all roots of the equation $\mathcal{Q}'_{1}(p)= 0$ are simple. But in the current three-wave interaction system (\ref{3WRIModel}), this algebraic equation given in (\ref{QurticeqQ1dp}) can admit a double root (see Sec. \ref{subsection_root}). How to derive bilinear rogue waves for this double root of the algebraic equation (\ref{QurticeqQ1dp}) is a new technical question which we will address in this section. Our treatment will make it clear how to bilinearly derive rogue waves for roots of arbitrary multiplicities in general. It turns out that in this double-root case, rogue waves are given through a $2\times 2$ block determinant, and this type of rogue waves has never been realized before. Even when this algebraic equation (\ref{QurticeqQ1dp}) admits only simple roots, a new feature of the three-wave interaction system (\ref{3WRIModel}) is that this equation (\ref{QurticeqQ1dp}) can admit two (unrelated) simple roots (see Sec. \ref{subsection_root}). This new feature gives rise to a new type of rogue waves corresponding to a mixing of these two simple roots, and its derivation requires a block-determinant bilinear solution as well as a new scaling to remove the exponential factors from this bilinear solution. This two-root case will also be treated in this section.

First, we introduce a variable transformation
\begin{eqnarray}
  && u_{1}(x,t)= \rho_{1}\frac{g_{1}}{f} e^{{\rm{i}} (k_1 x+\omega_{1} t)}, \nonumber \\
  && u_{2}(x,t)= \rho_{2}\frac{g_{2}}{f} e^{{\rm{i}} (k_{2}x + \omega_{2} t)}, \label{uftransform} \\
  && u_{3}(x,t)= {\rm{i}}\rho_{3}\frac{g_{3}}{f} e^{-{\rm{i}} [(k_1+k_{2})x + (\omega_{1}+\omega_{2}) t]},  \nonumber
\end{eqnarray}
where $f$ is a real function, and $g_1, g_2, g_3$ are complex functions. Using this transformation and parameter relations (\ref{Pararlation}), the three-wave system (\ref{3WRIModel}) is converted into the following  three  bilinear equations
\begin{eqnarray}
&& \left( D_{t}+c_{1} D_{x} -{\rm{i}}  \gamma_1 \right) g_{1}  \cdot f = -{\rm{i}}   \gamma_1  g_{2}^* g_{3}^*, \nonumber \\
&&   \left( D_{t}+c_{2} D_{x} -{\rm{i}}  \gamma_2   \right) g_{2}  \cdot f = -{\rm{i}}   \gamma_2 g_{1}^* g_{3}^*, \label{1stBilineform}  \\
&&   \left( D_{t} -{\rm{i}}   \gamma_3  \right) g_{3}  \cdot f = -{\rm{i}} \gamma_3 g_{1}^* g_{2}^*,  \nonumber
\end{eqnarray}
where $D$ is Hirota's bilinear differential operator defined by
\begin{eqnarray*}
&&P \left(D_{x}, D_{y}, D_{t},\cdots\right) F(x,y,t,\cdots) \cdot G(x,y,t,\cdots) \\
&& \equiv P\left(\partial_{x}-\partial_{x'}, \partial_{y}-\partial_{y'}, \partial_{t}-\partial_{t'}, \cdots \right) F(x,y,t,\cdots) G(x',y',t',\cdots)|_{x'=x, y'=y, t'=t,\cdots},
\end{eqnarray*}
with $P$ being a polynomial of $D_{x}$, $D_{y}$, $D_{t}$, $\dots$, and the constants $\gamma_1, \gamma_2, \gamma_3$ have been defined in Eq. (\ref{gamma123}).

Next, we introduce a coordinate transformation
\[ \label{VariaTrans1}
x=\frac{c_{1}}{ \gamma_1} r+\frac{c_{2}}{ \gamma_2} s, \quad
t=\frac{1}{ \gamma_1} r + \frac{1}{ \gamma_2} s,
\]
or equivalently,
\[ \label{x1x2}
r=\frac{ \gamma_1 }{c_{1}-c_{2}} \left( x-c_{2}t \right), \quad s=\frac{ \gamma_2}{c_{2}-c_{1}}\left( x-c_{1}t \right).
\]
Under this coordinate transformation, the bilinear equations (\ref{1stBilineform}) reduce to
\begin{eqnarray}
&& \left(  {\rm{i}} D_{r} + 1  \right) g_{1}  \cdot f = g_{2}^* g_{3}^*, \nonumber \\
&&   \left( {\rm{i}} D_{s} + 1  \right) g_{2}  \cdot f = g_{1}^* g_{3}^*, \label{2ndBilineform}  \\
&&   \left[ \frac{  \gamma_1 c_{2} }{ \gamma_3(c_{2}-c_{1})}{\rm{i}}D_{r}-\frac{  \gamma_2 c_{1} }{ \gamma_3(c_{2}-c_{1})}  {\rm{i}}D_{s}  +  1 \right] g_{3}  \cdot f = g_{1}^* g_{2}^*.  \nonumber
\end{eqnarray}
To derive solutions to these (1+1)-dimensional bilinear equations, we consider a higher-dimensional bilinear system
\begin{eqnarray}
&& \left( {\rm{i}} D_{r} +  1 \right) g_{1}  \cdot f =   h_{2} h_{3}, \nonumber \\
&&   \left({\rm{i}} D_{s} + 1 \right) g_{2}  \cdot f =   h_{1} h_{3}, \label{HiorderBiline}  \\
&&   \left({\rm{i}} D_{x_{1}} + 1\right)g_{3}  \cdot f =   h_{1} h_{2}.  \nonumber
\end{eqnarray}
We first construct a wide class of algebraic solutions to this higher-dimensional bilinear system. Then, we restrict these solutions so that they satisfy the dimension-reduction condition
\[\label{DimentionPara}
\left[ \frac{  \gamma_1 c_{2}}{  \gamma_3(c_{2}-c_{1})}   \partial_{r} -\frac{  \gamma_2 c_{1}}{  \gamma_3(c_{2}-c_{1})}   \partial_{s}  -\partial_{x_{1}} \right] \phi =C \phi,
\]
where $\phi$ is any of $f$ and $g_i$, and $C$ is some constant. In addition, we impose the complex conjugacy condition
\[\label{NonlocPara}
h_{i}^*=g_{i}, \quad 1\le i\le 3.
\]
Then, the higher-dimensional bilinear system (\ref{HiorderBiline}) would reduce to the bilinear system (\ref{2ndBilineform}) of the three-wave interaction equations, and the corresponding algebraic solutions would give rogue waves of the three-wave system.

Next, we follow the above outline to derive general rogue-wave solutions to the three-wave system (\ref{3WRIModel}).

\subsection{Gram determinant solutions for a higher-dimensional bilinear system}
From Ref. \cite{YangYangDNLS} and our additional calculations, we learn that if functions $m_{i,j}^{(n,k)}$, $\varphi_{i}^{(n,k)}$ and $\psi_{j}^{(n,k)}$ of variables ($x_{1}$, $r$,  $s$) satisfy the following differential and difference relations,
\begin{eqnarray}
&&\partial_{x_{1}}m_{i,j}^{(n,k)}=\varphi_{i}^{(n,k)}\psi_{j}^{(n,k)},  \nonumber  \\
&&\partial_{x_{1}}\varphi_{i}^{(n,k)}=\varphi_{i}^{(n+1,k)},\  \partial_{x_{1}}\psi_{j}^{(n,k)}=-\psi_{j}^{(n-1,k)}, \nonumber \\
&&\partial_{r}\varphi_{i}^{(n,k)}=\varphi_{i}^{(n,k-1)},\  \partial_{r}\psi_{j}^{(n,k)}=-\psi_{j}^{(n,k+1)},   \label{001} \\
&&\partial_{s}\varphi_{i}^{(n,k)}=\varphi_{i}^{(n-1,k)},\  \partial_{s}\psi_{j}^{(n,k)}=-\psi_{j}^{(n+1,k)},  \nonumber \\
&&\varphi_{i}^{(n+1,k)}= (a-b) \varphi_{i}^{(n,k)}+ \varphi_{i}^{(n,k+1)},\  \psi_{j}^{(n-1,k)}=(a-b)\psi_{j}^{(n,k)}+\psi_{j}^{(n,k-1)},  \nonumber
\end{eqnarray}
where $a$ and $b$ are arbitrary complex constants, then the $\tau$ function
\[\label{Mijdeterminants}
\tau_{n,k}=\det_{1\leq i,j \leq N} \left(m_{i,j}^{(n,k)}\right)
\]
would satisfy the following lowest-order bilinear equations in the extended Kadomtsev-Petviashvili (KP) hierarchy
\begin{eqnarray}
&& \left[ (b-a) D_{r} + 1 \right] \tau_{n+1, k} \cdot \tau_{n, k} = \tau_{n, k+1} \tau_{n+1, k-1},  \nonumber \\
&&  \left[ (b-a) D_{s} + 1 \right] \tau_{n, k-1} \cdot \tau_{n, k} = \tau_{n-1, k} \tau_{n+1, k-1}, \label{KPHBilineEq}\\
&& \left[  D_{x_{1}} + (a-b) \right] \tau_{n-1, k+1} \cdot \tau_{n, k} = (a-b) \tau_{n-1, k} \tau_{n, k+1}. \nonumber
\end{eqnarray}
Indeed, under the above differential and difference relations, these three bilinear equations all reduce to the Jacobi identity for determinants.

Now, we introduce functions $m^{(n,k)}$, $\varphi^{(n,k)}$ and $\psi^{(n,k)}$ as
\begin{eqnarray}
&& m^{(n, k)}= \frac{1}{p + q}\left(-\frac{p-a}{q+a}\right)^{k}\left(-\frac{p-b}{q+b}\right)^{n} \mathrm{e}^{\xi + \eta },  \label{mnk} \\
&& \varphi^{(n,k)}= (p-a)^k (p-b)^n e^{\xi}, \\
&&  \psi^{(n,k)}= \left[-(q+a)\right]^{-k}  \left[-(q+b)\right]^{-n}  e^{\eta},
\end{eqnarray}
where
\begin{eqnarray}
&& \xi =\frac{1}{p-a} r + \frac{1}{p-b} s + (p-b) x_{1}+\xi_{0} , \label{defxi} \\
&& \eta=\frac{1}{q+a} r + \frac{1}{q+b} s +  (q+b) x_{1} + \eta_{0},  \label{defeta}
\end{eqnarray}
and $p, q, \xi_{0}$ and $\eta_{0}$ are arbitrary complex constants. It is easy to see that these functions satisfy the differential and difference relations (\ref{001}) with indices $i$ and $j$ ignored. Then, by defining functions
\[
m_{ij}^{(n,k)}=\mathcal{A}_i \mathcal{B}_{j} m^{(n,k)}, \quad
\varphi_i^{(n,k)}=\mathcal{A}_i\varphi^{(n,k)}, \quad
\psi_j^{(n,k)}=\mathcal{B}_{j}\psi^{(n,k)},    \label{mijn}
\]
where $\mathcal{A}_{i}$ and $\mathcal{B}_{j}$ are differential operators with respect to $p$ and $q$ respectively as
\begin{eqnarray}\label{New003b}
\mathcal{A}_{i}=\frac{1}{ i !}\left[f_{1}(p)\partial_{p}\right]^{i}, \quad
\mathcal{B}_{j}=\frac{1}{ j !}\left[f_{2}(q)\partial_{q}\right]^{j},
\end{eqnarray}
and $f_{1}(p)$, $f_{2}(q)$ are arbitrary functions, these functions would also satisfy the differential and difference relations (\ref{001}) since operators $\mathcal{A}_{i}$ and $\mathcal{B}_{j}$ commute with differentials. Consequently, for an arbitrary sequence of indices $(i_1,i_2,\cdots,i_N)$ and $(j_1,j_2,\cdots,j_N)$, the determinant
\[ \label{tildetaun}
\tau_{n,k}=\det_{1\le\nu, \mu\le N}\left( m_{i_\nu,j_\mu}^{(n,k)}\right)
\]
satisfies the higher-dimensional bilinear system (\ref{KPHBilineEq}).

Next, we will reduce the higher-dimensional bilinear system (\ref{KPHBilineEq}) to the original bilinear system (\ref{2ndBilineform}), so that the higher-dimensional solutions (\ref{tildetaun}) become rogue waves in the three-wave interaction system (\ref{3WRIModel}). By comparing the system (\ref{KPHBilineEq}) with (\ref{HiorderBiline}), we see that we need to set $b-a=\rm{i}$. Our later analysis will show that constants $a$ and $b$ need to be purely imaginary as well. This means that one of these two constants is a free imaginary parameter. But this free imaginary constant can be removed by a parameter shift of $p$ and $q$ (such as $p-a\to p$ and $q+a\to q$), which will not affect rogue wave solutions. Thus, without loss of generality, we will choose
\[ \label{akappa}
a=0, \quad b=\rm{i}
\]
in the $\tau$ function (\ref{tildetaun}) in later analysis.

\subsection{A generalized dimensional reduction procedure} \label{sec:dimred}
Dimension reduction (\ref{DimentionPara}) is a crucial step in the bilinear KP-reduction procedure. This reduction will restrict the indices in the determinant (\ref{tildetaun}), and select the $[f_1(p), f_2(q)]$ functions in the differential operators (\ref{New003b}) as well as the $(p,q)$ values in the matrix element of the $\tau$ function (\ref{tildetaun}). There are at least two ways to perform this reduction, which result in different $\tau$-function expressions \cite{OhtaJY2012,JCChen2018LS,XiaoeYong2018,YangYangBoussi}. We will adopt a generalized version of the $\mathcal{W}$-$p$ treatment we developed in \cite{YangYangBoussi}, which gives simpler rogue-wave expressions. This generalization of our original treatment in \cite{YangYangBoussi} is necessary in order to deal with double roots in the underlying algebraic equation (\ref{QurticeqQ1dp}) for rogue-wave derivations.

Introducing the linear differential operator  $\mathcal{L}_{0}$  as
\[
\mathcal{L}_{0}  =\frac{  \gamma_1 c_{2}}{  \gamma_3(c_{2}-c_{1})}   \partial_{r} -\frac{  \gamma_2 c_{1}}{  \gamma_3(c_{2}-c_{1})}   \partial_{s} -\partial_{x_{1}},
\]
then the dimensional reduction condition (\ref{DimentionPara}) we impose is
\[ \label{dimenredc3WIR}
\mathcal{L}_{0} \tau_{n,k} = C \tau_{n,k},
\]
where $C$ is some constant. It is easy to see that
\[ \label{Lmijnk}
\mathcal{L}_{0} m_{i,j}^{(n,k)}= \mathcal{A}_ {i} \mathcal{B}_{j} \mathcal{L}_{0} m^{(n,k)} = \mathcal{A}_{i} \mathcal{B}_{j}\left[ \mathcal{Q}_{1}(p) + \mathcal{Q}_{2}(q) \right] m^{(n,k)},
\]
where
\[
\mathcal{Q}_{1}(p)= \left( \frac{  \gamma_{1} c_{2}}{\gamma_3(c_{2}-c_{1})} \right) \frac{1 }{p} - \left( \frac{\gamma_{2} c_{1}}{\gamma_3(c_{2}-c_{1})}  \right) \frac{1}{p-\rm{i}} - p,
\]
and
\[
\mathcal{Q}_{2}(q)= \left( \frac{  \gamma_{1} c_{2}}{\gamma_3(c_{2}-c_{1})} \right) \frac{1}{q} - \left( \frac{ \gamma_{2} c_{1}}{\gamma_3(c_{2}-c_{1})} \right) \frac{1}{q+\rm{i}}- q.
\]
Notice that the above $\mathcal{Q}_{1}(p)$ function is the same as that defined in Eq. (\ref{Q1polynomial}).

We should point out that the above choices of $\mathcal{Q}_{1}(p)$ and $\mathcal{Q}_{2}(q)$ functions are not unique. Indeed, for an arbitrary real constant $\chi$, the shifted functions $\mathcal{Q}_{1}(p)+\rm{i}\chi$ and $\mathcal{Q}_{2}(q)-\rm{i}\chi$ would also work (real $\chi$ is required so that the complex conjugacy condition (\ref{mijsym}) in later text can be met). Using such shifted $\mathcal{Q}_{1}(p)$ functions, Theorems 1-3 would also produce valid rogue wave solutions, where the series expansions of $p(\kappa)$ as well as those in Eqs. (\ref{schucoefalpha})-(\ref{schurcoeffsr}) and (\ref{schurcoeffcc}) will change due to this shift [note that this shift of $\mathcal{Q}_{1}(p)$ cannot be removed through a shift of $p$ since we have shifted $p$ to make $a=0$ in Eq. (\ref{akappa})]. However, we have examined some low-order rogue waves resulting from this $\mathcal{Q}_{1}(p)$ shift and found them to be equivalent to the ones without shift when free parameters (such as $a_r$) in those two sets of solutions are properly related. We believe that this equivalence of solutions under the $\mathcal{Q}_{1}(p)$ shift holds for rogue waves of all orders as well.

To meet the dimensional reduction condition (\ref{dimenredc3WIR}), we start with the general operator relation,
\[
\mathcal{A}_{i} \mathcal{Q}_{1}(p) =\sum_{l=0}^{i}  \frac{1}{ l !} \left[ \left( f_{1}\partial_{p} \right)^{l} \mathcal{Q}_{1}(p)\right] \mathcal{A}_{i-l}.
\]
This relation can be seen from the Leibnitz rule after we rewrite $f_{1}(p)$ as $\mathcal{W}_{1}(p)/\mathcal{W}_{1}'(p)$, so that $f_{1}\partial_{p}$ becomes $\partial_{\ln \mathcal{W}_{1}}$.
Note that on the right side of this relation, the operator $\left( f_{1}\partial_{p} \right)^{l}$ only applies to the function $\mathcal{Q}_{1}(p)$, not to the operator $\mathcal{Q}_{1}(p)\mathcal{A}_{i-l}$. Another relation similar to the above can also be written for $\mathcal{B}_{j} \mathcal{Q}_{2}(q)$. Using these relations, Eq. (\ref{Lmijnk}) gives
\begin{eqnarray}\label{dimentionreduc1}
\mathcal{L}_{0} \hspace{0.05cm} m_{i,j}^{(n,k)}=  \sum_{\mu=0}^i \frac{1}{\mu !} \left[ \left( f_{1}\partial_{p} \right)^{\mu} \mathcal{Q}_{1}(p)\right]   m_{i-\mu,j}^{(n,k)}+\sum_{l=0}^j \frac{1}{l!}
 \left[ \left( f_{2}\partial_{q} \right)^{l} \mathcal{Q}_{2}(q) \right] m_{i,j-l}^{(n,k)}.
\end{eqnarray}

In order to satisfy the dimensional reduction condition (\ref{dimenredc3WIR}), we need to select functions $[f_1(p), f_2(q)]$ as well as values of $(p, q)$ so that coefficients of certain indices on the right side of the above equation vanish \cite{OhtaJY2012}. For that purpose, we will select $p_0$ and $q_0$ values to be roots of the following algebraic equations
\[ \label{p0q0}
\mathcal{Q}'_{1}(p_0)=0, \quad \mathcal{Q}'_{2}(q_0)=0.
\]
At these $(p_0, q_0)$ values, the $\mu=l=1$ terms on the right side of Eq.~(\ref{dimentionreduc1}) will vanish.
Notice that the $\mathcal{Q}'_{1}(p_0)=0$ equation above is the same as (\ref{QurticeqQ1dp}), whose root structure has been delineated in Sec. \ref{subsection_root}. Roots of the $\mathcal{Q}'_{2}(q_0)=0$ equation are related to those of $\mathcal{Q}'_{1}(p_0)=0$ as
\[ \label{p0q0b}
q_0=p_0^*.
\]
Since the $m^{(n, k)}$ function in (\ref{mnk}) has a factor of $1/(p+q)$, in order for $m_{ij}^{(n,k)}$ in (\ref{mijn}) to be nonsingular when evaluated at $(p, q)=(p_0, q_0)$, the $p_0$ value cannot be purely imaginary.

To select $f_1(p)$ and $f_2(q)$ functions, we need to impose further conditions, and these conditions will depend on the multiplicity of the root $p_0$ in the $\mathcal{Q}'_{1}(p)=0$ equation.

\subsubsection{A simple root}
If $p_0$ is a simple root to the $\mathcal{Q}'_{1}(p)=0$ equation, the condition on $f_1(p)$ we impose will be
\[ \label{2ndordlinODE}
\left( f_{1} \partial_{p} \right)^{2} \mathcal{Q}_{1}(p) = \mathcal{Q}_{1}(p).
\]
Note that this is a differential equation, not an operator equation. The reason for this imposition is that under this condition, as well as the earlier condition (\ref{p0q0}), all odd-$\mu$ terms on the right side of Eq. (\ref{dimentionreduc1}), when evaluated at $p=p_0$, would vanish. To solve this differential equation (\ref{2ndordlinODE}), we put $f_1(p)$ in the form
\[ \label{functiondiff}
f_{1}(p) = \frac{\mathcal{W}_{1}(p)}{\mathcal{W}_{1}'(p)},
\]
where $\mathcal{W}_{1}(p)$ is to be determined. In this form, the condition (\ref{2ndordlinODE}) becomes
\[ \label{2ndordlinODE1}
\partial_{\ln \mathcal{W}_{1} }^{2}  \mathcal{Q}_{1}(p) = \mathcal{Q}_{1}(p).
\]
Scaling $\mathcal{W}_{1}(p_0)=1$, which does not affect the $f_1(p)$ function, the unique solution to the above equation under the condition of $\mathcal{Q}'_{1}(p_0)=0$ is
\[ \label{2ndodesoluQ1p}
\mathcal{Q}_{1}(p) = \frac{1}{2}\mathcal{Q}_{1}(p_0) \left(\mathcal{W}_{1}(p) + \frac{1}{\mathcal{W}_{1}(p)} \right).
\]
From this equation, we get
\[ \label{W1p}
\mathcal{W}_{1}(p)=\frac{\mathcal{Q}_{1}(p)\pm \sqrt{\mathcal{Q}_{1}^2(p)-\mathcal{Q}_{1}^2(p_0)}}{\mathcal{Q}_{1}(p_0)},
\]
and thus $f_1(p)$ can be obtained from (\ref{functiondiff}) as
\[ \label{diffoperaf1p}
f_{1}(p)=\pm \frac{\sqrt{\mathcal{Q}_{1}^2(p)-\mathcal{Q}_{1}^2(p_0)}}{\mathcal{Q}'_{1}(p)}.
\]
This new derivation of $f_1(p)$ reproduces that in the original $\mathcal{W}$-$p$ treatment of the Boussinesq equation in \cite{YangYangBoussi}. It also reproduces $f_1(p)=\pm p$ for the NLS equation in \cite{OhtaJY2012} and $f_1(p)=\pm (p+\rm{i}\alpha)$ for the generalized derivative NLS equations in \cite{YangYangDNLS}. Notice that even though $\mathcal{Q}'_{1}(p_0)=0$, $f_1(p)$ still has a limit when $p\to p_0$, and hence $f_1(p_0)$ is well-defined. This $f_1(p)$ function has two sign choices. But we can readily see that these two signs lead to equivalent rogue wave solutions. In fact, these two signs correspond to the two branches of $p(\kappa)$ solutions in Eq. (\ref{defpk}), which yield equivalent rogue waves (see Remark 3).

A similar treatment can be applied to the $q$ variable, and the results for $f_2(q)$ and $\mathcal{W}_{2}(q)$ are the same as (\ref{W1p})-(\ref{diffoperaf1p}), except that the variable subscript 1 changes to 2, and $(p, p_0)$ change to $(q, q_0)$.

Due to the condition (\ref{2ndordlinODE}) and $\mathcal{Q}'_{1}(p_0)=0$, as well as similar ones for the $q$ variable, we find from Eq. (\ref{dimentionreduc1}) that
\[ \label{contigurelati}
  \mathcal{L}_{0} \left. m_{i,j}^{(n,k)} \right|_{p=p_{0}, \ q=q_{0}}= \mathcal{Q}_{1}(p_0) \sum^i_{\begin{subarray}{c} \mu=0\\ \mu: \hspace{0.02cm} even \end{subarray}}
  \frac{1}{\mu !} \left.m_{i-\mu,j}^{(n,k)}\right|_{p=p_{0}, \ q=q_{0}}+ \mathcal{Q}_{2}(q_0) \sum_{\begin{subarray}{c} l=0\\ l: \hspace{0.02cm} even \end{subarray}}^{j}\frac{1}{l!} \left. m_{i,j-l}^{(n,k)}\right|_{p=p_{0}, \ q=q_{0}}.
\]
Then, when we restrict indices of the general determinant (\ref{tildetaun}) to
\[ \label{tausol}
\tau_{n,k} = \det_{1\leq i, j\leq N}\left(\left. m_{2i-1,2j-1}^{(n,k)} \right|_{p=p_{0}, \ q=q_{0}} \right),
\]
and use the above contiguity relation (\ref{contigurelati}) as was done in Ref. \cite{OhtaJY2012}, we get
\begin{eqnarray}\label{ReductionCondi}
\mathcal{L}_{0} \tau_{n,k} = \left[\mathcal{Q}_{1}(p_0) + \mathcal{Q}_{2}(q_0) \right] \hspace{0.02cm} N \hspace{0.04cm}  \tau_{n,k}.
\end{eqnarray}
Thus, the $\tau_{n,k}$ function (\ref{tausol}) satisfies the dimensional reduction condition (\ref{dimenredc3WIR}).

If we compare the above dimension reduction procedure with the original $\mathcal{W}$-$p$ method proposed in \cite{YangYangBoussi}, we can see that the current technique reproduces all results of the previous method. However, the current technique is more general. More importantly, it can be readily extended to treat roots of higher multiplicities in the $\mathcal{Q}'_{1}(p)=0$ equation, as we will see shortly in Sec. \ref{subsectriple}.

\subsubsection{Two simple roots} \label{sec:2roots}
If the $\mathcal{Q}'_{1}(p)=0$ equation admits two simple roots $(p_{0,1}, p_{0,2})$, then we can construct a more general $2\times 2$ block determinant
\[ \label{2x2blotausdet1}
\tau_{n,k} = \det \left( \begin{array}{cc}
                           \tau_{n,k}^{\left[1,1\right]} & \tau_{n,k}^{\left[1,2\right]} \\
                           \tau_{n,k}^{\left[2,1\right]} & \tau_{n,k}^{\left[2,2\right]}
                         \end{array}
 \right),
\]
where
\[ \label{2x2blotaudet2}
\tau_{n,k}^{\left[I,J\right]} = \mbox{mat}_{1\leq i \le N_{I}, 1\leq j\leq N_{J}}\left(\left. m_{2i-1,2j-1}^{(n,k)} \right|_{p=p_{0,I}, q=q_{0,J}} \right),\ \ \ 1\leq I,J \leq 2,
\]
$m_{i,j}^{(n,k)}$ is given by Eqs. (\ref{mnk})-(\ref{mijn}) with $[f_1(p), f_2(q)]$ replaced by $[f_1^{(I)}(p), f_2^{(J)}(q)]$, the function $f_1^{(I)}(p)$ is provided by Eq.~(\ref{diffoperaf1p}) with $p_0$ replaced by $p_{0,I}$, the function $f_2^{(J)}(q)$ is the same as (\ref{diffoperaf1p}) but with the variable subscript 1 changing to 2 and $(p, p_0)$ changing to $(q, q_{0,J})$, with
\[ \label{q0J}
q_{0,J}=p_{0,J}^*,
\]
$\xi_0$ is replaced by $\xi_{0,I}$, $\eta_0$ is replaced by $\eta_{0,J}$, and $N_1, N_2$ are arbitrary positive integers. This $2\times 2$ block determinant (\ref{2x2blotausdet1}) also satisfies the higher-dimensional bilinear system (\ref{KPHBilineEq}), and its proof will be provided in Appendix C.

Since the $m_{ij}^{(n,k)}$ function contains a factor of $1/(p+q)$ in view of (\ref{mnk}), the matrix elements in the block determinant (\ref{2x2blotausdet1}) would contain factors of $1/(p_{0,I}+q_{0,J})$ $(1\le I,J\le 2)$. In order for these factors to be nonsingular, we must require $(p_{0,1}, p_{0,2})$ non-imaginary and $p_{0,2}\ne -p_{0,1}^*$ in view of Eq. (\ref{q0J}).

In the present case, the contiguity relation (\ref{contigurelati}) becomes
\[  \label{newcontigrelat}
  \mathcal{L}_{0} \left. m_{i,j}^{(n,k)} \right|_{p=p_{0,I}, \ q=q_{0,J}}= \mathcal{Q}_{1}(p_{0,I}) \sum^i_{\begin{subarray}{c} \mu=0\\ \mu: \hspace{0.02cm} even \end{subarray}}
  \frac{1}{\mu !} \left.m_{i-\mu,j}^{(n,k)}\right|_{p=p_{0,I}, \ q=q_{0,J}}+ \mathcal{Q}_{2}(q_{0,J}) \sum_{\begin{subarray}{c} l=0,\\ l: \hspace{0.02cm} even\end{subarray}}^{j}\frac{1}{l!} \left. m_{i,j-l}^{(n,k)}\right|_{p=p_{0,I}, \ q=q_{0,J}}.
\]
Utilizing this contiguity relation similar to Ref. \cite{OhtaJY2012}, we get
\begin{eqnarray}\label{NewReducCondi}
\mathcal{L}_{0} \tau_{n,k} = \left\{\left[\mathcal{Q}_{1}(p_{0,1})+\mathcal{Q}_{2}(q_{0,1})\right]N_1+\left[\mathcal{Q}_{1}(p_{0,2})+\mathcal{Q}_{2}(q_{0,2})\right]N_2\right\}
\hspace{0.04cm}  \tau_{n,k}.
\end{eqnarray}
Thus, the $2\times 2$ block determinant (\ref{2x2blotausdet1}) also satisfies the dimensional reduction condition (\ref{dimenredc3WIR}).

\subsubsection{A double root} \label{subsectriple}
If $p_0$ is a double root to the $\mathcal{Q}'_{1}(p)=0$ equation, i.e.,
\[ \label{Q1p0cond}
\mathcal{Q}'_{1}(p_0)=\mathcal{Q}''_{1}(p_0)=0,
\]
then the previous condition (\ref{2ndordlinODE}) for $f_1(p)$ cannot be satisfied, because evaluation of that condition at $p=p_0$ would give $Q_1(p_0)=0$, which is not true. In this double-root case, the new condition on $f_1(p)$ will need to be
\[ \label{3rdordlinode}
\left( f_{1} \partial_{p} \right)^{3} \mathcal{Q}_{1}(p) = \mathcal{Q}_{1}(p).
\]
With $f_{1}$ in the same form as (\ref{functiondiff}), this condition is
\[ \label{3rdordlinODE}
\partial_{\ln\mathcal{W}_{1}}^{3} \mathcal{Q}_{1}(p) = \mathcal{Q}_{1}(p).
\]
Scaling $\mathcal{W}_{1}(p_0)=1$, the unique solution to this equation under conditions (\ref{Q1p0cond}) is
\[ \label{3odeGeneSoluQ1}
\mathcal{Q}_{1}(p) = \frac{\mathcal{Q}_{1}(p_0)}{3} \left( \mathcal{W}_{1}(p) +\frac{2}{\sqrt{\mathcal{W}_{1}(p)}}
\cos\left[ \frac{\sqrt{3}}{2} \ln \mathcal{W}_{1}(p) \right] \right).
\]
From this equation, one can solve for $\mathcal{W}_{1}(p)$ and then obtain $f_1(p)$ through (\ref{functiondiff}). Alternatively, one can derive $f_1(p)$ directly from the condition (\ref{3rdordlinode}) by expanding both $f_1(p)$ and $\mathcal{Q}_{1}(p)$ into Taylor series around $p=p_0$. Similar results can be obtained for $f_2(q)$.

Under conditions (\ref{Q1p0cond})-(\ref{3rdordlinode}) and similar ones for the $q$ variable, Eq. (\ref{dimentionreduc1}) can be simplified as
\[  \label{tricontigurel}
  \mathcal{L}_{0} \left. m_{i,j}^{(n,k)} \right|_{p=p_{0}, \ q=q_{0}}= \mathcal{Q}_{1}(p_0) \sum^i_{\begin{subarray}{c} \mu=0\\ \mu \equiv 0
  (\textbf{mod} 3) \end{subarray}}
  \frac{1}{\mu !} \left.m_{i-\mu,j}^{(n,k)}\right|_{p=p_{0}, \ q=q_{0}}+ \mathcal{Q}_{2}(q_0) \sum_{\begin{subarray}{c} l=0\\ l \equiv 0
  (\textbf{mod} 3) \end{subarray}}^{j}\frac{1}{l!} \left. m_{i,j-l}^{(n,k)}\right|_{p=p_{0}, \ q=q_{0}}.
\]
Using this contiguity relation, we can show as in Ref. \cite{OhtaJY2012} that the $2\times 2$ block determinant
\[ \label{cubicrwstype3p}
\tau_{n,k}=\det \left(
\begin{array}{cc}
  \tau^{[1,1]}_{n,k} & \tau^{[1,2]}_{n,k} \\
  \tau^{[2,1]}_{n,k} & \tau^{[2,2]}_{n,k}
\end{array} \right),
\]
where
\[ \label{Blockmatrixp}
\tau^{[I, J]}_{n,k}=\mbox{mat}_{1\leq i \le N_{I}, 1\leq j\leq N_{J}} \left(\left.
m_{3i-I, \, 3j-J}^{(n,k)}\right|_{p=p_{0}, \hspace{0.06cm} q=q_{0}, \hspace{0.06cm} \xi_0=\xi_{0I}, \hspace{0.06cm} \eta_0=\eta_{0J}}\right), \quad 1\leq I,J \leq 2,
\]
$m_{i,j}^{(n,k)}$ is given by Eqs. (\ref{mnk})-(\ref{mijn}), $q_0=p_0^*$, and $N_1, N_2$ are non-negative integers, satisfies the dimensional reduction condition (\ref{ReductionCondi}).
This $2\times 2$ block determinant (\ref{cubicrwstype3p}) clearly also satisfies the higher-dimensional bilinear system (\ref{KPHBilineEq}) for reasons similar to that given in Appendix C.

When the dimensional reduction condition is satisfied, we can use it to eliminate $x_{1}$ from the higher-dimensional bilinear system (\ref{KPHBilineEq}). Then, in view of the parameter choices in (\ref{akappa}), we get
\begin{eqnarray}
&& \left[ {\rm{i}} D_{r} + 1 \right] \tau_{n+1, k} \cdot \tau_{n, k} = \tau_{n, k+1} \tau_{n+1, k-1},  \nonumber \\
&&  \left[ {\rm{i}} D_{s} + 1 \right] \tau_{n, k-1} \cdot \tau_{n, k} = \tau_{n-1, k} \tau_{n+1, k-1}, \label{HdimBilineEq}\\
&& \left[ \frac{   \gamma_{1} c_{2}}{ \gamma_3(c_{2}-c_{1})} {\rm{i}}D_{r} -\frac{  \gamma_{2} c_{1}}{ \gamma_3(c_{2}-c_{1})} {\rm{i}}D_{s} +  1 \right] \tau_{n-1, k+1} \cdot \tau_{n, k} = \tau_{n-1, k} \tau_{n, k+1}.  \nonumber
\end{eqnarray}

\subsection{Complex conjugacy condition} \label{sec:reality}
We now impose the complex conjugacy condition
\[ \label{ComplConjuTaunf}
\tau_{-n,-k} = \tau_{n,k}^*.
\]
This condition can be satisfied by imposing the parameter constraint
\[
\xi_{0}=\eta_{0}^*
\]
in Eq. (\ref{tausol}) for a simple root, and $\xi_{0,I}=\eta_{0,I}^*$ in Eqs. (\ref{2x2blotausdet1}) and (\ref{cubicrwstype3p}) for two simple roots and a double root. Indeed, for a simple root under this constraint and in view that $q_{0}=p_{0}^*$, we can show that $[f_1(p_0)]^*=f_2(q_0)$, and
\[ \label{mijsym}
 \left. m_{j,i}^{(-n, -k)}\right|_{p=p_{0}, \ q=q_{0}} = \left. \left[m_{i,j}^{(n,k)}\right]^* \right|_{p=p_{0}, \ q=q_{0}}.
\]
Thus, the condition (\ref{ComplConjuTaunf}) holds. In the case of two simple roots, since $q_{0,I}=p_{0,I}^*$ and $\xi_{0,I}=\eta_{0,I}^*$, we can show that
\[
\left. m_{i,j}^{(n,k)} \right|_{p=p_{0,I}, \ q=q_{0,J}} = \left. \left[m_{j,i}^{(-n, -k)}\right]^* \right|_{p=p_{0,J}, \ q=q_{0,I}},
\]
so that
\[
\tau_{n,k}^{\left[I,J\right]}= \left[ \tau_{-n,-k}^{\left[J,I\right]}\right]^*.
\]
Thus, the complex conjugacy condition (\ref{ComplConjuTaunf}) holds as well. The proof for the double-root case is similar to the two-simple-roots case.

Lastly, we define
\[ \label{fgdef1}
f=\tau_{0,0}, \quad g_{1}=\tau_{1,0}, \quad  g_{2}=\tau_{0,-1}, \quad g_{3}=\tau_{-1,1},
\]
and
\[ \label{fgdefhi}
h_{1}=\tau_{-1,0}, \quad  h_{2}=\tau_{0,1}, \quad h_{3}=\tau_{1,-1},
\]
where $\tau_{n,k}$ is as defined in any of the equations (\ref{tausol}), (\ref{2x2blotausdet1}), and (\ref{cubicrwstype3p}). Then, due to the above complex conjugacy conditions, we see that in all these cases, $h_{i}^*=g_{i}$. Thus, these $f$ and $g_i$ functions satisfy the original bilinear system (\ref{2ndBilineform}), and they give rational solutions to the three-wave equations through the transformation (\ref{uftransform}).

\subsection{Introduction of free parameters}
Now, it is time to introduce free parameters into these rational solutions. As we did previously for the derivative NLS equations in \cite{YangYangDNLS}, we will introduce these free parameters through the arbitrary constant $\xi_0$ in Eqs. (\ref{defxi}) and (\ref{tausol}) for a simple root, and through $\xi_{0,I}$ in Eqs. (\ref{2x2blotausdet1}) and (\ref{cubicrwstype3p}) for two simple roots and a double root. Specifically, for the $\tau_{n,k}$ function in Eq. (\ref{tausol}) for a simple root $p_0$, we choose $\xi_0$ as
\[ \label{defxi0}
\xi_0=\sum _{r=1}^\infty \hat{a}_{r}  \ln^r \mathcal{W}_{1}(p),
\]
where $\mathcal{W}_{1}(p)$ is defined in Eq. (\ref{W1p}), and $\hat{a}_r$ are free complex constants. For the $\tau_{n,k}$ function in Eq. (\ref{2x2blotausdet1}) for two simple roots, we choose $\xi_{0,I}$ as
\[ \label{defxi0I}
\xi_{0,I}=\sum _{r=1}^\infty a_{r,I}  \ln^r \mathcal{W}_{1}^{(I)}(p), \quad I=1, 2,
\]
where $\mathcal{W}_{1}^{(I)}(p)$ is as defined in Eq. (\ref{W1p}) with $p_0$ replaced by $p_{0,I}$, and $a_{r,I}$ are free complex constants. And for the $\tau_{n,k}$ function in Eq. (\ref{cubicrwstype3p}) for a double root, we choose $\xi_{0,I}$ as
\[ \label{defxi0I2}
\xi_{0,I}=\sum _{r=1}^\infty \hat{a}_{r,I}  \ln^r \mathcal{W}_{1}(p), \quad I=1, 2,
\]
where $\mathcal{W}_{1}(p)$ is defined in Eq. (\ref{3odeGeneSoluQ1}), and $\hat{a}_{r,I}$ are free complex constants.

Compared to the old parameterization in Ref.~\cite{OhtaJY2012}, this new parameterization allows us to eliminate the summations in differential operators $\mathcal{A}_{i}$ and $\mathcal{B}_{j}$ in Eq. (\ref{New003b}). One may think that the above parameterization is difficult since the functions $\mathcal{W}_{1}(p)$ and $\mathcal{W}_{1}^{(I)}(p)$ from equations such as (\ref{2ndodesoluQ1p}) and
(\ref{3odeGeneSoluQ1}) are complicated. This may be so if one tries to derive the rogue solutions from the differential operator form (see Sec. \ref{sec_diff} below). However, these complications from the $\mathcal{W}_{1}(p)$ and $\mathcal{W}_{1}^{(I)}(p)$ functions will disappear when the rogue solutions are expressed through Schur polynomials, as we will see in Sec. \ref{SchurPolynoExp}.

\subsection{Regularity of solutions}
Using arguments very similar to that in \cite{OhtaJY2012}, we can show that these rational solutions are bounded for all signs of nonlinearity $(\epsilon_1, \epsilon_2, \epsilon_3)$, i.e., for all soliton-exchange, explosive and stimulated backscatter cases (\ref{solexchange})-(\ref{scatter2}). This regularity of solutions for the explosive case is noteworthy, since in this case localized initial conditions in the three-wave system (\ref{3WRIModel}) can explode to infinity in finite time \cite{Kaupreview1979}.

\subsection{Rational solutions in differential operator form}  \label{sec_diff}
Putting all the above results together and setting $x_{1}=0$, regular rational solutions to the three-wave interaction system (\ref{3WRIModel}) are given by the following theorems.
\begin{quote}
\textbf{Theorem 4} \hspace{0.05cm} \emph{If the algebraic equation (\ref{Q1polynomial2}) admits a non-imaginary simple root $p_0$, then the three-wave interaction system (\ref{3WRIModel}) admits regular rational solutions given by Eqs. (\ref{uftransform}) and (\ref{fgdef1}), where
\[ \label{sigmank12}
\tau_{n,k}=
\det_{
\begin{subarray}{l}
1\leq i, j \leq N
\end{subarray}
}
\left(
\begin{array}{c}
m_{2i-1,2j-1}^{(n,k)}
\end{array}
\right),
\]
the matrix elements in $\tau_{n,k}$ are defined by}
\[ \label{mij-diff}
m_{i,j}^{(n,k)}=\left. \mathcal{A}_i \mathcal{B}_{j}m^{(n,k)} \right|_{p=p_{0}, \ q=p_{0}^*},
\]
\begin{eqnarray}
m^{(n,k)}=\frac{1}{p + q}\left(-\frac{p}{q}\right)^{k}\left(-\frac{p-\rm{i}}{q+\rm{i}}\right)^{n} e^{\Theta(x,t)},  \label{mnk3}
\end{eqnarray}
\begin{eqnarray} \label{Thetaxt1}
&& \Theta(x,t)=  \frac{  \gamma_{1} \left( x-c_{2}t \right) }{c_{1}-c_{2}}  \left( \frac{1}{p} + \frac{1}{q} \right)+ \frac{ \gamma_{2} \left( x-c_{1}t \right) }{c_{2}-c_{1}}   \left( \frac{1}{p-\rm{i}} + \frac{1}{q+\rm{i}} \right) \nonumber  \\
&& \hspace{1.0cm} +  \sum _{r=1}^\infty  \hat{a}_{r} \ln^r \mathcal{W}_{1}(p) + \sum _{r=1}^\infty \hat{a}^*_{r}  \ln^r \mathcal{W}_{2}(q),
\end{eqnarray}
\emph{$\mathcal{A}_i$ and $\mathcal{B}_{j}$ are given in Eq. (\ref{New003b}), $f_{1}(p)$ and $\mathcal{W}_{1}(p)$ are given by Eqs. (\ref{W1p})-(\ref{diffoperaf1p}), $f_2(q)$ and $\mathcal{W}_{2}(q)$ are the same as (\ref{W1p})-(\ref{diffoperaf1p}) except that the variable subscript 1 changes to 2 and $(p, p_0)$ change to $(q, p_0^*)$, and $\hat{a}_{r} \hspace{0.05cm} (r=1, 2, \dots)$ are free complex constants.}
\end{quote}

\begin{quote}
\textbf{Theorem 5} \hspace{0.05cm} \emph{If the algebraic equation (\ref{Q1polynomial2}) admits two non-imaginary simple roots $(p_{0,1}, p_{0,2})$ with $p_{0,2}\ne -p_{0,1}^*$, then the three-wave interaction system (\ref{3WRIModel}) admits regular rational solutions given by Eqs. (\ref{uftransform}) and (\ref{fgdef1}), where $\tau_{n,k}$ is a $2\times 2$ block determinant}
\[ \label{2x2blotausdet1b}
\tau_{n,k} = \det \left( \begin{array}{cc}
                           \tau_{n,k}^{\left[1,1\right]} & \tau_{n,k}^{\left[1,2\right]} \\
                           \tau_{n,k}^{\left[2,1\right]} & \tau_{n,k}^{\left[2,2\right]}
                         \end{array} \right),
\]
\[ \label{sigmank2c}
\tau_{n,k}^{\left[I,J\right]} = \left( m_{2i-1,2j-1}^{(n,k,I,J)}   \right)_{1\leq i \le N_{I}, 1\leq j\leq N_{J}},
\]
\emph{$N_1$ and $N_2$ are positive integers, the matrix elements in $\tau_{n,k}^{\left[I,J\right]}$ are defined by}
\begin{eqnarray} \label{mijIJ-diff}
&& m_{i,j}^{(n,k,I,J)}=
  \frac{\left[f_{1}^{(I)}(p)\partial_{p}\right]^{i}}{ i !}\frac{\left[f_{2}^{(J)}(q) \partial_{q}\right]^{j}}{ j !} \left. m^{(n,k,I,J)}
   \right|_{p=p_{0,I}, \hspace{0.05cm} q=p_{0,J}^*},
\end{eqnarray}
\begin{eqnarray}
&& m^{(n,k,I,J)}=
\frac{1}{p + q}\left(-\frac{p}{q}\right)^{k}\left(-\frac{p-\rm{i}}{q+\rm{i}}\right)^{n} e^{\Theta_{I,J}(x,t)},
\end{eqnarray}
\begin{eqnarray}
&& \Theta_{I,J} (x,t) = \frac{  \gamma_{1} \left( x-c_{2}t \right) }{c_{1}-c_{2}}  \left( \frac{1}{p} + \frac{1}{q} \right)+ \frac{ \gamma_{2} \left( x-c_{1}t \right) }{c_{2}-c_{1}}   \left( \frac{1}{p-\rm{i}} + \frac{1}{q+\rm{i}} \right) \nonumber  \\
&& \hspace{1.0cm} +\sum _{r=1}^\infty  a_{r,I} \ln^r \mathcal{W}_{1}^{(I)}(p) + \sum _{r=1}^\infty a^*_{r,J}  \ln^r \mathcal{W}_{2}^{(J)}(q),
\end{eqnarray}
\emph{$f_{1}^{(I)}(p), f_{2}^{(J)}(q)$ are given in Sec. \ref{sec:2roots},
$\mathcal{W}_{1}^{(I)}(p)$ is defined in Eq. (\ref{W1p}) with $p_0$ replaced by $p_{0,I}$, $\mathcal{W}_{2}^{(J)}(q)$ is defined similar to Eq. (\ref{W1p}) except that the variable subscript 1 changes to 2 and $(p, p_0)$ change to $(q, p_{0,J}^*)$, and $a_{r,1}, a_{r,2} \hspace{0.05cm} (r=1, 2, \dots)$ are free complex constants.}
\end{quote}

\begin{quote}
\textbf{Theorem 6} \hspace{0.05cm}
\emph{If the algebraic equation (\ref{Q1polynomial2}) admits a double root $p_{0}$, then the three-wave interaction system (\ref{3WRIModel}) admits regular rational solutions given by Eqs. (\ref{uftransform}) and (\ref{fgdef1}), where}
 \[ \label{2x2blotausdet1c}
\tau_{n,k} = \det \left( \begin{array}{cc}
                           \tau_{n,k}^{\left[1,1\right]} & \tau_{n,k}^{\left[1,2\right]} \\
                           \tau_{n,k}^{\left[2,1\right]} & \tau_{n,k}^{\left[2,2\right]}
                         \end{array}
 \right),
\]
\[\label{Blockmatrixc}
\tau^{[I, J]}_{n,k}=
\left(
m_{3i-I, \, 3j-J}^{(n,k, \hspace{0.04cm} I, J)}
\right)_{1\leq i \leq N_{I}, \, 1\leq j \leq N_{J}},
\]
\emph{$N_1$ and $N_2$ are non-negative integers, the matrix elements in $\tau_{n,k}^{\left[I,J\right]}$ are defined by}
\begin{eqnarray} \label{mijIJ-diff3}
&& m_{i,j}^{(n,k,I,J)}=
  \frac{\left[f_{1}(p)\partial_{p}\right]^{i}}{ i !}\frac{\left[f_{2}(q) \partial_{q}\right]^{j}}{ j !} \left. m^{(n,k,I,J)}
   \right|_{p=p_{0}, \hspace{0.05cm} q=p_{0}^*},
\end{eqnarray}
\begin{eqnarray}
&& m^{(n,k,I,J)}=
\frac{1}{p + q}\left(-\frac{p}{q}\right)^{k}\left(-\frac{p-\rm{i}}{q+\rm{i}}\right)^{n} e^{\Theta_{I,J}(x,t)},
\end{eqnarray}
\begin{eqnarray}
&& \Theta_{I,J} (x,t) = \frac{  \gamma_{1} \left( x-c_{2}t \right) }{c_{1}-c_{2}}  \left( \frac{1}{p} + \frac{1}{q} \right)+ \frac{ \gamma_{2} \left( x-c_{1}t \right) }{c_{2}-c_{1}}   \left( \frac{1}{p-\rm{i}} + \frac{1}{q+\rm{i}} \right) \nonumber  \\
&& \hspace{1.0cm} +\sum _{r=1}^\infty  a_{r,I} \ln^r \mathcal{W}_{1}(p) + \sum _{r=1}^\infty a^*_{r,J}  \ln^r \mathcal{W}_{2}(q),
\end{eqnarray}
\emph{$\mathcal{W}_{1}(p)$ is given by Eq. (\ref{3odeGeneSoluQ1}), $f_1(p)$ is given through $\mathcal{W}_{1}(p)$ by Eq. (\ref{functiondiff}), $\mathcal{W}_{2}(q)$ and $f_2(q)$ are given by the same equations (\ref{functiondiff}) and (\ref{3odeGeneSoluQ1}) but with the variable subscript 1 changing to 2 and $(p, p_0)$ changing to $(q, p_0^*)$, and $\hat{a}_{r,1},\hat{a}_{r,2} \hspace{0.05cm} (r=1, 2, \dots)$ are free complex constants.}
\end{quote}

\subsection{Rogue wave solutions through Schur polynomials} \label{SchurPolynoExp}
In this subsection, we derive more explicit expressions for rational solutions in Theorems 4-6 and prove Theorems 1-3.

We first introduce the generator $\mathcal{G}$ of differential operators $\left[ f_{1} \partial_{p} \right]^{i}  \left[ f_{2} \partial_{q} \right]^{j}$ as
\[ \label{GeneratorG}
\mathcal{G}= \sum_{i=0}^\infty \sum_{j=0}^{\infty} \frac{\kappa^i}{i!} \frac{\lambda^j}{j!} \left[ f_{1} \partial_{p} \right]^{i}  \left[ f_{2} \partial_{q} \right]^{j}.
\]
The main benefit of introducing functions $\mathcal{W}_1$ and $\mathcal{W}_2$ through equations such as (\ref{functiondiff}) is that we can rewrite the above generator as
\[ \label{GeneratorGWp}
\mathcal{G}= \sum_{i=0}^\infty \sum_{j=0}^{\infty} \frac{\kappa^i}{i!} \frac{\lambda^j}{j!} \left[\partial_{\ln\mathcal{W}_1}\right]^{i}  \left[\partial_{\ln\mathcal{W}_2}\right]^{j}=\exp\left(\kappa \partial_{\ln\mathcal{W}_1} + \lambda \partial_{\ln\mathcal{W}_2}\right).
\]
Then, for any function $F(\mathcal{W}_1,\mathcal{W}_2)$, we have \cite{OhtaJY2012}
\[ \label{Gdef}
\mathcal{G} F(\mathcal{W}_1,\mathcal{W}_2)=F(e^{\kappa}\mathcal{W}_1,e^{\lambda}\mathcal{W}_2).
\]
Since $p$ is related to $\mathcal{W}_{1}$, and $q$ related to $\mathcal{W}_{2}$, we can write
\[
p = p\left(\mathcal{W}_{1}\right),\ \ \ q = q\left(\mathcal{W}_{2}\right).
\]
The specifics of these relations depend on the root structure of $p_0$ in Eq. (\ref{Q1polynomial2}). If $p_0$ is a simple root, then $p$ and $\mathcal{W}_{1}$ are related by Eq. (\ref{2ndodesoluQ1p}). If $p_0$ is a double root, then $p$ and $\mathcal{W}_{1}$ are related by Eq. (\ref{3odeGeneSoluQ1}). In both cases, $q$ and $\mathcal{W}_{2}$ are related by similar equations.

From Eqs. (\ref{2ndodesoluQ1p}), (\ref{3odeGeneSoluQ1}) and similar ones for the $q$ function, we see that when $p=p_0$ and $q=q_0$, $\mathcal{W}_{1}=\mathcal{W}_{2}=1$. Thus, for $m^{(n,k)}$ in Eq. (\ref{mnk3}) of Theorem~4,
\begin{eqnarray}
&&\left. \mathcal{G} m^{(n,k)}\right|_{ p= p_{0},\  q =q_{0}} =
\frac{(-1)^{k+n}}{ p \left(\kappa\right)+ q \left( \lambda \right)}
\left(\frac{p \left(\kappa\right)}{ q \left( \lambda \right)} \right)^k  \left( \frac{p \left(\kappa\right)-\rm{i}}{q \left( \lambda \right)+\rm{i}} \right)^n \ \exp \left(\sum _{r =1}^\infty (\hat{a}_{r} \kappa^{r} +\hat{a}^*_{r} \lambda^{r})\right)\times    \nonumber \\
&& \hspace{2.6cm} \exp\left[   \frac{  \gamma_{1} \left( x-c_{2}t \right) }{c_{1}-c_{2}}  \left( \frac{1}{p \left(\kappa\right)} + \frac{1}{q \left( \lambda \right)} \right)+ \frac{ \gamma_{2} \left( x-c_{1}t \right) }{c_{2}-c_{1}}   \left( \frac{1}{p \left(\kappa\right)-\rm{i}} + \frac{1}{q \left( \lambda \right)+\rm{i}} \right)  \right],     \label{Gmnk0}
\end{eqnarray}
where
\[
p(\kappa) \equiv \left. p\left(\mathcal{W}_{1}\right)\right|_{\mathcal{W}_{1}=\exp(\kappa)}, \quad  q(\lambda)\equiv
\left. q\left(\mathcal{W}_{2}\right)\right|_{\mathcal{W}_{2}=\exp(\lambda)}.
\]
When $p_0$ is a simple root of Eq. (\ref{Q1polynomial2}) as in Theorem 4, this $p(\kappa)$ function is obtained by substituting $\mathcal{W}_{1}=e^\kappa$ into Eq. (\ref{2ndodesoluQ1p}), which results in Eq. (\ref{defpk}) in Theorem 1. Since $q_0=p_0^*$ from Eq. (\ref{p0q0b}), we can see that the $q(\lambda)$ function can be obtained from $p(\kappa)$ as
\[
q(\lambda)=p^*(\lambda),
\]
where $\lambda$ is treated as a real variable.

From Eq. (\ref{Gmnk0}), we get
\begin{eqnarray}
&& \frac{1}{m^{(n,k)}}\left. \mathcal{G} m^{(n,k)}\right|_{p=p_{0},\  q=q_{0}}=\frac{p_{0}+q_{0}}{p(\kappa)+q(\lambda)} \left( \frac{p(\kappa)}{p_{0}} \right)^k  \left( \frac{q(\lambda)}{q_{0}}  \right)^{-k} \left( \frac{p(\kappa)-\rm{i}}{p_{0}-\rm{i}} \right)^n \left( \frac{q(\lambda)+\rm{i}}{q_{0}+\rm{i}}  \right)^{-n} \exp \left(\sum _{r =1}^\infty (\hat{a}_{r} \kappa^{r} +\hat{a}^*_{r} \lambda^{r})\right) \times  \nonumber \\
&&  \exp\left[   \frac{  \gamma_{1} \left( x-c_{2}t \right) }{c_{1}-c_{2}}  \left( \frac{1}{p \left(\kappa\right)} - \frac{1}{p_{0}}+ \frac{1}{q \left( \lambda \right)}-\frac{1}{q_{0}} \right) +  \frac{ \gamma_{2} \left( x-c_{1}t \right) }{c_{2}-c_{1}}  \left( \frac{1}{p \left(\kappa\right)-\rm{i}} - \frac{1}{p_{0}-\rm{i}}+ \frac{1}{q \left( \lambda \right)+\rm{i}}-\frac{1}{q_{0}+\rm{i}} \right)  \right].   \label{AlgeExpress1}
\end{eqnarray}
Now, we expand the right side of the above equation into power series of $\kappa$ and $\lambda$. Its first term can be treated by the techniques of \cite{OhtaJY2012,YangYangBoussi} as
\begin{eqnarray*}
&& \frac{p_{0}+q_{0}}{p +q}= \frac{\left(p_{0}+q_{0}\right)^2}{(p_{0}+q_{0})(p+q)} = \frac{\left(p_{0}+q_{0}\right)^2}{(p+q_{0})(q+p_{0})} \sum_{\nu=0}^{\infty} \left[ \frac{(p-p_{0})(q-q_{0})}{(p+q_{0})(q+p_{0})} \right]^{\nu} \\
&& =  \frac{\left(p_{0}+q_{0}\right)^2}{(p+q_{0})(q+p_{0})} \sum_{\nu=0}^{\infty} \left( \frac{p_{1} q_{1} }{(p_{0}+q_{0})^2} \kappa \lambda  \right)^{\nu}
\left( \frac{p_{0} + q_{0} }{p_{1} \kappa}  \frac{p-p_{0}}{p+q_{0}}  \right)^{\nu}
\left( \frac{p_{0} + q_{0} }{q_{1} \lambda}  \frac{q-q_{0}}{q+p_{0}}  \right)^{\nu} \\
&&= \sum_{\nu=0}^{\infty} \left( \frac{p_{1} q_{1} }{(p_{0}+q_{0})^2} \kappa \lambda  \right)^{\nu} \exp\left(  \sum_{r=1}^{\infty}
\left(  \nu s_{r}- b_{r}\right) \kappa^r  + \left(  \nu s^*_{r}- b^*_{r}\right) \lambda^{r} \right),
\end{eqnarray*}
where $p_1=(dp/d\kappa)|_{\kappa=0}$, $q_1=(dq/d\lambda)|_{\lambda=0}=p_1^*$, $b_r$ is the Taylor coefficient of $\kappa^r$ in the expansion of
\[
\ln \left[ \frac{p \left(\kappa\right)+q_{0}}{p_{0}+q_{0}} \right]= \sum_{r=1}^{\infty}b_{r} \kappa^r,
\]
and $s_r$ is the Taylor coefficient of $\kappa^r$ in the expansion of (\ref{schurcoeffsr}) in Theorem 1. Using the expansions (\ref{schucoefalpha})-(\ref{schucoeflambda}) in Theorem 1 and similar ones for the $q(\lambda)$ function through the functional relation $q(\lambda)=p^*(\lambda)$, we can rewrite the rest of the terms on the right side of Eq. (\ref{AlgeExpress1}) as
\begin{equation*}
\exp \left\{ \sum_{r=1}^{\infty} \kappa^r \left[ \left( \alpha_{r} - \beta_{r} \right) x +\left( c_{1}\beta_{r}-c_{2}\alpha_{r} \right)t + n \theta_{r}+ k \lambda_{r} \right] +  \sum_{r=1}^{\infty}  \lambda^r  \left[ \left( \alpha^*_{r} - \beta^*_{r} \right) x +\left( c_{1}\beta^*_{r}-c_{2}\alpha^*_{r} \right)t - n \theta^*_{r} - k \lambda^*_{r} \right]  +\sum _{r =1}^\infty (\hat{a}_{r}\kappa^{r}+\hat{a}^*_{r} \lambda^{r})  \right\}.
\end{equation*}
Combining these results, Eq. (\ref{AlgeExpress1}) becomes
\begin{eqnarray} \label{Gmnk}
&& \frac{1}{m^{(n,k)}}\left. \mathcal{G} m^{(n,k)}\right|_{p=p_{0},\  q=q_{0}} =
 \sum_{\nu=0}^{\infty} \left( \frac{p_{1} q_{1} \  \kappa \lambda }{(p_{0}+q_{0})^2}  \right)^{\nu}
\exp\left(  \sum_{r=1}^{\infty} \left(x_{r}^{+}+\nu s_{r}\right) \kappa^r + \sum_{r=1}^{\infty} \left(x_{r}^{-}+\nu s_{r}^* \right) \lambda^{r} \right),
\end{eqnarray}
where $x_{r}^{\pm}(n,k)$ are as defined in Eqs. (\ref{defxrp})-(\ref{defxrm}) with
\[
a_{r}\equiv \hat{a}_{r}-b_{r}.
\]
Taking the coefficients of $\kappa^i \lambda^j$ on both sides of the above equation, we get
\begin{equation*}
\frac{m_{i,j}^{(n,k)}}{\left. m^{(n,k)}\right|_{p=p_{0}, q=q_{0}}}
=\sum_{\nu=0}^{\min(i,j)} \left( \frac{p_{1} q_{1} }{(p_{0}+q_{0})^2}  \right)^{\nu}  S_{i-\nu}\left( \textbf{\emph{x}}^{+}+\nu \textbf{\emph{s}} \right)
S_{j-\nu}\left( \textbf{\emph{x}}^{-}+\nu \textbf{\emph{s}}^* \right),
\end{equation*}
where $m_{i,j}^{(n,k)}$ is the matrix element given in Eq. (\ref{mij-diff}). Notice that the above function is the matrix element in the determinant $\sigma_{n,k}$ of Theorem 1. This matrix element of $\sigma_{n,k}$ is only a polynomial function of $x$ and $t$, since the exponential factors in the matrix element $m_{i,j}^{(n,k)}$ of $\tau_{n,k}$ in Eq. (\ref{mij-diff}) are eliminated by the above scaling of $\left. m^{(n,k)}\right|_{p=p_{0}, q=q_{0}}$. The $\sigma_{n,k}$ determinant in Theorem 1 is related to the determinant $\tau_{n,k}$ in Theorem 4 by
\begin{equation} \label{sigmatau}
\sigma_{n,k}=\frac{\tau_{n,k}}{\left(\left. m^{(n,k)}\right|_{p=p_{0}, q=q_{0}}\right)^N}.
\end{equation}
Since the $f$ and $g_i$ functions given through $\tau_{n,k}$ in Eq. (\ref{fgdef1}) satisfy the bilinear equations (\ref{1stBilineform}), and those bilinear equations are invariant when $\tau_{n,k}$ is divided by an exponential of a linear and real function in $x$ and $t$, it is easy to see from the above relation that the $f$ and $g_i$ functions given through $\sigma_{n,k}$ in Eq. (\ref{SchpolysolufN}) satisfy those bilinear equations as well. Thus, Schur polynomial expressions of rational solutions in Theorem 1 for a simple root $p_0$ of Eq. (\ref{Q1polynomial2}) are proved.

Following very similar approaches, Schur polynomial expressions of rational solutions in Theorem 3 for a double root $p_0$ of Eq. (\ref{Q1polynomial2}) can also be proved. In this case, the parameters $\{a_{r,1}, a_{r,2}\}$ in Theorem 3 are related to parameters $\{\hat{a}_{r,1}, \hat{a}_{r,2}\}$ in Theorem 6 through $a_{r,I}\equiv \hat{a}_{r,I}-b_{r}$.

To derive Schur polynomial expressions of rational solutions in Theorem 5 for two simple roots, some modifications to the above treatment need to be made. In this case, a counterpart scaling of Eq.~(\ref{Gmnk}) would not work. The reason is that such a scaling function, which is $\left. m^{(n,k,I,J)}\right|_{p=p_{0,I}, q=q_{0,J}}$ now, would contain a factor of $1/(p_{0,I}+q_{0,J})$, which takes on different values in different blocks. Because of this, the block determinant $\sigma_{n,k}$ so scaled and the original block determinant $\tau_{n,k}$ could not be related by a factor as in Eq. (\ref{sigmatau}), and hence the scaled determinant would not satisfy the underlying bilinear equations. Since the difficulty arises from the factor $1/(p_{0,I}+q_{0,J})$ in $\left. m^{(n,k,I,J)}\right|_{p=p_{0,I}, q=q_{0,J}}$, the way to overcome this difficulty is to use the new scaling of $\left. (p+q)m^{(n,k,I,J)}\right|_{p=p_{0,I}, q=q_{0,J}}$, where the factor $1/(p_{0,I}+q_{0,J})$ is eliminated. In this case, we have
\begin{eqnarray*}
\hspace{-0.7cm} && \frac{1}{(p+q)m^{(n,k,I,J)}}\left. \mathcal{G} m^{(n,k,I,J)}\right|_{p=p_{0,I},\  q=q_{0,J}}=\frac{1}{p_I(\kappa)+q_J(\lambda)} \left( \frac{p_I(\kappa)}{p_{0,I}} \right)^k  \left( \frac{q_J(\lambda)}{q_{0,J}}  \right)^{-k} \left( \frac{p_I(\kappa)-\rm{i}}{p_{0,I}-\rm{i}} \right)^n \left( \frac{q_J(\lambda)+\rm{i}}{q_{0,J}+\rm{i}}  \right)^{-n} \exp \left(\sum _{r =1}^\infty (a_{r,I} \kappa^{r} +a^*_{r,J} \lambda^{r})\right) \times  \nonumber \\
\hspace{-0.7cm} && \hspace{1.5cm} \exp\left\{   \frac{ \gamma_{1} \left( x-c_{2}t \right) }{c_{1}-c_{2}}  \left( \frac{1}{p_I \left(\kappa\right)} - \frac{1}{p_{0,I}}+ \frac{1}{q_J \left( \lambda \right)}-\frac{1}{q_{0,J}} \right) +
\frac{ \gamma_{2} \left( x-c_{1}t \right) }{c_{2}-c_{1}}  \left( \frac{1}{p_I \left(\kappa\right)-\rm{i}} - \frac{1}{p_{0,I}-\rm{i}}+ \frac{1}{q_J \left( \lambda \right)+\rm{i}}-\frac{1}{q_{0,J}+\rm{i}} \right)  \right\},
\end{eqnarray*}
where functions $p_I(\kappa)$ are defined in Theorem 2, and $q_J(\lambda)=p_J^*(\lambda)$. Then, following a similar procedure as above, we can expand the right side of the above equation into power series of $\kappa$ and $\lambda$ and get
\begin{eqnarray*}
\frac{1}{(p+q)m^{(n,k,I,J)}} \left. \mathcal{G} m^{(n,k,I,J)}\right|_{p=p_{0,I},\  q=q_{0,J}}=  \sum_{\nu=0}^{\infty} \left( \frac{1}{p_{0,I}+q_{0,J}} \right) \left( \frac{p_{1,I} q_{1,J} \  \kappa \lambda }{(p_{0,I}+q_{0,J})^2}  \right)^{\nu}
\exp\left(  \sum_{r=1}^{\infty} \left(x_{r,I,J}^{+}+\nu s_{r,I,J}\right) \kappa^r + \sum_{r=1}^{\infty} \left(x_{r,I,J}^{-}+\nu s_{r,J,I}^*\right) \lambda^{r} \right),
\end{eqnarray*}
where $x_{r,I,J}^{\pm}(n,k)$ and $s_{r,I,J}$ are defined in Theorem 2. Taking the coefficients of $\kappa^i \lambda^j$ on both sides of this equation, we get
\begin{equation*}
\frac{m_{i,j}^{(n,k,I,J)}}{\left.(p+q)m^{(n,k,I,J)}\right|_{p=p_{0,I}, q=q_{0,J}}}=
\sum_{\nu=0}^{\min(i,j)} \left( \frac{1}{p_{0,I}+q_{0,J}} \right) \left[ \frac{p_{1,I} q_{1,J} }{(p_{0,I}+q_{0,J})^2}  \right]^{\nu} \hspace{0.06cm} S_{i-\nu}\left(\textbf{\emph{x}}^{+}_{I,J}(n,k) +\nu \textbf{\emph{s}}_{I,J}\right)  \hspace{0.06cm} S_{j-\nu}\left(\textbf{\emph{x}}^{-}_{I,J}(n,k) + \nu \textbf{\emph{s}}_{J,I}^*\right),
\end{equation*}
where $m_{i,j}^{(n,k,I,J)} $ is the matrix element defined in Eq. (\ref{mijIJ-diff}) of Theorem 5 in view that $q_{0,J}=p^*_{0,J}$ [see (\ref{q0J})]. The above scaled function is the matrix element in the block determinant $\sigma_{n,k}$ in Theorem 2. The benefit of the above scaling is that the scaled block determinant $\sigma_{n,k}$ is now related to the original block determinant $\tau_{n,k}$ in Eq. (\ref{2x2blotausdet1b}) by a factor similar to Eq. (\ref{sigmatau}), and thus this scaled block determinant remains a solution to the underlying bilinear equations.

Regarding boundary conditions of these rational solutions, using Schur polynomial expressions of these solutions and the same technique as in Ref. \cite{OhtaJY2012}, we can show that for solutions in Theorems 1 and 3, when $x$ or $t$ approaches infinity, $f(x,t)$ and $g_{i}(x,t)$ functions have the same leading term. For solutions in Theorem 2, we can use a generalization of the technique in Ref. \cite{OhtaJY2012} to show the same fact (see the end of Appendix A for some details). Thus, rational solutions in these three theorems satisfy the boundary conditions (\ref{BoundaryCond}) and are rogue waves. Theorems 1-3 are then proved.

\section{Conclusion and Discussion}
In this article, we have derived general rogue waves in (1+1)-dimensional three-wave resonant interaction systems by the bilinear method. Our solutions are divided into three families, which correspond to a simple root, two simple roots and a double root of the quartic equation (\ref{Q1polynomial2}) and presented in Theorems 1-3 respectively. We have shown that while the first family of solutions associated with a simple root exist for all signs of the nonlinear coefficients in the three-wave interaction equations, the other two families of solutions associated with two simple roots and a double root can only exist in the soliton-exchange case (\ref{solexchange}), where the nonlinear coefficients have certain signs. Dynamics of the derived rogue waves has also been examined, and many new rogue patterns have been exhibited (see Figs. 1-7). In addition, relations between our bilinear rogue waves and those derived earlier by Darboux transformation are explained.

Technically, our main contribution of the paper is a generalization of the dimension reduction procedure in the bilinear derivation of rogue waves. This generalization is necessary to treat the double-root case of the algebraic equation (\ref{QurticeqQ1dp}) during dimension reduction. We have shown that the function $f_1(p)$ in the differential operator $\mathcal{A}_{i}$ of Eq. (\ref{New003b}) needs to be selected judiciously depending on the root multiplicity of the algebraic equation (\ref{QurticeqQ1dp}). For simple and double roots which are encountered in the three-wave system (\ref{3WRIModel}), that function is selected by conditions (\ref{2ndordlinODE}) and (\ref{3rdordlinode}) respectively. It is then clear that, should this root have multiplicity higher than two, which does not occur in the present three-wave system but may arise in other situations, the function $f_1(p)$ would be selected by a condition similar to (\ref{3rdordlinode}), but with the exponent 3 in that equation replaced by the multiplicity of the root plus one. Because of this, we have laid out the most general dimension reduction procedure for the bilinear derivation of rogue waves, and this procedure can be applied to a wide range of integrable systems beyond the three-wave interaction system.

\section*{Acknowledgement}

This material is based upon work supported by the Air Force Office of Scientific
Research under award number FA9550-18-1-0098 and the National Science Foundation under award number DMS-1910282.

\begin{center}
\textbf{Appendix A}
\end{center}

In this appendix, we derive the polynomial degree of the block-determinant $\sigma_{n,k}$ for the $(N_1, N_2)$-th order rational solutions in Theorem 2, and show that these rational solutions satisfy the rogue-wave boundary conditions (\ref{BoundaryCond}).

The $2\times 2$ block determinant $\sigma_{n,k}$ in Eq. (\ref{sigmaTheorem2}) of Theorem 2 can be rewritten as the determinant of a product between two larger matrices,
\begin{eqnarray}  \label{sigmaA}
\sigma_{n,k}=\det \left(\Phi\Psi\right),
\end{eqnarray}
where
\begin{eqnarray*}
\Phi=\left( \begin{array}{cccc}
              \Phi^{[1,1]}_{N_1\times 2N_1} & \Phi^{[1,2]}_{N_1\times 2N_2} & \textbf{O}_{N_{1}\times 2N_{1}}  & \textbf{O}_{N_{1}\times 2N_{2}}   \\
              \textbf{O}_{N_{2}\times 2N_{1}}   & \textbf{O}_{N_{2}\times 2N_{2}}
              & \Phi^{[2,1]}_{N_2\times 2N_1} & \Phi^{[2,2]}_{N_2\times 2N_2}
            \end{array}\right),
\end{eqnarray*}
\begin{eqnarray*}
\Psi=\left( \begin{array}{cccc}
\Psi^{[1,1]}_{N_1\times 2N_1} &  \textbf{O}_{N_{1}\times 2N_{2}} & \Psi^{[2,1]}_{N_{1}\times 2N_{1}} & \textbf{O}_{N_{1}\times 2N_{2}}  \\
\textbf{O}_{N_{2}\times 2N_{1}}    & \Psi^{[1,2]}_{N_{2}\times 2N_{2}}  &   \textbf{O}_{N_{2}\times 2N_{1}}  & \Psi^{[2,2]}_{N_{2}\times 2N_{2}}
\end{array} \right)^{T},
\end{eqnarray*}
the matrix elements are defined by
\begin{eqnarray*}
&& \Phi^{[I,J]}_{i,j}= \left( \frac{1}{p_{0,I}+p^*_{0,J}} \right)^{\frac{1}{2}}
\left( \frac{p_{1,I}}{p_{0,I}+p^*_{0,J}} \right)^{j-1}  S_{i-(j-1)}\left(\textbf{\emph{x}}^{+}_{I,J}(n,k) + (j-1) \textbf{\emph{s}}_{I,J}\right), \\
&& \Psi^{[I,J]}_{i,j}=   \left( \frac{1}{p_{0,I}+p^*_{0,J}} \right)^{\frac{1}{2}}
\left( \frac{q_{1,J}^*}{p_{0,I}+p^*_{0,J}} \right)^{j-1}  S_{i-(j-1)}\left(\textbf{\emph{x}}^{-}_{I,J}(n,k) + (j-1) \textbf{\emph{s}}_{I,J}^*\right),
\end{eqnarray*}
and $S_i\equiv 0$ if $i<0$. Eq. (\ref{sigmaA}) is a generalization of that used in Ref.~\cite{OhtaJY2012}, but expressed in a new way. According to the Cauchy-Binet formula, we can further rewrite $\sigma_{n,k}$ in Eq. (\ref{sigmaA}) as
\begin{eqnarray} \label{sigmaA2}
\sigma_{n,k}=\det \left(\Phi\Psi\right)= \sum_{1\le \mu_1 <\mu_2 < \cdots < \mu_N\le 4N} \det \left(\Phi_{\mathbf{\mu}}\right) \hspace{0.05cm} \det\left(\Psi_{\mathbf{\mu}}\right),
\end{eqnarray}
where $N=N_1+N_2$, $\Phi_{\mathbf{\mu}}$ is a square matrix made up by the $(\mu_1, \mu_2,  \cdots, \mu_N)$-th columns of the larger matrix $\Phi$, and $\Psi_{\mathbf{\mu}}$ is another square matrix made up by the $(\mu_1, \mu_2, \cdots, \mu_N)$-th rows of the larger matrix $\Psi$.

When calculating the polynomial degrees of $\det \left(\Phi_{\mathbf{\mu}}\right)$, one slight complication is that, when $\Phi_{\mathbf{\mu}}$ contains columns from both $\Phi^{[1,1]}$ and $\Phi^{[1,2]}$ matrices, and/or from both $\Phi^{[2,1]}$ and $\Phi^{[2,2]}$ matrices, this $\Phi_{\mathbf{\mu}}$ matrix would involve
different Schur polynomials $S_{i}\left(\textbf{\emph{x}}^{+}_{I,J}(n,k) + \nu \hspace{0.06cm} \textbf{\emph{s}}_{I,J}\right)$ due to different $J$ indices. This complication can be overcome since we can relate these Schur polynomials with different $J$ indices in a simple way. To do so, we notice that
\begin{equation*}
x_{r,I,2}^{+}(n,k)+ \nu s_{r,I,2}= x_{r,I,1}^{+}(n,k) + \nu s_{r,I,1}  + \left( b_{r,I,1}-b_{r,I,2}+  \nu s_{r,I,2}-\nu s_{r,I,1}\right).
\end{equation*}
Then, using the definition of Schur polynomials (\ref{defSchur}), we can relate $S_{i}\left(\textbf{\emph{x}}^{+}_{I,1}(n,k) + \nu \hspace{0.06cm} \textbf{\emph{s}}_{I,1}\right)$ and
$S_{i}\left(\textbf{\emph{x}}^{+}_{I,2}(n,k) + \nu \hspace{0.06cm} \textbf{\emph{s}}_{I,2}\right)$ as
\[ \label{S-relation-1}
S_{i}\left(\textbf{\emph{x}}^{+}_{I,2}(n,k) +\nu \textbf{\emph{s}}_{I,2}\right)=\sum_{j=0}^i d_j \hspace{0.05cm} S_{i-j}\left(\textbf{\emph{x}}^{+}_{I,1}(n,k) +\nu \textbf{\emph{s}}_{I,1}\right),
\]
where $\{d_{j}\}$ are constants depending on $\{b_{r,I,1}-b_{r,I,2}+  \nu s_{r,I,2}-\nu s_{r,I,1}\}$.

Another small complication in calculating the polynomial degrees of $\det \left(\Phi_{\mathbf{\mu}}\right)$ is that different columns inside each of the block matrices $\Phi^{[I,J]}$ are Schur polynomials of the type
$S_{i}\left(\textbf{\emph{x}}^{+}_{I,J}(n,k) + \nu \hspace{0.06cm} \textbf{\emph{s}}_{I,J}\right)$ with different $\nu$ values. But once again, we can relate $S_{i}\left(\textbf{\emph{x}}^{+}_{I,J}(n,k) + \nu \hspace{0.06cm} \textbf{\emph{s}}_{I,J}\right)$ with different $\nu$ values, say $\nu_{1}$ and $\nu_{2}$, in a simple way as
\[\label{S-relation-2}
S_{i}\left(\textbf{\emph{x}}^{+}_{I,J}(n,k) +\nu_{2} \textbf{\emph{s}}_{I,J}\right)=
\sum_{j=0}^i \hat{d}_j \hspace{0.05cm}  S_{i-j}\left(\textbf{\emph{x}}^{+}_{I,J}(n,k) +\nu_{1} \textbf{\emph{s}}_{I,J}\right),
\]
where $\{\hat{d}_{j}\}$ are constants depending on $\{(\nu_{2}-\nu_{1}) s_{r,I,J}\}$.

Now, we examine the highest polynomial degree of $\det \left(\Phi_{\mathbf{\mu}}\right)$. Utilizing the above two Schur polynomial relations and applying simple column manipulations, we can easily see by techniques of Ref.~\cite{OhtaJY2012} that the highest polynomial degree of $\det \left(\Phi_{\mathbf{\mu}}\right)$ can be reached by multiple choices of the $(\mu_1, \mu_2,  \cdots, \mu_N)$ column indices in the larger matrix $\Phi$. For example, the indices of $\left[1, 2, \cdots, N_1, 4N-(N_2-1), 4N-(N_2-2), \cdots, 4N\right]$ and $[2, 3, \cdots, N_1, N_1+1, 4N-(N_2-1), 4N-(N_2-2), \cdots, 4N]$ yield the same polynomial degree of $\left[N_1(N_1+1)+N_2(N_2+1)\right]/2$ in both $x$ and $t$ for $\det \left(\Phi_{\mathbf{\mu}}\right)$. However, these Schur polynomial relations (\ref{S-relation-1})-(\ref{S-relation-2}) and column manipulations also make it clear that the polynomial degree of $\det \left(\Phi_{\mathbf{\mu}}\right)$ cannot be higher than $\left[N_1(N_1+1)+N_2(N_2+1)\right]/2$.

Using similar techniques, we can show that the highest polynomial degree of $\det \left(\Psi_{\mathbf{\mu}}\right)$ is also $[N_1(N_1+1)+N_2(N_2+1)]/2$. Combining these two results, the highest polynomial degree of $\sigma_{n,k}$ in Theorem 2 can be derived from Eq.~(\ref{sigmaA2}) as $N_{1}(N_{1}+1) +  N_{2}(N_{2}+1)$ in both $x$ and $t$.

A closer examination of the above polynomial-degree analysis for $\sigma_{n,k}$ further reveals that the highest-degree terms of $x$ and $t$ in $\sigma_{n,k}$ come from
\begin{equation*}
w_0 \left[x_{1,1,1}^{+}(n,k) \hspace{0.05cm} x_{1,1,1}^{-}(n,k)\right]^{N_1(N_1+1)/2} \left[x_{1,2,2}^{+}(n,k) \hspace{0.05cm} x_{1,2,2}^{-}(n,k)\right]^{N_2(N_2+1)/2},
\end{equation*}
where $w_0$ is a $(n,k)$-independent constant. Thus,
\begin{eqnarray*}
&& \sigma_{n,k} = w_0 \left|\left( \alpha_{1,1} - \beta_{1,1} \right) x +\left( c_{1}\beta_{1,1}-c_{2}\alpha_{1,1} \right)t \right|^{N_1(N_1+1)} \left|\left( \alpha_{1,2} - \beta_{1,2} \right) x +\left( c_{1}\beta_{1,2}-c_{2}\alpha_{1,2} \right)t \right|^{N_2(N_2+1)}  \\
&& \hspace{1cm} + \hspace{0.06cm} \mbox{lower degree terms of} \hspace{0.05cm} x \hspace{0.08cm} \mbox{and} \hspace{0.08cm} t.
\end{eqnarray*}
This relation shows that the rational solutions in Theorem 2 satisfy the boundary conditions (\ref{BoundaryCond}), and are thus rogue waves in the three-wave system.

\vspace{0.2cm}
\begin{center}
\textbf{Appendix B}
\end{center}

In this appendix, we present explicit expressions of second-order rogue waves for a non-imaginary simple root in Theorem 1.
These rogue waves are given as
\begin{equation*}
|u_{i,2}(x,t)|=\left| \rho_{i} \frac{g_{i,2}}{f_{2}} \right|,\ \ \ i=1,2,3,
\end{equation*}
where
\begin{equation*}
f_{2} =\sigma_{0,0},\ \ g_{1,2}=\sigma_{1,0},\ \ g_{2,2}=\sigma_{0,-1},\ \ g_{3,2}=\sigma_{-1,1},
\end{equation*}
\begin{eqnarray*}
&& \sigma_{n,k}=  \left[\left( x_{1,0}^+\right)^3-3 \left(\left(x_{1,1}^+\right)^2-2 x_{2,0}^+ + 2 x_{2,1}^+\right) x_{1,0}^+ +
6 x_{3,0}^+\right]\\
&& \hspace{0.70cm}  \times \left[\left( x_{1,0}^-\right)^3-3 \left(\left(x_{1,1}^-\right)^2-2 x_{2,0}^- + 2 x_{2,1}^-\right) x_{1,0}^- +6 x_{3,0}^-\right]\\
&& \hspace{0.7cm}  + 36  \zeta_{0}  \left( x_{1,0}^+   x_{1,0}^-  \right) \left(x_{1,2} ^+ x_{1,2}^-\right)+36 \zeta_{0}^2 \left(  x_{1,0} ^+ x_{1,0} ^-  + x_{1,2} ^+x_{1,2} ^-   \right)+6 \zeta_{0}^3,
\end{eqnarray*}
\begin{eqnarray*}
&& x_{j,\nu}^+ \equiv x_{j}^{+}(n,k) + \nu s_{j}
=\left( \alpha_{j} - \beta_{j} \right) x +\left( c_{1}\beta_{j}-c_{2}\alpha_{j} \right)t + n \theta_{j} + k \lambda_{j}+a_{j}+\nu s_{j},\\
&&  x_{j,\nu}^-\equiv x_{j}^{-}(n,k) + \nu s_{j}=\left( \alpha_{j}^* - \beta_{j}^* \right) x +\left( c_{1}\beta_{j}^*-c_{2}\alpha_{j}^* \right)t - n \theta_{j}^* - k \lambda_{j}^*
+ a_{j}^* +\nu s_{j}^*,
\end{eqnarray*}
$a_{1}=a_{2}=0$, and coefficients in the above expressions are
\begin{eqnarray*}
&&\zeta_{0} =\frac{|p_{1}|^2 }{(p_{0}+p_{0}^*)^2},\ \alpha_1=-\frac{p_1 \epsilon_1 \rho_{2} \rho_{3}}{p_{0}^2  (c_{1}-c_{2})\rho_{1}},\ \alpha_2=-\frac{  \epsilon_1 \rho_{2} \rho_{3} (p_1^2-p_{0}p_{2})}{ p_{0}^3  (c_{1}-c_{2})\rho_{1}},\ \alpha_3=\frac{  \epsilon_1 \rho_{2} \rho_{3}(p_1^3-2 p_0 p_2 p_1+p_0^2 p_3)}{ p_{0}^4  (c_{1}-c_{2})\rho_{1}}, \\
&& \beta_{1}=-\frac{p_1 \epsilon_2 \rho_{1} \rho_{3}}{(p_{0}-\rm{i})^2  (c_{1}-c_{2})\rho_{2}}, \ \beta_{2}=-\frac{ (p_1^2-p_{0}p_{2}) \epsilon_2 \rho_{1} \rho_{3}}{(p_{0}-\rm{i})^3  (c_{1}-c_{2})\rho_{2}}, \ \beta_{3}=-\frac{ (p_1^3-2 p_0 p_2 p_1+p_0^2 p_3) \epsilon_3 \rho_{1} \rho_{3}}{(p_{0}-\rm{i})^4  (c_{1}-c_{2})\rho_{2}}, \\
&&\theta_{1}=\frac{p_{1}}{p_{0}-\rm{i}}, \  \theta_{2}=\frac{1}{2} \left(\frac{2 p_2}{p_0-i}-\frac{p_1^2}{\left(p_0-i\right){}^2}\right), \ \theta_{3}=\frac{p_1^3-3 \left(p_0-i\right) p_2 p_1+3 \left(p_0-i\right){}^2 p_3}{3 \left(p_0-i\right)^3}, \\
&& \lambda_{1}=\frac{p_{1}}{p_{0}}, \  \lambda_{2}=\frac{2 p_0 p_2-p_1^2}{2 p_0^2},\ \lambda_{3}=\frac{p_1^3-3 p_0 p_2 p_1+3 p_0^2 p_3}{3 p_0^3},\\
&& s_{1}=\frac{p_2 q_0-p_1^2+p_0 p_2}{p_1 \left(p_0+q_0\right)}, s_{2}=\frac{p_1^4-2 p_2 p_1^2 \left(p_0+q_0\right)+2 p_3 p_1 \left(p_0+q_0\right)^2-p_2^2 \left(p_0+q_0\right)^2}{2 p_1^2 \left(p_0+q_0\right)^2}.
\end{eqnarray*}

\vspace{0.2cm}
\begin{center}
\textbf{Appendix C}
\end{center}

In this appendix, we prove that the $2\times 2$ block determinant (\ref{2x2blotausdet1}) satisfies the higher-dimensional bilinear system~(\ref{KPHBilineEq}).

First of all, we point out that the $\tau_{n,k}$ function (\ref{tildetaun}), with matrix elements given by Eqs. (\ref{mnk})-(\ref{New003b}), is very special. One can introduce much broader $\tau_{n,k}$ functions which can still satisfy the higher-dimensional bilinear system (\ref{KPHBilineEq}). Indeed, let us introduce more general functions $m_{ij}^{(n,k)}$, $\varphi_i^{(n,k)}$ and $\psi_j^{(n,k)}$ as
\begin{eqnarray*}
&& m_{ij}^{(n, k)}= \frac{1}{p_{i} + q_{j}}\left(-\frac{p_{i}-a}{q_{j}+a}\right)^{k}\left(-\frac{p_{i}-b}{q_{j}+b}\right)^{n} \mathrm{e}^{\xi_i + \eta_j },   \\
&& \varphi_{i}^{(n,k)}= (p_{i}-a)^k (p_{i}-b)^n e^{\xi_{i}}, \\
&& \psi_{j}^{(n,k)}= \left[-(q_{j}+a)\right]^{-k}  \left[-(q_{j}+b)\right]^{-n}  e^{\eta_{j}},
\end{eqnarray*}
where
\begin{eqnarray*}
&& \xi_{i} =\frac{1}{p_{i}-a} r + \frac{1}{p_{i}-b} s + (p_{i}-b) x_{1}+\hat{\xi}_{0,i} , \\
&& \eta_{j}=\frac{1}{q_{j}+a} r + \frac{1}{q_{j}+b} s + (q_{j}+b) x_{1}+\hat{\eta}_{0,j},
\end{eqnarray*}
and $p_{i}, q_{j}, \hat{\xi}_{0,i}, \hat{\eta}_{0,j}$ are arbitrary complex constants. It is easy to see that these functions satisfy the differential and difference relations (\ref{001}), a phenomenon similar to that reported in Ref. \cite{OhtaJKY2012}. Then, by defining new functions
\[
m_{ij}^{(n,k)}=\mathcal{A}_i \mathcal{B}_{j} m_{ij}^{(n,k)}, \quad
\varphi_i^{(n,k)}=\mathcal{A}_i\varphi_i^{(n,k)}, \quad
\psi_j^{(n,k)}=\mathcal{B}_{j}\psi_j^{(n,k)},    \label{mijn5}
\]
where $\mathcal{A}_{i}$ and $\mathcal{B}_{j}$ are differential operators with respect to $p_i$ and $q_j$ respectively as
\begin{eqnarray*}
\mathcal{A}_{i}=\frac{1}{ n_i !}\left[f_{1}(p_{i})\partial_{p_{i}}\right]^{n_{i}}, \quad
\mathcal{B}_{j}=\frac{1}{ n_j !}\left[f_{2}(q_{j})\partial_{q_{j}}\right]^{n_{j}},
\end{eqnarray*}
$n_i, n_j$ are arbitrary positive integers, and $f_{1}(p_i)$, $f_{2}(q_j)$ are arbitrary functions, these new functions would also satisfy the differential and difference relations (\ref{001}). Consequently, for an arbitrary sequence of indices $(i_1,i_2,\cdots,i_N)$ and $(j_1,j_2,\cdots,j_N)$, the much broader determinant
\[ \label{tildetaun5}
\tau_{n,k}=\det_{1\le\nu, \mu\le N}\left( m_{i_\nu,j_\mu}^{(n,k)}\right),
\]
with $m_{ij}^{(n,k)}$ given in Eq. (\ref{mijn5}), also satisfies the higher-dimensional bilinear system (\ref{KPHBilineEq}).

To reduce this broader determinant (\ref{tildetaun5}) to the $2\times 2$ block determinant (\ref{2x2blotausdet1}), we set $N=N_1+N_2$,
\begin{eqnarray*}
&& p_1=p_2=\cdots p_{N_1}, \quad p_{N_1+1}=p_{N_1+2}=\cdots=p_{N},  \\
&& q_1=q_2=\cdots q_{N_1}, \quad q_{N_1+1}=q_{N_1+2}=\cdots=q_{N},  \\
&& \hat{\xi}_{0,1}=\hat{\xi}_{0,2}=\cdots \hat{\xi}_{0,N_1}\equiv \xi_{0,1}, \quad \hat{\xi}_{0,N_1+1}=\hat{\xi}_{0,N_1+2}=\cdots=\hat{\xi}_{0,N}\equiv \xi_{0,2}, \\
&& \hat{\eta}_{0,1}=\hat{\eta}_{0,2}=\cdots \hat{\eta}_{0,N_1}\equiv \eta_{0,1}, \quad \hat{\eta}_{0,N_1+1}=\hat{\eta}_{0,N_1+2}=\cdots=\hat{\eta}_{0,N}\equiv \eta_{0,2}, \\
&& n_i=\left\{\begin{array}{l} i, \quad \hspace{0.75cm} 1\le i\le N_1, \\  i-N_1, \quad N_1+1\le i\le N, \end{array} \right. \\
&& i_\nu=\left\{\begin{array}{l} 2\nu-1, \quad \hspace{0.99cm} 1\le \nu \le N_1, \\  2(\nu-N_1)-1, \quad N_1+1\le \nu \le N, \end{array} \right. \quad j_\mu=\left\{\begin{array}{l} 2\mu-1, \quad \hspace{0.98cm} 1\le \mu \le N_1, \\  2(\mu-N_1)-1, \quad N_1+1\le \mu \le N. \end{array} \right.
\end{eqnarray*}
Then, the resulting determinant (\ref{tildetaun5}) would be of $2\times 2$ block type (\ref{2x2blotausdet1}), which clearly also satisfies the higher-dimensional bilinear system (\ref{KPHBilineEq}). For it to satisfy the dimensional reduction condition (\ref{dimenredc3WIR}), we take $p_1=p_2=\cdots p_{N_1}=p_{0,1}$, $p_{N_1+1}=p_{N_1+2}=\cdots=p_{N}=p_{0,2}$, $q_1=q_2=\cdots q_{N_1}=p_{0,1}^*$, and  $q_{N_1+1}=q_{N_1+2}=\cdots=q_{N}=p_{0,2}^*$ in its matrix element $m_{i_\nu,j_\mu}^{(n,k)}$, where $(p_{0,1}, p_{0,2})$ are two simple roots of the $\mathcal{Q}'_{1}(p)=0$ equation. The determinant (\ref{tildetaun5}) with these $(p, q)$ parameter choices then becomes the $2\times 2$ block determinant (\ref{2x2blotausdet1}).

\end{document}